\documentclass[namedreferences,hyperref,optionalrh,solaromanenum]{spr-sola}
\usepackage{amsmath} 
\usepackage{graphicx}                    % For eps figures, newer & more powerfull
\usepackage{xcolor}                       % For color text: \color command
\usepackage{ulem}
\usepackage[printonlyused,nohyperlinks]{acronym}
\usepackage{textcomp}

\newcommand{\sunrise}{\textsc{Sunrise}}
\newcommand{\sunriseone}{\textsc{Sunrise~i}}
\newcommand{\sunrisetwo}{\textsc{Sunrise~ii}}
\newcommand{\sunriseonetwo}{\textsc{Sunrise~i/ii}}
\newcommand{\sunrisethree}{\mbox{\textsc{Sunrise~iii}}}

\newcommand{\spc}{\hbox{S$^3$PC}}

\newcommand{\degC}{$^\circ$C}
\newcommand{\arcsec}{"}

\begin{document}

\begin{article}

\begin{opening}

\title{The Sunrise Ultraviolet Spectropolarimeter and Imager: Instrument description}

% Main co-authors
\author[addressref={mps},corref,email={feller@mps.mpg.de}] {\inits{A.}\fnm{Alex}~\lnm{Feller}\orcid{0009-0009-4425-599X}}
\author[addressref={mps}]{\inits{A.}\fnm{Achim}~\lnm{Gandorfer}\orcid{0000-0002-9972-9840}}
\author[addressref={mps}]{\inits{B.}\fnm{Bianca}~\lnm{Grauf}}
\author[addressref={mps}]{\inits{J.}\fnm{Johannes}~\lnm{H\"olken}\orcid{0000-0001-6029-7529}}
\author[addressref={mps,mendoza}]{\inits{F.~A.}\fnm{Francisco~A.}~\lnm{Iglesias}\orcid{0000-0003-1409-1145}}
\author[addressref={mps,aalto}]{\inits{A.}\fnm{Andreas}~\lnm{Korpi-Lagg}\orcid{0000-0003-1459-7074}}
\author[addressref={mps}]{\inits{T.~L.}\fnm{Tino~L.}~\lnm{Riethmüller}\orcid{0000-0001-6317-4380}}
\author[addressref={mps}]{\inits{J.}\fnm{Jan}~\lnm{Staub}\orcid{0000-0001-9358-5834}} 
\author[addressref={mps}]{\inits{G.}\fnm{German}~\lnm{Fernandez-Rico}\orcid{0000-0002-4792-1144}}
\author[addressref={mps}]{\inits{J.~S.}\fnm{Juan Sebasti\'an}~\lnm{Castellanos Dur\'an}\orcid{0000-0003-4319-2009}}
\author[addressref={mps}]{\inits{S.~K.}\fnm{Sami K.}~\lnm{Solanki}\orcid{0000-0002-3418-8449}}
\author[addressref={mps}]{\inits{H.~N.}\fnm{H.~N.}~\lnm{Smitha}\orcid{0000-0003-3490-6532}}
\author[addressref={mps}]{\inits{K.}\fnm{Kamal}~\lnm{Sant}\orcid{0009-0003-3842-5557}}
\author[addressref={mps}]{\inits{P.}\fnm{Peter}~\lnm{Barthol}} 

% Other members of SUSI team, in alphabetical order
\author[addressref={mps}]{\inits{M.}\fnm{Montserrat}~\lnm{Bayon Laguna}}  
\author[addressref={mps}]{\inits{M.}\fnm{Melani}~\lnm{Bergmann}}
\author[addressref={mps}]{\inits{J.}\fnm{Jörg}~\lnm{Bischoff}}
\author[addressref={mps}]{\inits{J.}\fnm{Jan}~\lnm{Bochmann}}
\author[addressref={ist}]{\inits{S.}\fnm{Stefan}~\lnm{Bruns}}
\author[addressref={mps}]{\inits{W.}\fnm{Werner}~\lnm{Deutsch}}
\author[addressref={mps}]{\inits{M.}\fnm{Michel}~\lnm{Eberhardt}}
\author[addressref={mps}]{\inits{R.}\fnm{Rainer}~\lnm{Enge}}
\author[addressref={mps}]{\inits{S.}\fnm{Sam}~\lnm{Goodyear}}
\author[addressref={mps}]{\inits{K.}\fnm{Klaus}~\lnm{Heerlein}}
\author[addressref={mps}]{\inits{J.}\fnm{Jan}~\lnm{Heinrichs}}
\author[addressref={mps}]{\inits{D.}\fnm{Dennis}~\lnm{Hirche}}
\author[addressref={mps}]{\inits{S.}\fnm{Stefan}~\lnm{Meining}} 
\author[addressref={mps}]{\inits{R.}\fnm{Roland}~\lnm{Mende}}
\author[addressref={mps}]{\inits{S.}\fnm{Sabrina}~\lnm{Meyer}}
\author[addressref={mps}]{\inits{M.}\fnm{Maria}~\lnm{Mühlhaus}}
\author[addressref={mps}]{\inits{M.~F.}\fnm{Marc Ferenc}~\lnm{Müller}}
\author[addressref={mps}]{\inits{K.}\fnm{Markus}~\lnm{Monecke}} 
\author[addressref={mps}]{\inits{D.}\fnm{Dietmar}~\lnm{Oberdorfer}} 
\author[addressref={mps}]{\inits{I.}\fnm{Ioanna}~\lnm{Papagiannaki}}

\author[addressref={mps}]{\inits{S.}\fnm{Sandeep}~\lnm{Ramanath}} 
\author[addressref={ist}]{\inits{M.}\fnm{Michael}~\lnm{Vergöhl}}
\author[addressref={mps}]{\inits{D.}\fnm{Du\v{s}an}~\lnm{Vukadinovi\'{c}}\orcid{0000-0003-1971-5551}}
\author[addressref={mps}]{\inits{S.}\fnm{Stephan}~\lnm{Werner}}
\author[addressref={mps}]{\inits{K.}\fnm{Andreas}~\lnm{Zerr}}

% Lead-CoIs
\author[addressref={kis}]{\inits{T.}\fnm{Thomas}~\lnm{Berkefeld}}
\author[addressref={apl}]{\inits{P.}\fnm{Pietro}~\lnm{Bernasconi}\orcid{0000-0002-0787-8954}}
\author[addressref={naoj}]{\inits{Y.}\fnm{Yukio}~\lnm{Katsukawa}\orcid{0000-0002-5054-8782}}
\author[addressref={iaa,s3pc}]{\inits{J.~C.}\fnm{Jose Carlos}~\lnm{del Toro Iniesta}\orcid{0000-0002-3387-026X}}

% CoIs
\author[addressref={kis}]{\inits{A.}\fnm{Alexander}~\lnm{Bell}}
\author[addressref={apl}]{\inits{M.}\fnm{Michael}~\lnm{Carpenter}}
\author[addressref={inta,s3pc}]{\inits{A.}\fnm{Alberto}~\lnm{\'{A}lvarez Herrero}\orcid{0000-0001-9228-3412}}
\author[addressref={naoj}]{\inits{M.}\fnm{Masahito}~\lnm{Kubo}\orcid{0000-0001-5616-2808}}
\author[addressref={iac}]{\inits{V.}\fnm{Valent\'{i}n}~\lnm{Mart\'{i}nez Pillet}\orcid{0000-0001-7764-6895}}
\author[addressref={iaa,s3pc}]{\inits{D.}\fnm{David}~\lnm{Orozco Su\'{a}rez}\orcid{0000-0001-8829-1938}}
\address[id={mps}]{Max-Planck-Institut für Sonnensystemforschung, Justus-von-Liebig-Weg 3, 37077 Göttingen, Germany}
\address[id={mendoza}]{Grupo de Estudios en Heliofísica de Mendoza, CONICET, Universidad de Mendoza, Boulogne sur Mer 683, 5500 Mendoza, Argentina}
\address[id={naoj}]{National Astronomical Observatory of Japan, 2-21-1 Osawa, Mitaka, Tokyo 181-8588, Japan}
\address[id={kis}]{Institut für Sonnenphysik (KIS), Georges-Köhler-Allee 401a, 79110 Freiburg, Germany}
\address[id={iaa}]{Instituto de Astrofísica de Andalucía, Glorieta de la Astronomía s/n, 18008 Granada, Spain}
\address[id={apl}]{Johns Hopkins Applied Physics Laboratory, 11100 Johns Hopkins Road, Laurel, Maryland, USA}
\address[id={inta}]{Instituto Nacional de Técnica Aeroespacial, Carretera de Ajalvir km4, 28850 Torrejón de Ardoz, Madrid, Spain}
\address[id={ist}]{Fraunhofer-Institut für Schicht- und Oberflächentechnik, Riedenkamp 2, 38108 Braunschweig, Germany}
\address[id={iac}]{Instituto de Astrof\'{\i}sica de Canarias, V\'{\i}a L\'actea, s/n, E-38205 La Laguna, Spain}
\address[id={aalto}]{Aalto University, Department of Computer Science, Konemiehentie 2, 02150 Espoo, Finland}
\address[id={s3pc}]{Spanish Space Solar Physics Consortium (\href{https://s3pc.es}{\spc})}

\begin{abstract}
 The third science flight of the balloon-borne solar observatory \sunrise{} carries three entirely new post-focus science  instruments with spectropolarimetric capabilities, concurrently covering an extended spectral range from the near ultraviolet to the near infrared. Sampling a larger height range, from the low photosphere to the chromosphere, with the sub-arcsecond resolution provided by the 1-m \sunrise{} telescope, is key in understanding critical small-scale phenomena which energetically couple different layers of the solar atmosphere. The \ac{susi} operates between 309\,nm and 417\,nm. A key feature of \ac{susi} is its capability to record up to several hundred spectral lines simultaneously without the harmful effects of the Earth's atmosphere. The rich \ac{susi} spectra can be exploited in terms of many-line inversions. Another important innovation of the instrument is the synchronized 2D context imaging which allows to numerically correct the spectrograph scans for residual optical aberrations. In this work we describe the main design aspects of \ac{susi}, the instrument characterization and testing, and finally its operation, expected performance and data products.      
\end{abstract}


\end{opening}

\acresetall

\section{Introduction}\label{sec_intro}

\sunrise{} is a unique solar observatory: a 1-m optical telescope and scientific instrumentation are carried by a zero-pressure balloon into the stratosphere on a multi-day flight \cite[]{Lagg25}. Operating in the stratosphere at an altitude of about 37 km allows \sunrise{} to access the \ac{nuv} part of the solar spectrum, without the adverse effects of atmospheric seeing. The primary mission of \sunrise{} is the investigation of physical processes in the solar atmosphere involving magnetic fields and plasma flows on small spatial scales. The first scientific flight \citep[\sunriseone, see][]{Solanki10, Barthol11} took place in June 2009 during a period of low solar activity, while the second flight \citep[\sunrisetwo,][]{Solanki17} in June 2013 allowed \sunrise{} to study a more active Sun. These missions have been highly successful, yielding a wealth of scientific findings. The third science flight (\sunrisethree) took place between 10--16 July 2024. The low-resolution telemetry data received during flight operation suggest an excellent performance of the observatory, including all science instruments. An in-depth performance assessment will be available after the analysis of the full-resolution data from the recovered onboard data storage. At the time of writing, this process has just begun. 

The scientific payload of \sunriseonetwo{} had limited capabilities for studying the chromosphere. In addition to a new gondola \citep{Bernasconi25} and an improved image stabilization system \citep{Berkefeld25}, 
\sunrisethree{} carries three new scientific instruments, all with spectropolarimetric capabilities, designed to sample a larger height range from the lower photosphere to the chromosphere. One of the instruments onboard \sunrisethree~is the \ac{susi}, which covers the wavelength range from 309\,nm to 417\,nm, thus extending the observational capabilities of \sunrisethree{} into the \ac{nuv}. 
The other two instruments are the \aclu{scip} \cite[\acsu{scip}:][]{Katsukawa25} and the \aclu{tumag} \cite[\acsu{tumag}:][]{delToroIniesta25}.

Here we provide a general overview of \ac{susi}, covering instrument design, characterization, calibration and operation, as well as the data products and expected performance.

\section{Scientific Motivation and Design Drivers}
\label{sec_reqs}

The spectrum across most of \ac{susi}'s wavelength range has largely remained underutilized in solar physics so far. It includes thousands of spectral lines among which are more than 100 chromospheric lines, covering a height range of more than 1300\,km in the solar atmosphere, seamlessly from deep photospheric to mid-upper chromospheric layers. Although the solar \ac{nuv} intensity spectrum has been well observed, albeit at low spatial resolution, the polarized spectrum is poorly known. A challenge in observing the solar spectrum in the \ac{nuv}, in particular with the increased signal-to-noise ratio required for spectropolarimetry, is the extinction by the Earth's atmosphere. From experience with the \aclu{sufi} \cite[\acsu{sufi}:][]{Gandorfer11}, the predecessor instrument flown on \sunriseonetwo, \ac{susi} observations are expected to be comparable to observations from space, in terms of the incoming photon flux, for wavelengths longer than roughly 300\,nm. 

\subsection{Previous Work}

\paragraph*{Low spatial resolution \ac{nuv} spectra with no polarization information}
Pioneering observations of the solar intensity spectrum, including the \ac{nuv} range, have been carried out at the International Scientific Station of the Jungfraujoch, located in the Swiss alps at an altitude of 3580 meters \citep[cf. the review by ][]{Delbouille95}. The measurements are published in the atlas by \cite{Delbouille73}, which covers the spectral range between 300\,nm and 1000\,nm. The resolving power of the measurements reaches up to $1.25 \cdot 10^6$ at its blue end. The work is widely accepted as a standard high resolution solar disk center mean spectrum, in particular  in the \ac{nuv}. The Jungfraujoch atlas data do not provide any information about spatial intensity distributions. Instead the measurements are averaged along the spectrograph slit positioned around disk center. 

Another remarkable reference work, extending into the \ac{nuv} as well, are the spectra obtained by Brault and co-workers with the \ac{fts} at the McMath-Pierce solar telescope of the Kitt Peak observatory, at an altitude of about 2100 m. The corresponding spectral atlas, denoted as Kitt Peak atlas in the following, was prepared in the 1980s, and later published in digital form \citep{Neckel99}. The atlas consists of disk-center and disk-averaged intensity spectra, covering the range between 329\,nm and 1251\,nm\footnote{This is the common wavelength range covered by both datasets. The disk-averaged data reach even further into the UV to 296\,nm, cf. \cite{kurucz84}.}. The spectral resolution is of order $3.5 \cdot 10^5$. Like the Jungfraujoch atlas, the Kitt Peak atlas does not offer spatially resolved information. 

\paragraph*{Low spatial resolution observations of scattering polarization}
A systematic recording of the spectrum of scattering polarization at the solar limb, extending into the \ac{nuv}, has been conducted with the UV-sensitive version of the ZIMPOL II polarimeter \citep{Gandorfer04}. The third volume of the second solar spectrum atlas by \cite{Gandorfer05}, recorded at the vertical grating spectrograph of the McMath-Pierce solar telescope, covers the range between 316\,nm and 392\,nm. The measurements show an unparalleled consistent polarimetric noise level well below 0.1\%, and high spectral resolution ($R = 3.3 - 3.7 \cdot 10^6$, depending on wavelength). The increased polarimetric sensitivity is required to detect the faint scattering polarization signatures. While the Hanle effect acts mainly around the dark line cores, other effects related to scattering polarization, like partial frequency redistribution and magneto-optical effects, can be observed in the line wings where the noise level is lower. Like the other reference works mentioned above, the data of the second solar spectrum atlas does not provide any information on spatial 
variations, being restricted to a single spatial pixel. In order to reach the high polarimetric sensitivity, combined with high spectral resolution, averaging along the spectrograph slit was unavoidable, despite the high photometric efficiency of the observatory when compared to other ground-based facilities.

\subsection{Unlocking the Potential of Spatially Resolved NUV Spectropolarimetry}

Thanks to the significantly increased photon efficiency compared to ground-based instrumentation, \ac{susi} can record spatially resolved \ac{nuv} spectra at sub-arcsecond resolution, including the full polarization information, and at a polarimetric noise level compatible with a large number of science cases (see Section~\ref{sec_characterization} for a detailed performance discussion). The spectral field-of-view of \ac{susi}'s spectrograph cameras encompasses hundreds of spectral lines, all observed in strict simultaneity.

More than 60 observing ideas collected by the \sunrisethree{} science working group formed the basis for the planning of the observation program. More than 80\% of these proposals require \ac{susi} data. We highlight a few of these ideas, where \ac{susi} is a key instrument of the observation.

\paragraph*{Many-line inversions}
A novel many-line inversion technique combines the spectropolarimetric information from individual lines within the \ac{susi} spectral \ac{fov}. \cite{Riethmueller19} first explored this technique with the help of numerical simulations. They have shown that compared to traditional inversions of few spectral lines in the visible range, the many-line approach in the \ac{nuv} significantly enhances the sensitivity, in particular to temperature and \ac{los} velocity. Further, despite the weaker Zeeman effect in individual lines, the magnetic-field sensitivity reaches levels comparable to classical inversion results in the visible range. Additionally, the accuracy of inferred solar atmospheric parameters in many-line inversions is less affected by uncertainties in atomic transitions associated with individual spectral lines. The many-line approach may even aid in refining uncertain transition properties of individual lines, a strategy vital for solar atmospheric diagnostics, where many lines remain poorly identified or frequently blended \citep{vukadinovic24}. By consistently combining the information of many lines formed at different heights, the height-dependence of solar atmospheric parameters is better constrained as well. Further in-depth explorations of the many-line inversion technique in different regions of the \ac{susi} spectral range are currently underway and the theoretical predictions will be discussed in later publications. 

\paragraph*{Unresolved magnetic fields using Hanle diagnostics}
Besides Zeeman diagnostics, \ac{susi} is capable of exploring the spatial distribution of scattering polarization signatures. The \ac{susi} spectral range is particularly well suited for such observations as the scattering amplitudes generally increase with shorter wavelengths. Unlike the Zeeman effect, the Hanle effect does not suffer from potential signal cancellation in the presence of unresolved magnetic fields, and can thus serve as a complementary diagnostic tool for solar magnetism on sub-arcsecond spatial scales. So far, quantitative Hanle diagnostics have been frequently limited due to the combined effects of collisional and Hanle depolarization, which are difficult to disentangle. The wealth of new \ac{susi} data, covering many lines with different collisional cross-sections and Hanle sensitivities, allow to improve on the diagnostic potential of scattering effects in the solar atmosphere in the presence of small-scale magnetic fields. 

\paragraph*{Full spectral scan} \ac{susi} is the first instrument to observe the \ac{nuv} spectral range from 309 to 417\,nm  with high spatial resolution and full polarization information. The detailed characterization of this spectral region, including the temperature and magnetic field dependence retrieved from the high-resolution spectra, has a huge exploratory potential and provides unprecedented information about the spectral line formation. 

\paragraph*{Wave propagation and micro flares in the lower solar atmosphere}
The unique combination of temporal stability, high spatial resolution and height resolution makes \ac{susi} the ideal instrument to trace dynamic effects through the various layers of the solar atmosphere. By performing sit \& stare observations, or by repeatedly scanning small regions of 0.5 to 1 arcsecond with high cadence, we will get the temporal evolution of atmospheric parameters across a large frequency range. \ac{susi} will thus see the propagation of acoustic and magnetic waves through the photosphere, and capture the details of small-scale reconnection events.

\paragraph*{Combining \acs{nuv} and \acs{nir} observations} Only the seeing-free and atmospheric refraction-free environment in the stratosphere allows for the strictly simultaneous observation of the chromosphere in the Calcium lines in the \ac{nuv} (Ca\, {\sc ii} H\&K, 393 - 397 \,nm) and the \ac{nir} (Ca\,{\sc ii} IR, 854\,nm). The slits of both instruments, \ac{susi} and \ac{scip} can be closely aligned (tolerance 0.3") despite the large wavelength difference, and can synchronously scan the solar surface. The Ca\,{\sc ii} lines in both spectral regions are used in today's \ac{nlte} inversions to determine the physical conditions in the chromosphere. So far, only one of these lines has been observed at a time. However, the \ac{nlte} inversions of just one of these lines requires the computation of all the five Ca\,{\sc ii} lines, the H\&K and the IR triplet, as their energy levels form a five-level quintessential Ca\,{\sc ii} model atom used in inversions. The combination of \ac{susi} and \ac{scip} will for the first time deliver this information in the Ca\,{\sc ii} H\&K and Ca\,{\sc ii} 854 nm lines simultaneously, enhancing the accuracy of the inferred chromospheric diagnostics.

\section{Instrument Design}\label{sec_design}

The design rationale for \ac{susi} is driven by the specific demands of a remote-controlled stratospheric flight, as well as by stringent schedule and resource limitations. The financial budget of \sunrise{} is at least an order of magnitude lower than for a comparable space mission. To comply with these constraints, a conservative hardware strategy has been adopted, based on well-established solar instrumentation concepts, and prioritizing off-the-shelf commercial components wherever possible. The bulk of in-house engineering efforts has been dedicated to developing custom \ac{susi} cameras, a custom grating mechanism, instrument control hardware and software, as well as to system integration, calibration and testing. 

The general project philosophy was to put reliability over optimum performance. More innovative approaches are reserved for the domain of post-facto data processing and analysis, with a particular focus on novel concepts like image restoration of \ac{sp} scans and many-line inversions.  

\begin{figure}
\centerline{\includegraphics[width=\textwidth]{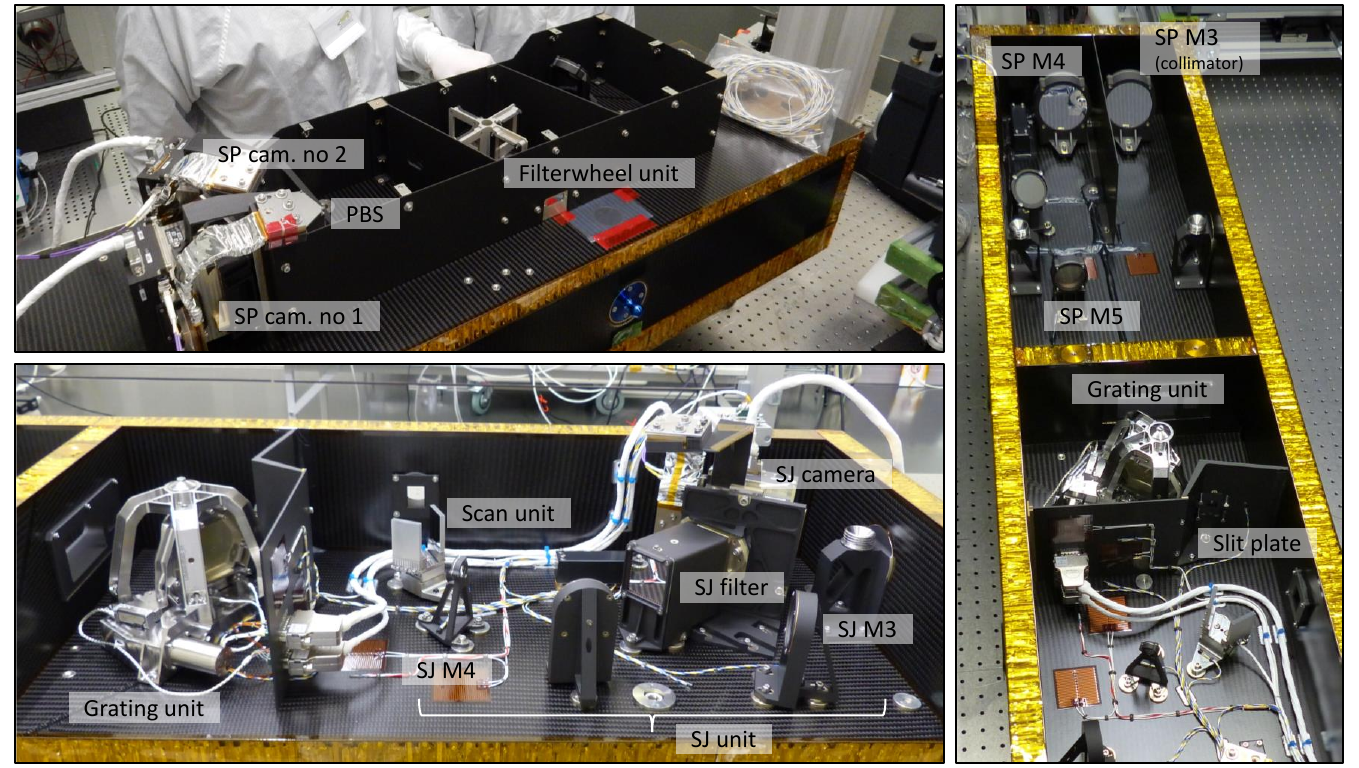}}
\small
\caption{\ac{susi} optics unit during assembly. \textit{Lower left and right panels:} open lower deck with the scan unit, the \acs{sp} optics, and the \acs{sj} imager. \textit{Upper left panel:} upper deck on top of the now closed lower deck, with the filterwheel unit, the \acs{pbs}, and the \acs{sp} cameras. Major optical components and subsystems are labelled.}
\label{fig_susi}
\end{figure}

Figure~\ref{fig_susi} shows some interior views of the \ac{susi} optics unit, recorded during the assembly phase. \ac{susi} can be split into four functional units. (1) The scan unit shifts the \ac{2d} solar image across the \ac{sp} entrance slit. (2) The \ac{sp} unit, with its diffraction grating and \acp{osf}, generates a \ac{2d} spectrum image on each of the two \ac{sp} cameras, with the wavelength information projected on one image dimension and the spatial information along the slit projected on the second image dimension. (3) The \ac{pmu}, consists of the rotating waveplate and the polarizing beamsplitter. The \ac{pmu} encodes the polarization state of the incoming light into a periodic intensity modulation which is recorded by the \ac{sp} cameras in phase to the rotating waveplate. (4) The \ac{sj} unit is fed by the beam reflected off the slit-plate. The \ac{sj} camera provides \ac{2d} context imaging around the \ac{sp} slit. Like its predecessor instrument \ac{sufi}, the \ac{sj} unit also includes a \ac{pid} which allows for continuous monitoring of the \ac{psf}, and thus for a post-facto image restoration of the \ac{sp} scans, following the technique developed by \citet{vanNoort17}.         

In the following sections we describe the instrument design, including optics for imaging, spectrum and polarization analysis, mechanisms, cameras, structural and environmental aspects, as well as instrument control. Most subsystems of \ac{susi} have been developed at the \ac{mps}, in particular the cameras, the opto-mechanics, the scan, grating and filterwheel units, some thermal and structural components, the instrument electronics including harness, as well as the instrument software. 

\subsection{Optical Design}
\label{sec_optdesign}

\begin{figure}
\centerline{\includegraphics[width=\textwidth]{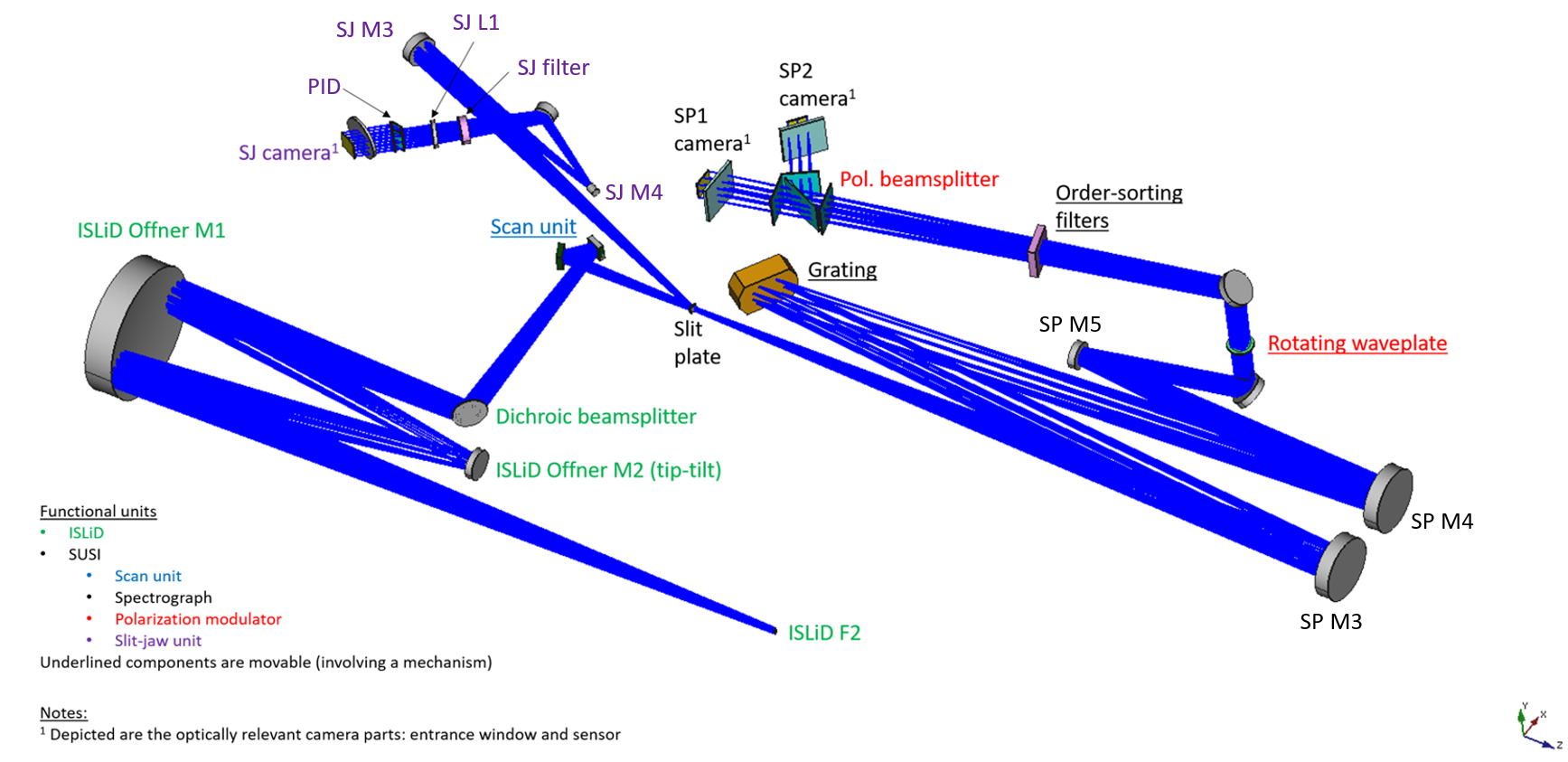}}
\small
\caption{Optical layout of \ac{susi} and of the relevant transfer optics within the \ac{islid}.}
\label{fig_optdesign_layout}
\end{figure}

\begin{figure}
\centerline{\includegraphics[width=\textwidth]{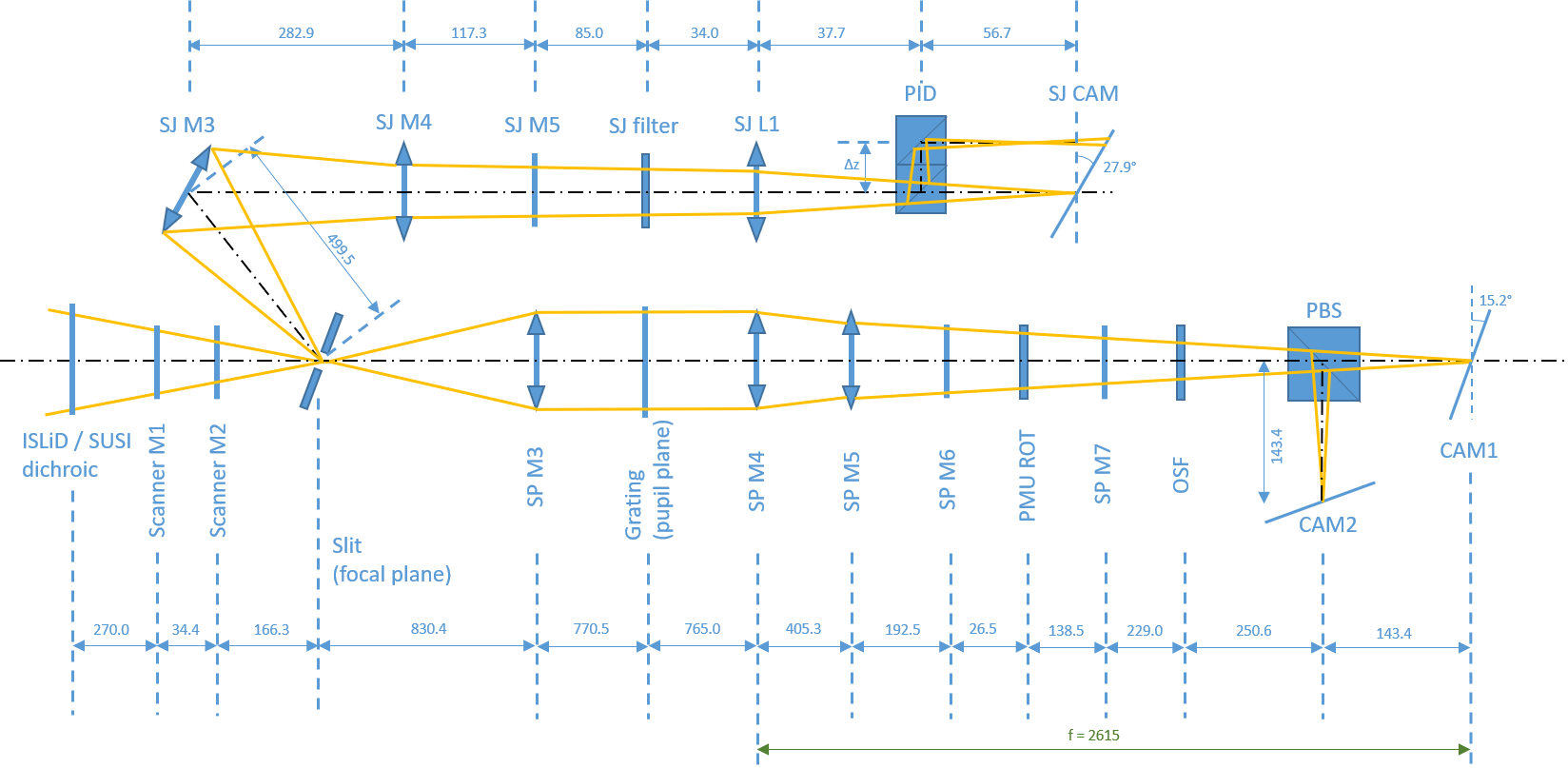}}
\small
\caption{Sketch of the \ac{susi} beam path (not to scale). The \textit{blue elements} with \textit{double arrows} denote powered optical components (curved mirrors or lenses). The \textit{rectangular blue elements} denote components with flat optical surfaces (e.g. folding mirrors, filters, camera detectors). The image beam path is depicted in \textit{yellow}. Distances are given in units of mm.}
\label{fig_optdesign_sketch}
\end{figure}

In this section we describe components and design aspects whose main purpose is optical imaging and guiding of the beam path within the \ac{susi} \ac{o-unit}. The optical layout of \ac{susi}, and part of the upstream beam path inside the \ac{islid} unit feeding \ac{susi} \citep{Lagg25}, is shown in Figures~\ref{fig_optdesign_layout}~and~\ref{fig_optdesign_sketch}. 

\paragraph*{\ac{islid} transfer optics feeding \ac{susi}} The first Offner system of the \ac{islid} unit performs a 1:1 re-imaging of the telescope F2 focal plane onto the \ac{susi} \ac{sp} slit plate. The F/24.2 white-light beam of the \sunrise{} telescope is reflected off a dichroic element at a 90° angle and fed into \ac{susi}. For shorter wavelengths, the dichroic element acts as a mirror with a high-reflectivity band below the cut-off of about 470\,nm.\footnote{The transition between high and low reflectivity spans the region between 450\,nm and 482\,nm. These wavelengths correspond to 90\% and 10\% of the maximum reflectivity respectively. The point of 50\% reflectivity, which is reached at about 470\,nm, is used as cut-off reference wavelength.} The longer wavelengths are transmitted by the dichroic element and further distributed within the \ac{islid} unit to the other optical units inside the \ac{pfi}. 

\paragraph*{Scan unit} The scan unit consists of two folding mirrors, mounted on a commercial piezo-driven translation mechanism from SmarAct GmbH. The 45° configuration of the mirrors allows to shift the solar image perpendicularly to the mechanically static \ac{sp} slit via a simple linear motion of the scan unit and without altering the focal distance. Projected into the Sky, the scan direction corresponds to the azimuth direction of the telescope pointing system, i.e. the direction parallel to the Earth's horizon. Due to the field rotation of the altitude-azimuth telescope pointing the scan direction in a heliographic reference system is time dependent. The unit covers a scan range up to $\pm$ 35\arcsec{} around the F2 center position. The scanning movement can be continuous with constant speed, or stepped, depending on the requirements of a given observing program. The scan speed can be chosen between 0.01\arcsec{} and 1.4\arcsec{} per second, again depending on the observing requirements. The deviation of the scan movement from an ideal linear path is below $5 \cdot 10^{-3}$\arcsec{} \ac{rms}, which is well below the optical resolution limit.

\paragraph*{Slit plate} The \ac{sp} slit plate is a planar fused-silica glass plate. The front side of the plate (i.e. the side facing the incoming light) has a reflective aluminum coating. A central rectangular uncoated area of 7 µm width (0.06\arcsec{} on the sky) represents the slit. A mechanical field stop in front of the slit plate limits the slit length to 7.0 mm (60\arcsec{}). The axis of the slit is aligned with the grooves of the diffraction grating. The slit width covers about two camera pixels. The selected width is the result of a trade-off between spectral resolving power, signal-to-noise ratio and diffraction effects. The slit plate is slightly rotated around the slit axis, with the surface normal being at an angle of 11° with respect to the optical axis of the incoming beam path. A \ac{2d} context image of dimensions 22\arcsec{} by 60\arcsec{}, centered around the slit and delimited by the field stop, is back-reflected sidewards to the \ac{sj} unit.  

\paragraph*{\ac{sp} optics} The \ac{sp} optics follows the basic concept of a Czerny-Turner configuration. It consists of three powered mirrors and the grating. The \ac{sp} accepts a larger focal ratio (F/16) than the incoming beam, to limit efficiency losses due to diffraction effects at the slit. The powered mirrors all have a spherical shape. The collimator mirror (\ac{sp} M3) has a focal length of 830 mm and sends the image plane to infinity while projecting a pupil image of 35 mm diameter onto the grating. 

The grating is mounted on a rotating mechanism which allows tuning the angle of incidence within the range $29.2 \pm 4.3$°. This angle range encompasses, with some 0.1° margin at each end, the entire spectral working range of \ac{susi} (see Section~\ref{sec_grating} for more details on the grating properties). The axis of rotation of the grating is aligned with the grating grooves and coincides with the reflective plane of the grating. This minimizes changes in the imaging properties while tuning across the spectral working range. During instrument characterization (see also Section~\ref{sec_optchar}) no significant changes in imaging quality vs. wavelength have been found. Optical alignment errors have been reduced such that residual shifts of the spectrum perpendicular to the dispersion direction do not cause any wavelength-dependent clipping of the 60\arcsec{} spatial field. 

The \ac{sp} camera optics consists of a concave mirror (\ac{sp} M4) and a convex mirror (\ac{sp} M5) in a Schiefspiegler configuration with an effective focal length of 2615 mm. The camera optics re-images the slit, now dispersed into a spectrum, onto the two \ac{sp} cameras. Together with the focal length of the collimator optics this results in a magnification of 3.15:1. The magnification factor ensures a sampling close to the spatial diffraction limit of the telescope at the shortest wavelength of the \ac{susi} working range. At $\lambda$ = 309\,nm, the diffraction cut-off scale $\lambda F$, where $F$ = 76.4 is the effective focal ratio on the \ac{sp} cameras, is slightly oversampled with 2.14 pixels. The absolute focal lengths of the \ac{sp} optics are set by the grating size and determine the spectral resolution (cf. Section~\ref{sec_grating}).

\begin{figure}
\centerline{\includegraphics[width=\textwidth]{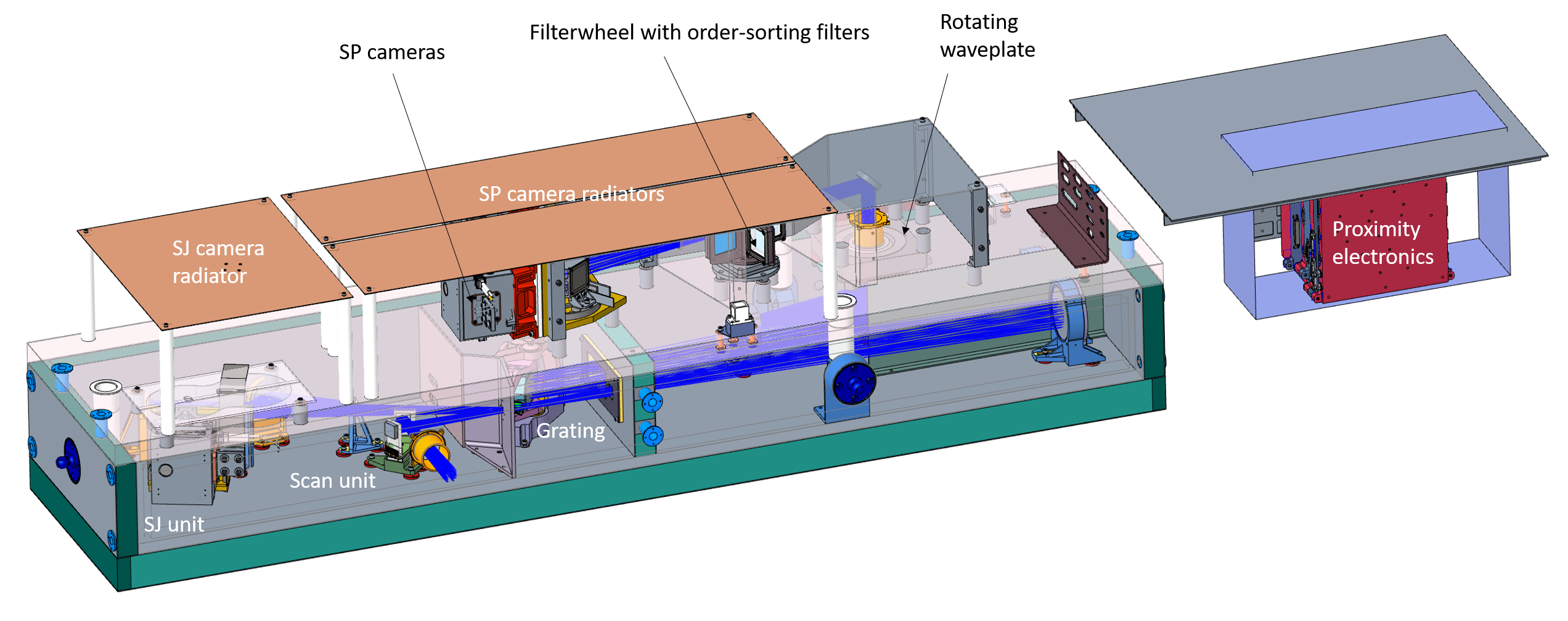}}
\small
\caption{Mechanical layout of \ac{susi}, with the \ac{o-unit} on the left and the adjacent proximity electronics. To enhance the visibility of key components, some outer structural elements, as well as harness and insulation are hidden from the CAD view.}
\label{fig_mech_layout}
\end{figure}

The remaining \ac{sp} imaging optics consists of two folding mirrors (\ac{sp} M6, M7), directing the beam to the upper level of the \ac{susi} optical unit (cf. Section~\ref{sec_mechdesign}, Figure~\ref{fig_mech_layout}) with the \ac{pmu} and \ac{osf} units and the \ac{sp} camera unit. These units as well as the grating unit are described in more detail below. 

\paragraph*{Slit-jaw optics}

The main role of the \ac{sj} unit is context imaging and wavefront sensing. A \ac{2d} image is reflected from the slit plate towards the \ac{sj} unit. The image is magnified by a factor 3.14:1 and re-imaged onto the \ac{sj} camera.  

Similar to the \ac{sp} camera optics, the \ac{sj} imaging optics consists of a Schiefspiegler configuration with a concave mirror (\ac{sj} M3) and a convex mirror (\ac{sj} M4). An additional lens (\ac{sj} L1) completes the imaging optics. The lens reduces the curvature of the focal plane as well as distortions. The slight difference in the magnification factor between the \ac{sp} and \ac{sj} optics is due to engineering constraints and can be compensated by a numerical re-scaling (interpolation) of the images during data reduction.    

The spectral range transmitted to the \ac{sj} camera is limited by a bandpass filter. The bandpass is centered on 325.0\,nm and has a \ac{fwhm} of 0.9\,nm. In combination with the dichroic element of the \ac{islid} transfer optics, the residual out-of-band transmission is negligible. The out-of-band blocking has been verified during instrument characterization (cf. Section~\ref{sec_optchar}). The selected spectral window essentially represents the solar \ac{nuv} continuum formed in the low photosphere. This facilitates the co-alignment with other instruments that record continuum radiation, both within \sunrise{} and as part of other observatories. The integrated spectrum across the filter bandpass is not dominated by any particular spectral lines. In the trade-off process, priority was given to a central wavelength in the lower part of the \ac{susi} spectral working range, which is expected to be beneficial for the restoration quality of the spectrographic scans. 

As already described in the introduction to this section, the \ac{pid} allows for continuous \ac{psf} monitoring and thus provides the required information for the image restoration of the \ac{sp} scans. The \ac{susi} specific restoration technique, an adaptation of the initial ground-based application \citep{vanNoort17}, is currently under development and will be described in a future publication.

The optical design concept of a \ac{pid} is described in detail in \cite{Gandorfer11}. The \ac{susi} \ac{pid} projects two images of the identical scene, delimited by the slit-plate field stop, next to each other on the \ac{sj} camera detector. For one image, the focal plane is conjugated to the focal plane on the slit plate. For the other image, the \ac{pid} introduces a fixed amount of defocus (0.95~$\lambda$ peak-to-valley). This corresponds to a nominal focus shift of the \ac{susi} \ac{o-unit} with respect to F2 of about 1.4~mm, which has been verified during instrument alignment (cf. Section~\ref{sec_optchar}, Figure~\ref{fig_contrast_curve}). Based on experience, in particular from the predecessor instrument \ac{sufi}, a defocus in the range between half a wave and one wave allows for a reliable application of the phase-diversity technique \citep[see e.g.][and references therein]{loefdahl94}, which is part of the \ac{momfbd} code \citep{vanNoort05} used for wavefront sensing. In addition to the \ac{pid} based continuous monitoring of optical aberrations, occasional extended phase-diversity measurements are foreseen during flight. During these measurements the telescope M2 mirror is moved in the axial direction to five different positions covering a focus range of about $\pm 1\, \lambda$. Based on the wavefront sensing experiences with the \ac{sophi} \citep{Bailen24}, these measurements will allow to further constrain the different aberration terms.

\paragraph*{Tilted focal planes}

The Schiefspiegler design of the \ac{sp} and \ac{sj} optics offers an excellent compensation of optical aberrations with simple spherical mirrors, which is a significant advantage in terms of optics manufacturing and alignment. However the design leads to a tilted focal plane. To match the focal plane, the \ac{sp} and \ac{sj} cameras have to be tilted by 15.2° and 27.9° around the Y-axis (in the \sunrise{} coordinate system) to match the focal plane. This results in a foreshortening effect in one image dimension (0.96 for the \ac{sp} cameras along the spectral dimension, and 0.88 for the \ac{sj} camera along the short field dimension). The foreshortening in the spectral dimension has no consequences in terms of data reduction, and will simply be taken into account in the spectral calibration. In the trade-off process for the \ac{sj} optics, priority was given to a simple and efficient optical design, while accepting the disadvantages of different resolutions in both image dimensions and the need for numerical scale correction.    

\subsection{Mechanical and Thermal Design}
\label{sec_mechdesign}

\paragraph*{Mechanical structure}

As part of the optimization process to achieve a very compact instrument design a ``double-decker" concept has been selected for the instrument structure, composed of stiff \ac{cfrp} baseplates manufactured by Schütze GmbH: 
\begin{enumerate}
    \item The ``lower deck" hosts the \ac{sp} optics and the slit-jaw imager.
    \item The ``upper deck" hosts the polarization modulation unit, the order sorting pre-filters and the \ac{sp} cameras.
\end{enumerate}

The allocation of the functional units to the instrument ``decks"  follows the optical design and its alignment and integration concept, allowing for a sequential integration that preserves the positional accuracy of the components while assembling the two parts.
The two decks are joined together through three vertical frames and two sidewalls, further improving the overall stiffness of the instrument structure.

\paragraph*{Thermal aspects}

The main goal of the thermal design of the \ac{susi} \ac{o-unit} is to guarantee the thermal stability of the camera detectors during data acquisition. Also, the temperatures of the components inside the double deck structure must be kept close to room temperature during operation.

Like the other scientific instruments on board \sunrisethree{}, the \ac{susi} \ac{o-unit} is located inside the \ac{pfi} structure. The \ac{pfi} is covered with \ac{lldpe}, which is mostly transparent to infrared and visible radiation, allowing the radiators to view the cold sky, but also protecting the units inside the \ac{pfi} against the cold tropopause during the ascent phase. 

The \ac{susi} double deck structure is covered with Styrofoam pieces, wrapped in \ac{vda} coated Mylar\texttrademark, to maximize the insulation of the unit from the environment. The only exposed surfaces on top of the unit are the radiators connected to the camera detectors through graphite thermal straps. The thermal stability of the detectors is ensured by heaters located on the cameras' cold fingers. 

The \ac{susi} \ac{o-unit} is also equipped with heaters. A set of non-operational heaters is powered autonomously when the temperature falls below 0\degC. In addition, several operational heaters are manually controlled to compensate for the varying thermal conditions during flight, mainly due to the changes in the Sun elevation.  

For more details about the thermal design of \ac{susi} \ac{o-unit}, and the thermal analysis results, the reader is referred to \citet{FernandezSoler20}.

\subsection{Grating and Order-sorting Filters}
\label{sec_grating}

\paragraph*{Grating}
The grating is a commercial plane-ruled diffraction grating from Newport Inc., with a groove density of 600 lines per mm and a blaze angle of 33.1°. The spread, i.e. the difference between the angle of diffraction and the angle of incidence is fixed by the optical design to 8.2°. The grating is thus operated around its blaze angle in the diffraction orders four to six, depending on wavelength. The relatively low diffraction orders reduce the number of required \acp{osf}, and allow for larger filter bandwidths which significantly relaxes the requirements on the \ac{osf} unit.

\begin{figure}
\centerline{\includegraphics[width=0.7\textwidth]{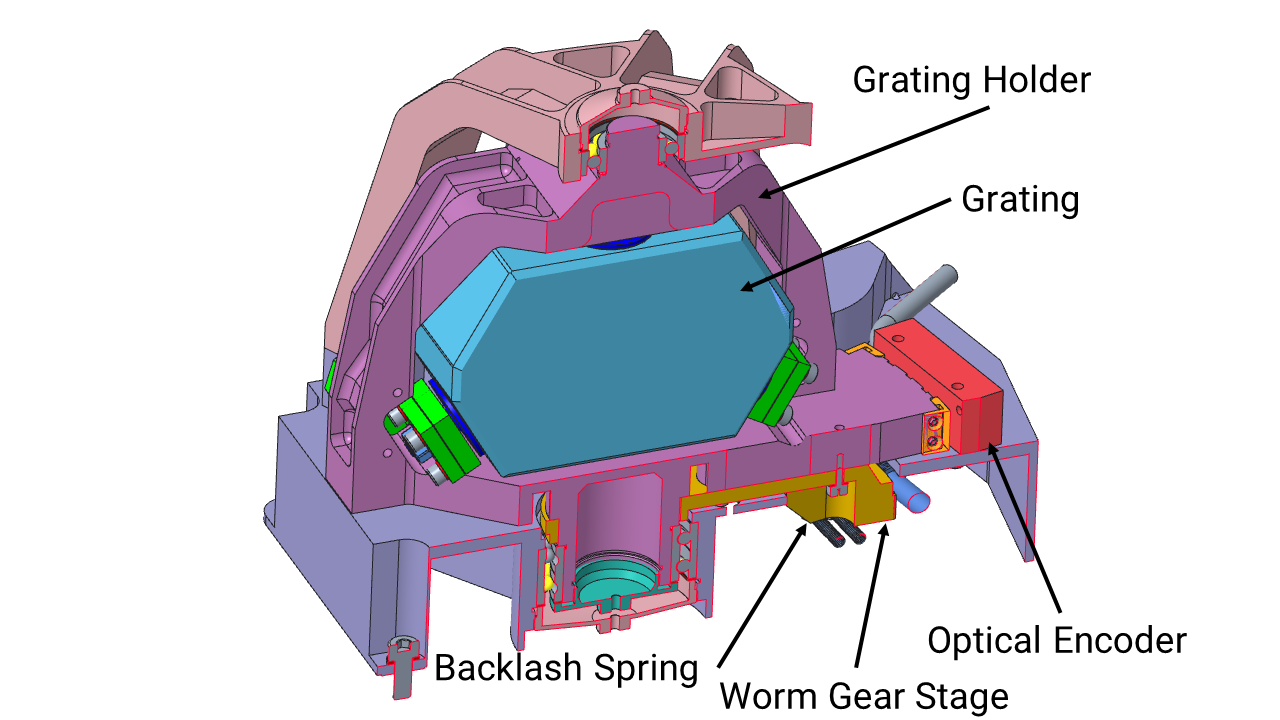}}
\small
\caption{Sectional cut through the \ac{susi} grating mechanism, showing its major components: grating, worm gear stage, optical position encoder. Here, the stepper motor is hidden by the mechanism's structure.}
\label{fig_mech_grating}
\end{figure}

The grating mechanism has been exclusively designed and built at the \ac{mps}. Figure~\ref{fig_mech_grating} shows a sectional cut through the grating mechanism. Conceptually, the mechanism is split into two parts: The ``clean" optical part with the grating and the 
``dirty" part with the moving part and the electronics. The latter is enclosed in the mechanism's mounting box, in order to prevent potential contamination of the sensitive \ac{uv} optics.  

Functionally, the mechanism reaches a positioning accuracy of 3.5\arcsec, which corresponds to 9.4 pm or 8.8 pixels at 370\,nm. Note that the grating positioning accuracy in wavelength units is wavelength dependent, cf. Figure~\ref{fig_specres}. The high accuracy is achieved through a worm gear stage with a reduction ratio of 1:2000 and a stepper motor with a resolution of 200 steps per rotation. The worm gear stage is spring loaded in order to reduce the backlash introduced into the drive train. An optical position encoder is used to measure the absolute position of the grating with an angular accuracy of less than 1\arcsec. The mechanism control software ensures that a commanded grating angle is always approached from the same direction and settles on the final position via a series of iterations based on encoder feedback. The positioning speed including fine-tuning allows centering on any wavelength with the above-mentioned accuracy and within two minutes.

\paragraph*{\aclp{osf}}

\begin{figure}
\centerline{\includegraphics[width=\textwidth]{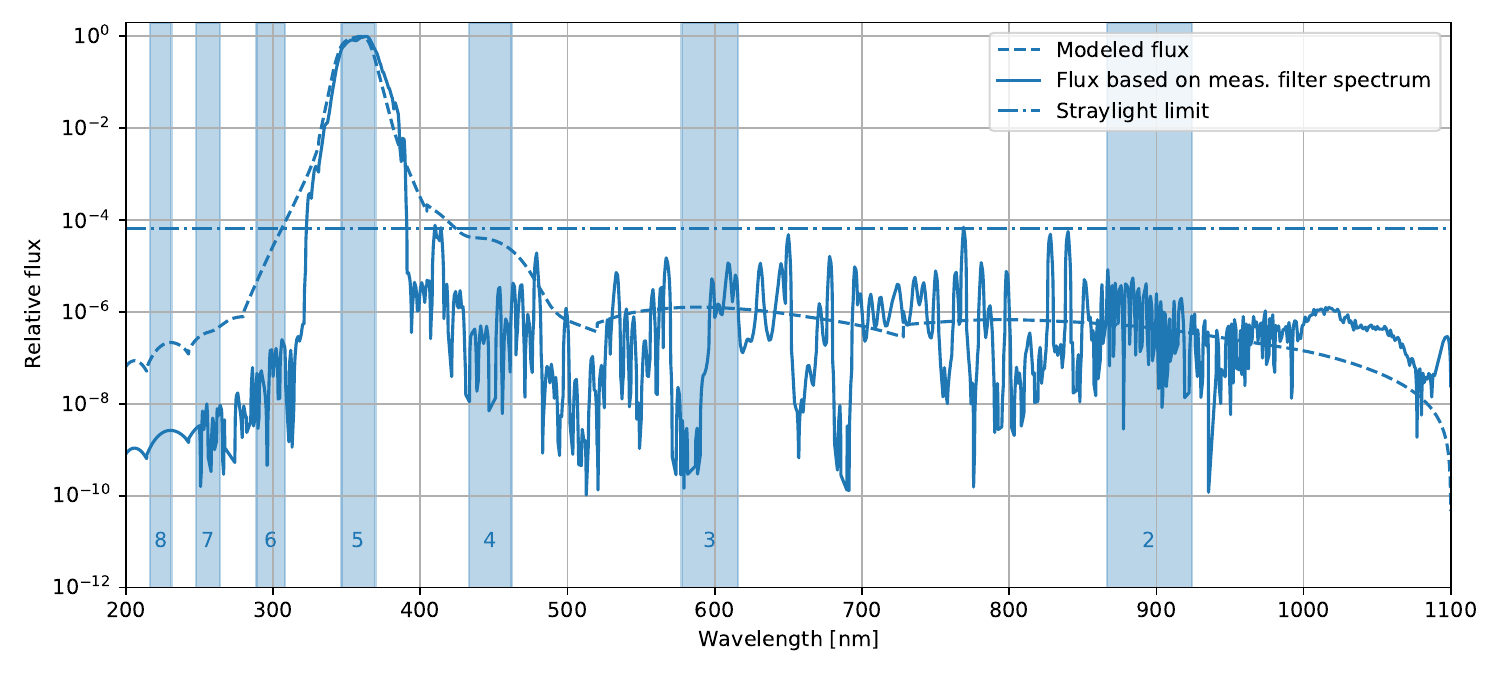}}
\small
\caption{Expected relative solar flux, corresponding to an \ac{sp} configuration with \ac{osf} no.~2 in the beam path. The wavelength range spans the relevant camera sensitivity range. The modelled flux is based on a 3-cavity filter model with nominal \ac{cwl} and \ac{fwhm} (358\,nm and 23\,nm respectively), and with an out-of-band transmission of $10^{-5}$. The \textit{shaded areas} represent the flux seen by the \ac{sp} cameras in the different diffraction orders (labels 1 to 8), when the grating is tuned through the filter working range, defined here as \mbox{\ac{cwl} $\pm$ \ac{fwhm} / 2}. The flux in the grating orders not containing the science window is required to remain below the stray-light limit denoted by the \textit{horizontal dot-dashed line}.}
\label{fig_osf}
\end{figure}

The purpose of the \acp{osf} is to isolate a given grating diffraction order and to suppress the light of unwanted orders to an acceptable level. \ac{susi} has four \acp{osf} with a size of 40 mm $\cdot$ 40 mm, mounted at 90° angles in the filterwheel unit. The unit is based  on a commercial piezo-driven rotation stage (SmarAct SR-7012), and is located between the \ac{pmu-rot} and the \ac{pbs}. Its rotation speed of 15°~s$^{-1}$ allows to position any \ac{osf} in the beam path within about 18~s. The filter passbands cover the \ac{susi} spectral working range with some gaps, as shown in Figure~\ref{fig_phot_budget}. Figure~\ref{fig_osf} shows, for \ac{osf} no.~2 as an example, the expected relative solar flux within the camera spectral sensitivity range. The flux is based on the measured filter transmission, multiplied with a smoothed ATLAS~3 spectrum \citep{Thuillier04}. The ATLAS~3 spectrum is also used as a reference for the photon budget estimates, cf. Section~\ref{sec_throu_char}. The effect of other components with a strong wavelength dependence is taken into account as well: the reflectivity of the \ac{islid} dichroic element, the grating reflectivity, and the camera quantum efficiency. Above its cutoff around 470\,nm (cf. Section~\ref{sec_optdesign}), the dichroic element significantly contributes to the out-of-band suppression. 

Table~\ref{tab_osf} shows the main \ac{osf} specifications. The out-of-band transmission is specified as an integer value referring to the upper \ac{od} limit. For example, an \ac{od} value of 3 means that the out-of-band transmission stays below $10^{-3}$ for all contributing grating diffraction orders.  

\begin{table}
    \begin{tabular}{lllll}
    \hline
    Filter No. & 1 & 2 & 3 & 4 \\
    \hline
    \ac{cwl} [nm] & 317 & 327 & 358 & 401 \\
    \ac{fwhm} & 16.7 & 19.6 & 25.0 & 32.7 \\
    Peak transm. & 0.83 & 0.83 & 0.80 & 0.89 \\
    Out-of-band transm. & 3 & 2 & 4 & 3 \\
    \hline
   \caption{Main \ac{osf} specifications.\label{tab_osf}}
    \end{tabular}
\end{table}

The out-of-band suppression is the most critical \ac{osf} requirement. The integrated contribution of polarized and unpolarized spectral stray light from other grating diffraction orders must stay below the noise limit of $10^{-3}$ imposed by the polarimetric sensitivity requirement. The stray-light requirement must also hold for the strongest spectral lines with a line core intensity as low as 1\% of the nearby continuum. Based on this requirement, and on the expected relative flux, a safe operating range can be defined for each \ac{osf}, where the integrated stray-light contribution stays below $5 \cdot 10^{-4}$ times the flux measured at any given wavelength position. It has been verified by analysis that for all OSFs, the range \mbox{\ac{cwl} $\pm$ \ac{fwhm} / 2} is safe in terms of stray light, with a few 0.1\,nm margin on each side.           

The \acp{osf} have been custom designed and manufactured at the Fraunhofer Institute for Surface Engineering and Thin Films \citep{Bruns21}. They are made of fused-silica glass plates, with the bandpass coating on one side, and an anti-reflection coating on the other side. The main-purpose of the anti-reflection coating is to suppress fringes and ghost images, as the OSFs are used close to normal incidence. Haze and polarization properties of the coatings have to be kept under control as well, to ensure compatibility with the stray light and polarimetric accuracy requirements.    

\subsection{Polarization Modulation Unit}
\label{sec_pmu}

The \ac{pmu-rot} and the \ac{pbs} are the key components of the \ac{pmu}. Our decision to use a rotating waveplate for polarization modulation was primarily motivated by the limited applicability of liquid crystals in the \ac{nuv}. The \ac{pmu-rot} mechanism is based on heritage of the CLASP sounding rocket missions \cite[cf.][]{Ishikawa15}. The mechanism has been provided by our Japanese partners and its design is practically identical to the \ac{pmu-rot} mechanism of the \ac{scip}. The mechanism is described in more detail in \citet{Katsukawa25}. The nominal rotation period, corresponding to two polarization modulation cycles, is fixed to 512~ms. Each modulation cycle is sampled with 12 frames. All three \ac{susi} cameras are synchronized to the phase of the waveplate rotation via a periodic trigger signal emitted by the \ac{pmu-rot} mechanism. The signal is generated by an internal clock which is also used as the reference for the closed-loop control of the waveplate rotation speed. The random synchronization error between the \ac{pmu-rot} and the cameras has been verified to be below 0.21 ms, which is compatible with the polarimetric accuracy requirement (cf. Section~\ref{sec_polcal}). 

The waveplate is a zero-order, air-gapped, double quartz plate with a retardation of 118\,nm at a design wavelength of 335\,nm. The retardation is wavelength dependent, scaling with the dispersion of the refractive indices of quartz. In units of wavelength, the retardation covers the range between $\lambda / 2.59$ at 309 nm and $\lambda / 3.65$ at 417 nm. The design choice is the result of a trade-off between optical and polarimetric performance criteria. The value for the retardation has been chosen as a trade-off between wavelength dependent polarimetric efficiency and available solar photon flux. Both surfaces of the waveplate have an anti-reflective coating with less than 0.2\% residual reflectivity across the \ac{susi} spectral working range. This ensures that fringes and ghost images are suppressed to a level that is compatible with the imaging and polarimetric accuracy requirements.        

The \ac{pbs} acts as analyzer for both \ac{sp} cameras, with one camera receiving the transmitted and the other camera the reflected light. The \ac{pbs} is a fused silica prism composed of two halves. The inner surfaces of the prism halves, acting as the polarizing layer, are bonded at an angle of 55° with respect to the incoming optical axis. The key specifications for the \ac{pbs} are the extinction ratios and the flux balance between reflected and transmitted channels. The extinction ratios in transmission and in reflection are better than 1:950 and 1:20 respectively, which is compatible with the polarimetric accuracy requirement. For unpolarized light the relative flux difference between transmitted and reflected beams is below 5\%. This ensures balanced noise properties for both channels which is important for dual-beam polarimetry.         

\subsection{Cameras}
\label{sec_cameras_design}

All \ac{susi} cameras are custom-built at MPS based on the GSENSE400BSI-VIS, a back-illuminated \ac{cmos} detector manufactured by Gpixel Inc. This detector has a size of 2048$ \cdot $2048\,pixels, 11\,$\mu$m pixel pitch, a 12-bit analog to digital converter, a maximum frame rate of 48\,fps, high quantum efficiency in the \ac{nuv} (40\,\% at 300\,nm and 88\,\% at 400\,nm), and very low readout noise, see Sect.~\ref{sec_characterization}. The main camera design features are described in the following paragraphs.

\begin{figure}
    \centering
    \includegraphics[scale=0.3]{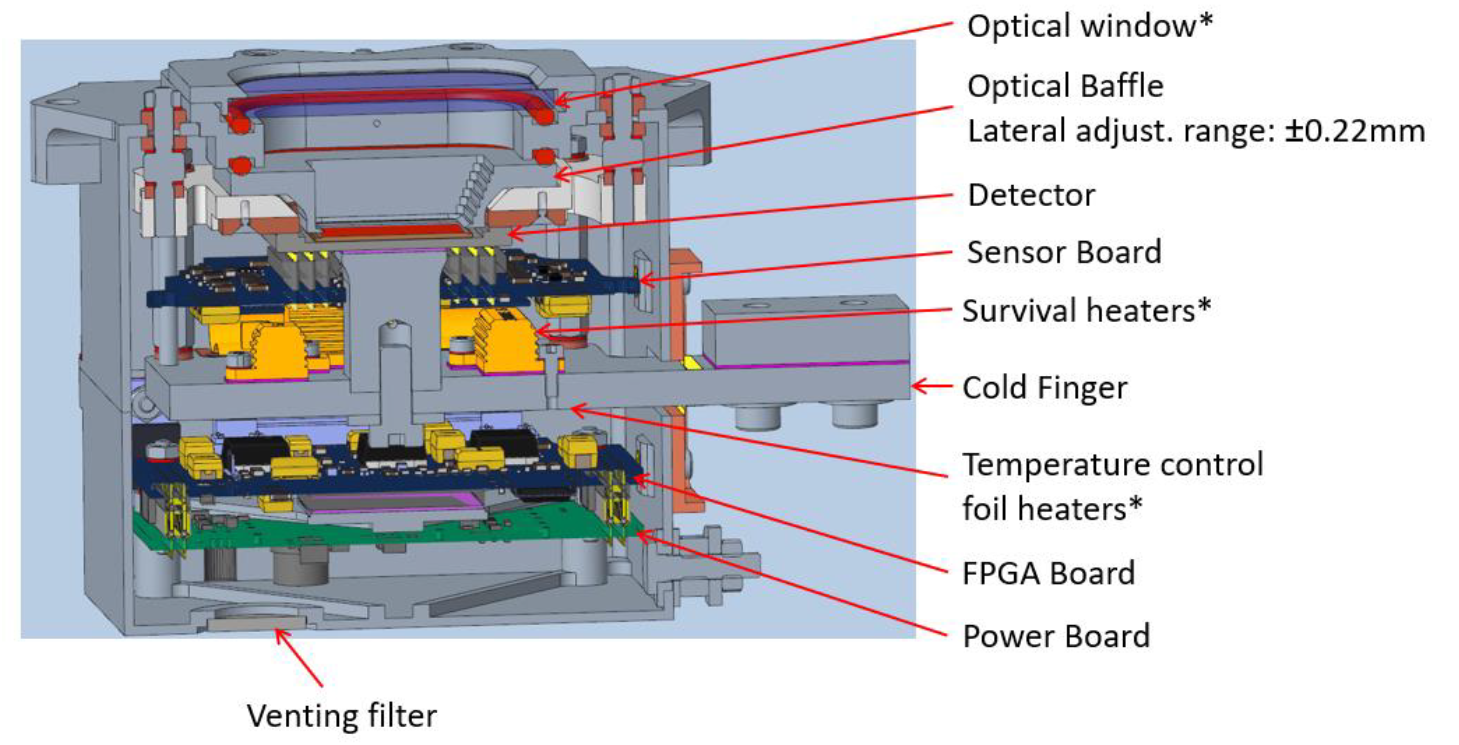}
    \vspace{1em}
    \includegraphics[scale=0.4]{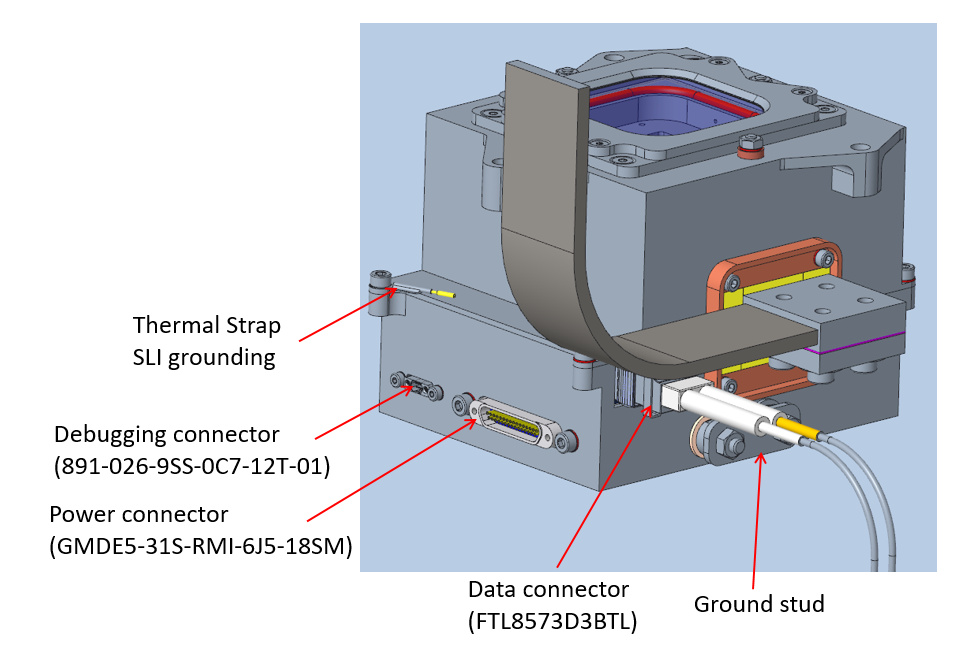}    
    \caption{CAD cut (\textit{top}) detailing the main components of the \ac{susi} cameras, including the main devices used for thermal control. CAD view of the camera housing (\textit{bottom}) including the bent thermal strap that connects the detector cold finger to the external radiator, and the main power, control, and optical data connectors.}
    \label{fig_cam_cut}
\end{figure}
   
\paragraph*{Electronics}

The camera front-end electronics is distributed over three printed-circuit boards (Figure~\ref{fig_cam_cut}): detector, \ac{fpga} and power board. The  general camera control logic, including detector clocking, is implemented in a Xilinx Kintex Ultrascale \ac{fpga} operating at 25~MHz. All \ac{susi} cameras operate at a framerate of 46.9 \,fps in strict synchronization with the \ac{pmu}, via a dedicated \ac{lvds} trigger line. This implies that 12 camera frames sample each polarization modulation cycle. 
    
The detector has an electronic rolling shutter and is exposed continuously with practically no dead time (99.65\,\% duty cycle). The time difference between the start of the exposure of two consecutive detector rows is 10.34\,$\mu$s. As a consequence, different detector rows sample different position angles of the \ac{pmu} waveplate (14.74\,deg difference between the exposures of rows no.~1 and 2048). 

The camera control and data transfer is done via a high-speed fiber-optical interface, using the Channel Link High Speed protocol. Under nominal operating conditions, each camera generates about 2.2~Gbps of image data.

\paragraph*{Detector} 

The photon flux and stability requirements of the \ac{sp} cameras are considerably stricter than those of the \ac{sj} camera. The expected mean signal level at the \ac{sp} cameras for the continuum at the blue limit of the spectral working range, and in a quiet solar region, is of order $10^2$ \,e$^{-}$ per pixel and frame (cf. Section~\ref{sec_throu_char}). In the cores of strong spectral lines the signal level can get as low as a few e$^{-}$ per pixel and frame. Therefore, the \ac{sp} cameras operate with the longest possible exposure time (21.3\,ms) and with the highest detector conversion gain setting of 1.5~DN/e$^{-}$ for all \ac{susi} spectral bands, except for the one near 408\,nm, which uses a gain setting of  1.2~DN/e$^{-}$. DN stands for digital number and denotes the numerical value generated by the cameras \acp{adc}, based on the photo-charges collected in the detector pixels. There are in addition three extra lower-gain settings available for use with high-flux solar targets, such as flares or active regions. The high gain setting and the selected operating temperature (see thermal aspects below) yields a total noise in dark conditions of only about 1.8\,e$^{-}$ \ac{rms}, which is dominated by readout noise. 

The expected mean flux at the \ac{sj} camera is up to two orders of magnitude larger compared to the above-stated continuum flux seen by the \ac{sp} cameras. Thus, the detector operates with a much shorter exposure time (nominal 1\,ms) and lower conversion gain (0.054~DN/e$^{-}$), yielding an expected mean signal of order $2 \cdot 10^3$~DN per pixel and frame, for a quiet region of the Sun. The mean flux estimation of the \ac{sj} channel has a large uncertainty. Therefore, we have foreseen and tested a range of possible exposure times (0.03 - 4.0~ms) and four extra conversion gains for the \ac{sj} detector operation (0.02, 0.03, 0.07, and 0.11~DN/e$^{-}$). During the \ac{susi} commissioning performed before the science phase, the optimal values for exposure time and gain will be selected, based on the actual flux detected at the floating altitude of \sunrisethree.
    
\paragraph*{Detector optics} 

The \ac{sp} camera detectors are protected by a fused-silica entrance window. The \ac{sj} camera entrance window is a Schott UG-11 glass which acts as an additional bandpass filter (in combination with the \ac{islid} dichroic element and the main \ac{sj} filter) to further suppress the out-of-band transmission. 

All \ac{susi} cameras include an optical baffle and a detector mask designed to minimize stray light reaching the tilted focal planes, and to shield a portion of the detector's external border area (about 50 pixels on each side of the detector). The shielded pixel values are used during data reduction to calibrate for detector-related artifacts, such as a variable bias.
    
\paragraph*{Mechanics} 

The camera housing (Figure~\ref{fig_cam_cut}) is almost identical for all three \ac{susi} cameras. It has a volume of about 112 x 95 x 84~mm$^3$. Its main purpose is to provide mechanical support, thermal insulation, and shielding against stray light and dust.
    
\paragraph*{Thermal aspects} 

The thermal regulation of the cameras is achieved by combining passive radiative cooling and active resistive heating. The cooling is done by connecting the aluminum detector cold finger to external camera radiators using thermal straps (Figure~\ref{fig_cam_cut}). The operational heating is achieved using controlled foil heaters, attached to the bottom of the detector cold finger (Figure~\ref{fig_cam_cut}). The selected working temperatures are in the range of 10\degC{} to 25\degC{} for the \ac{sp} and \ac{sj} cameras, respectively. These were chosen to ensure a dark noise contribution of less than 10\,\% of the total noise at the lowest signal level in the 323\,nm window.

The temperature stability requirement is more strict for the \ac{sp} cameras because a variable dark signal can introduce polarimetric artifacts. We simulated the polarimetric artifact added by an uncorrected additive term to the images before dual-beam polarimetric demodulation. For the case of Zeeman diagnostics at 50\.\% (line wing) of the flux level in the dimmest window (323\,nm) and a 5\,\% polarization signal, the maximum variation allowed in the calibrated dark level is $\approx 0.6$~DN. This requirement, along with the thermal dependence of the dark current, translates into the required sensor thermal stability (cf. Section~\ref{sec_camera_characterization}).   

In addition, there is a set of manually operated survival heaters attached to the top of the detector cold finger (Figure~\ref{fig_cam_cut}). These heaters support the operational heaters during the Sunrise ascend phase, particularly when flying through the tropopause, where external temperatures can drop below -50~C$^\circ$ and where heat dissipation by convection is still significant.

\section{Understanding the Instrument and Getting the Best out of the Data}
\label{sec_characterization}

\subsection{Optical Verification}
\label{sec_optchar}

\paragraph*{Focus alignment}

\begin{figure}
\centerline{\includegraphics[width=\textwidth]{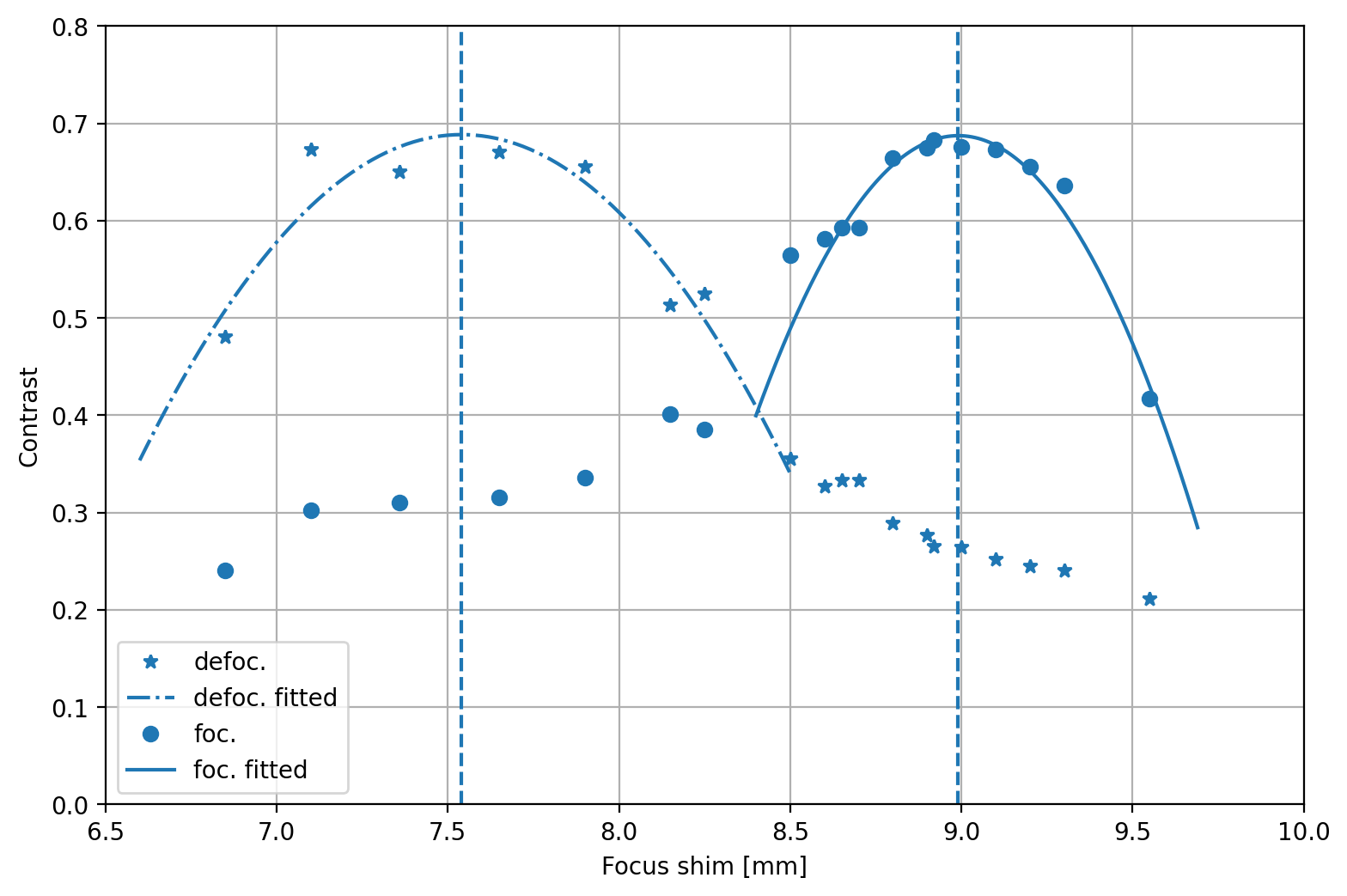}}
\small
\caption{Focus series recorded during the alignment of the \ac{susi} \ac{o-unit} within the \ac{pfi}. The data points denote the contrast of the F2 random-dot target, measured in the focused and defocused \ac{sj} channels (see plot legend for the identification of the respective channel). The \textit{curves} represent simple polynomials of degree 2, fitted to the data points around maximum contrast. The \textit{vertical lines} mark the position of optimum focus for each channel, as derived from the fitted polynomials.}
\label{fig_contrast_curve}
\end{figure}

During the alignment of the \ac{susi} \ac{o-unit} within the \ac{pfi}, the optimum focus position is determined by means of a focus series (Figure~\ref{fig_contrast_curve}). With the random-dot target inserted in F2 \cite[see \ac{islid} and \ac{pfi} sections in ][]{Lagg25}, the contrast of the target image in the focused and defocused \ac{sj} channels is measured at different distances of the \ac{o-unit} with respect to the F2 focal plane. We denote by focused and defocused \ac{sj} channels the two images projected by the \ac{pid} onto the \ac{sj} camera (cf. Section~\ref{sec_design}). One of the images, called the focused channel, is conjugated to the focal plane on the \ac{susi} slit plate. The other image, differing from the focused channel by a fixed amount of \ac{pid}-generated defocus, is called the defocused channel. The \ac{sj} \ac{o-unit} is aligned such that the slit-plate focal plane is conjugated to F2. This position is found by maximizing the contrast of the target image in the \ac{sj} focused channel. 

In addition, the actual amount of \ac{pid} generated defocus is determined by finding the position of the \ac{pid} which maximizes the contrast of the target image in the defocused channel. The difference of about 1.5~mm, measured between the optimum focus positions of the focused and defocused \ac{sj} channels, differs slightly from the nominal design value of 1.2~mm. This acceptably small difference, which is not of any practical relevance for wavefront sensing, can be attributed to measurement uncertainties, as well as alignment and optical manufacturing tolerances.  

\paragraph*{Imaging performance}

The imaging performance of an optical system can be characterized in different ways. On individual components such as mirrors or lenses we typically measure wavefront errors. For the end-to-end characterization of the fully assembled instrument we rely on images of test targets. This type of image performance characterization has the advantage that it can be easily performed in flight configuration without the need of any auxiliary measuring devices like interferometers. The test targets are positioned in the F2 focal plane by a filterwheel mechanism. They are made of glass plates with a chromium surface coating forming the test patterns.

\begin{figure}
\centerline{\includegraphics[width=\textwidth]{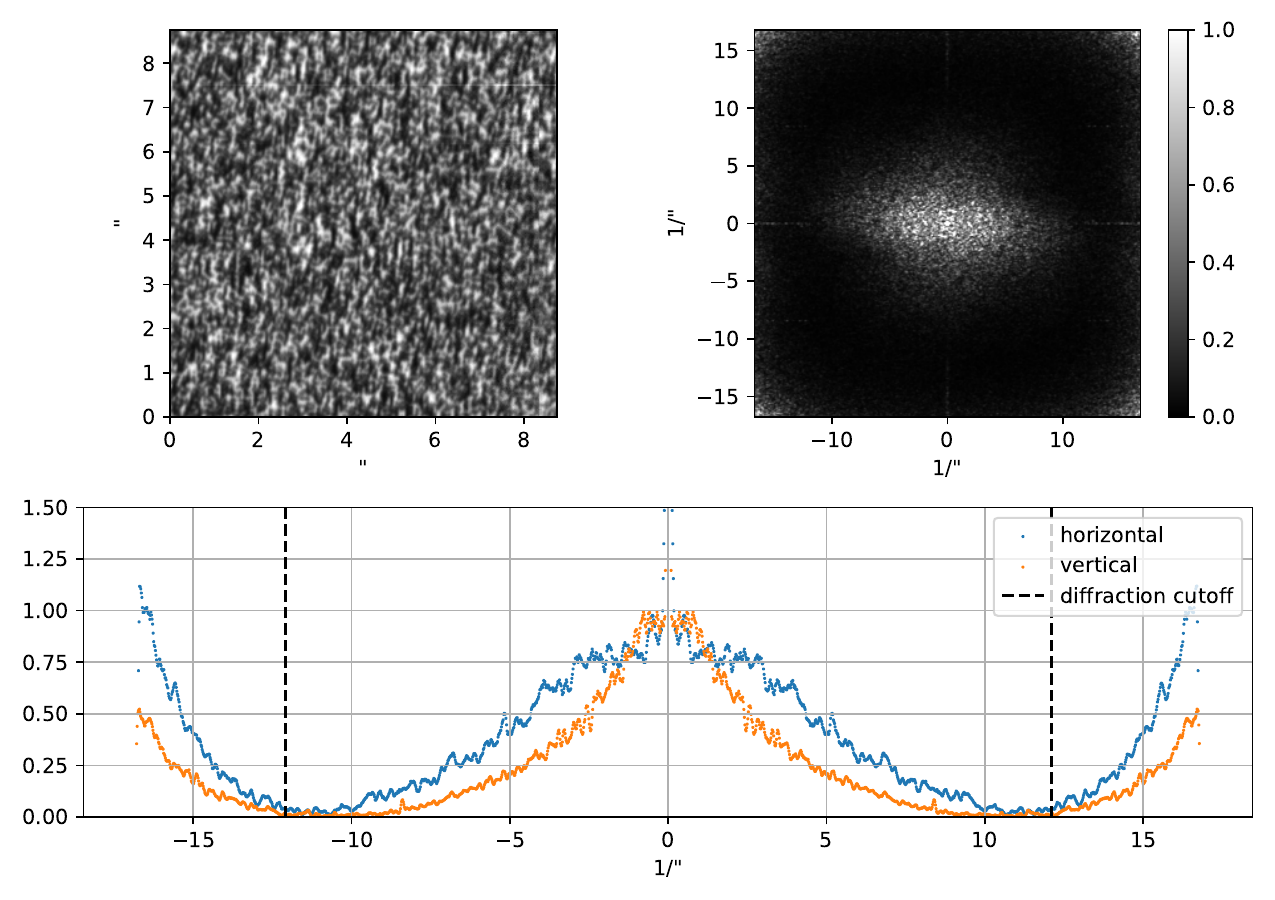}}
\small
\caption{MTF derived from an \ac{sp} scan of the random-dot target image, recorded by \ac{sp} camera no.~1. The scan has been performed during a pre-flight test of the fully integrated gondola, with Sun pointing, at a wavelength of 401 nm. \textit{Upper left:} Small cutout of the target image. The scan and slit directions are along the \textit{horizontal and vertical axes} respectively. \textit{Upper right:} derived \ac{2d} \ac{mtf} \textit{Bottom:} horizontal and vertical averages of the \ac{2d} \ac{mtf}. The spatial frequencies in units of 1/\arcsec{} represent the angular resolution in the Sky. The \textit{vertical lines} denote the theoretical diffraction cutoff. The \ac{mtf} has been normalized to 1 at frequency 0\arcsec$^{-1}$ and a white noise offset has been subtracted.}  
\label{fig_mtf_rdot}
\end{figure}

The random-dot target is well suited to estimate the \ac{mtf} because of its balanced spectrum of spatial frequencies. It consists of a regular grid of 1000 x 1000 squares, spanning an area of 7 x 7 mm$^2$. The squares are chosen randomly to be either dark (i.e. masked by the coating) or bright (uncoated). The \ac{mtf} is derived from the random-dot target by computing the ratio of the Fourier transform modulus (i.e. the square root of the power spectrum) of a measured random-dot image and a nominal random-dot image unaffected by optical aberrations. An example result for \ac{sp} camera no.~1 is shown in Figure~\ref{fig_mtf_rdot}. Similar results are obtained for the other \ac{susi} cameras.

Different factors, such as manufacturing tolerances in the chromium masks, field-dependent aberrations and diffraction effects at the target, impose limits on the accuracy of the derived \acp{mtf}. Nevertheless the measured \acp{mtf} are adequate metrics to evaluate the imaging performance of the instrument and to detect major optical deficiencies. The measurements demonstrate that both the \ac{sp} and \ac{sj} optics are capable of transmitting image contrast with an acceptable \ac{s2n} ratio, up to frequencies close to the diffraction cutoff. Some astigmatism is present, leading to different contrast in the horizontal and vertical directions. 

Due to the high sensitivity of the random-dot target contrast to small changes in the \ac{mtf}, target contrast measurements are also used as a simple metric for tracking relative changes in the imaging performance (e.g. during focus alignment, see previous paragraph).

\begin{figure}
\centerline{\includegraphics[width=\textwidth]{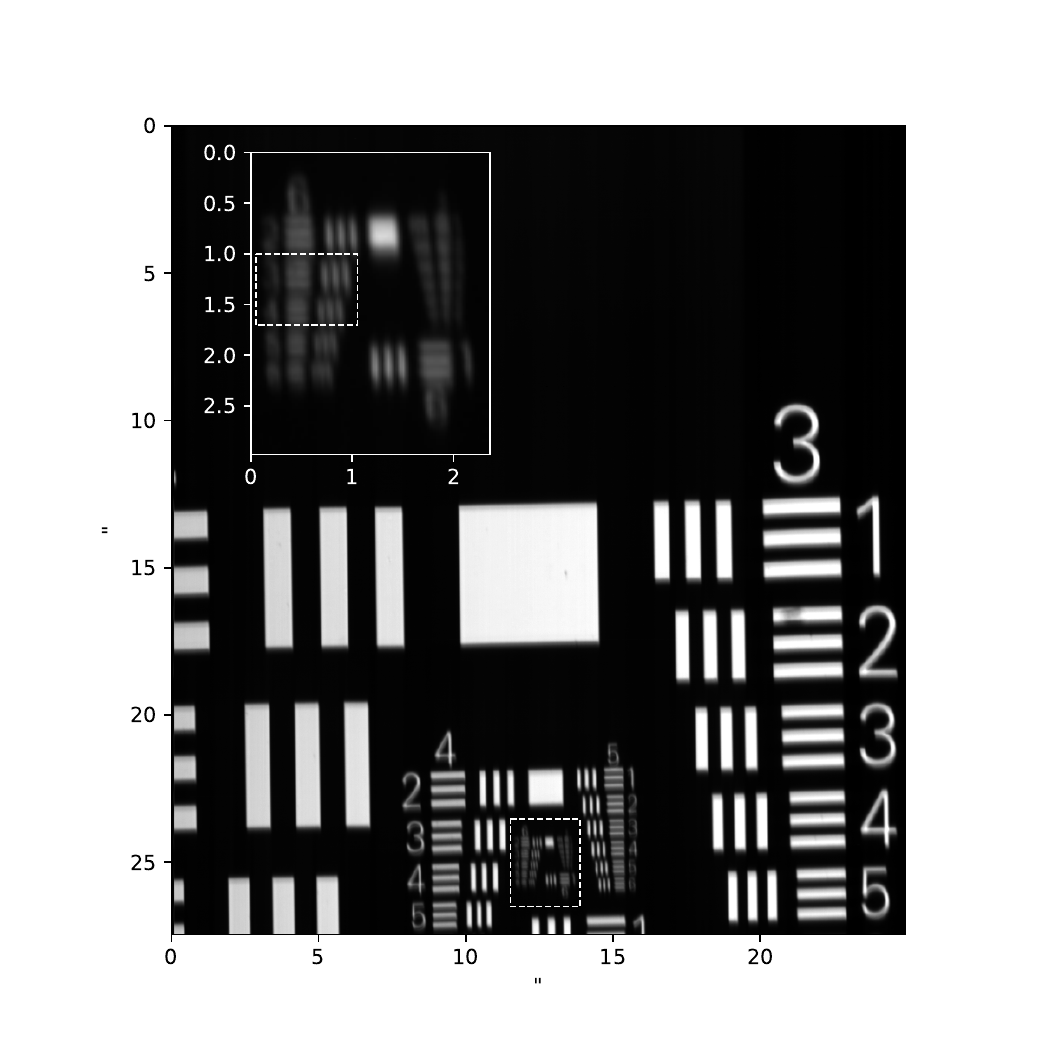}}
\small
\caption{Example scan across the \ac{usaf} target, recorded by \ac{sp} camera no.~1 at 407.3 nm. As in Figure~\ref{fig_mtf_rdot}, the scan and slit directions are along the \textit{horizontal and vertical axes} respectively. A small \ac{roi} around groups six and seven is shown as a magnified inset in the \textit{upper left} part of the image. Patterns up to group six, element three to four are resolved (\textit{dashed rectangle} in the inset), which is consistent with a diffraction-limited resolution of 0.10\arcsec{} at this wavelength, based on the Rayleigh criterion.}
\label{fig_usaf}
\end{figure}

The \ac{usaf} target, carrying a test pattern originally defined by the MIL-STD-150A standard, offers a quick visual performance assessment in terms of the smallest resolved Ronchi pattern group. An example is shown in Figure~\ref{fig_usaf}. The astigmatism shows here as a difference in contrast between the horizontal and vertical Ronchi patterns.

\subsection{Camera Characterization and Verification}
\label{sec_camera_characterization}

The flight models of the \ac{susi} cameras have been subject to various characterization measurements. Selected results are shown in Table~\ref{tab_cam_char} and described in the paragraphs below.

\paragraph*{Dark current} 
The dark current requirements described in Section~\ref{sec_cameras_design} limit the thermal electron flux in the camera pixels to a maximum of approximately 40\,e$^-$/s and 560\,e$^-$/s for the \ac{sp} and \ac{sj} cameras, respectively. As seen in Table~\ref{tab_cam_char}, this is satisfied for detector temperatures below 15\degC~and 25\degC, respectively, which are thus adopted as the maximum operational temperatures during flight. The nominal operational temperatures are 10\degC~and 20\degC, respectively. 
    
\paragraph*{Thermal stability} 
The detector thermal control system provides a temperature stability better than  $\pm$0.05\degC, as verified in thermal-vacuum tests involving operational thermal cycles between 0\degC~and 30\degC. The strict stability requirement of the \ac{sp} camera dark level introduced in Section~\ref{sec_cameras_design}, has been verified by evaluating the difference in the signal levels of dark frames acquired one hour, four hours, and 72 hours apart. In all cases, the differences were below 0.4 $\pm$ 0.3~DN. We note that the in-flight calibration schedule included in the nominal observing program ensures that science and dark calibration data are not more than about six hours apart. 
    
\paragraph*{Conversion gain}
The measured gain values are within the manufacturer specifications and show a low thermal dependence ($<$ 0.5\%/\degC). In view of the achieved detector temperature stabilization, drifts in the conversion gain are negligible. 

\paragraph*{Readout noise} 
The total camera noise in dark conditions is dominated by readout noise. The measured values are within manufacturer specifications and also present a low thermal dependence ($<$ 0.5\%/\degC). We note that the very low readout noise in the high gain settings ($<$ 2\,e$^-$ \ac{rms}), is crucial to avoid integration time overhead in very low-flux spectral windows, see e.g., \cite{Iglesias19}.
    
\paragraph*{Linearity and bias} 
Non-linear detector response is a major concern in high-precision polarimetry \citep[see e.g.,][]{Keller96} and a detailed study of the response of various \ac{cmos} detectors including the GSENSE400BSI-VIS detector used for \ac{susi}, has been carried out in house \citep{Sant22}. In the context of \ac{susi}, two types of non-linearities have been found to be of practical relevance: (1) the traditional non-linear response to the amount of photo-electric charges collected by an individual pixel (pixel non-linearity), and (2) a dependence of the bias level on the amount of photo-electric charges collected in a given detector row (banding effect).

Using representative synthetic spectra, we have run instrumental simulations to derive a total camera non-linearity upper limit of 0.5\%, in order to keep polarimetric artifacts below the noise level of $10^{-3}$ imposed by the \ac{susi} polarimetric sensitivity requirement, see \cite{Sant22} for details. The values reported in Table~\ref{tab_cam_char} correspond to the pixel non-linearity measured with flat illumination and varying exposure time, and are defined as the maximum deviation from a linear fit over a 1000~DN dynamic range. This type of non-linearity can be typically calibrated to an accuracy better than some 0.5\% using a look-up table based on low-order polynomial fits of the residual deviations from a linear response, see e.g., \cite{Dewijn2019}. 

The bias level corresponds to the DN value reached in a given pixel in the limiting case of the exposure time approaching 0. The bias level is thus in principle a property of the camera electronics only, but in practice the charges collected in the detector pixels can have an influence. For the \ac{susi} detector, changes in the measured bias level due to the banding effect are of order 20 DN to 40 DN \citep{Sant22}, which is significant in view of the required polarimetric accuracy, in particular under low-flux conditions (cf. Section~\ref{sec_throu_char}). The banding effect can be corrected to sufficient accuracy by evaluating the shielded pixel columns close to the detector edges.

\paragraph*{Effect of the cameras on polarimetric accuracy}    
To directly evaluate the effect of non-linearities, and any other camera related artifacts on the polarimetric accuracy of \ac{susi}, we have developed a specific characterization measurement. The \ac{sp} cameras are illuminated with a pulsed light source that emits a periodic signal composed of a set of 12 pulses, corresponding to the 12 polarimetric modulation states. Each pulse has a specific width such that the associated 12 images recorded by the camera emulate the modulated flux levels corresponding to a polarization degree of approximately 10\%. The value represents the level of instrumental polarization of order 4\%, including some margin. 

A triangular optical target is added to the beam path \cite[see][for a target description]{Sant22}. The expected demodulated dataset corresponds to a high-contrast Stokes I image with a sharp triangular edge, and Stokes Q/I, U/I, and V/I images with the aforementioned spatially constant offset of about 10\%. In addition to providing an edge in the intensity distribution, the triangular target also allows to assess the effect of residual errors in the banding correction.

Approximately $2 \cdot 10^4$ frames are acquired in this configuration to study the dependence of the spatial \ac{rms} value in Q/I, U/I, and V/I with respect to the number of accumulated frames. We observe a reduction approximately compatible with pure photon noise statistics, namely, proportional to the square-root of the number of accumulated frames, down to the level of about $6 \cdot 10^{-4}$. This demonstrates that camera-related polarimetric artifacts are below the required polarimetric sensitivity limit of \ac{susi}.

\paragraph*{Synchronization}
The measured exposure time windows of the \ac{susi} cameras show \ac{rms} fluctuations of about 11\,$\mu$s. In the case of the \ac{sp} cameras with a nominal exposure time of 21.3 ms, this ensures a relative synchronization stability of $5 \cdot 10^{-4}$ which is compatible with the polarimetric accuracy requirement. The detector clocking is designed such that the only way to lose a frame is by an overload of the \ac{susi} frame buffer on the \ac{icu} (cf. Section~\ref{sec_software}).

\begin{table}
    \centering
    \begin{tabular}{c c c c c}
    \hline   
    \hline        
    \textbf{Meas. ID}& \textbf{Gain} & \textbf{Dark curr.} & \textbf{R/O noise}  & \textbf{Non-lin.} \\
    Cam\_Gain & DN/e$^{-}$  & e$^-$/s & e$^{-}$  & \%  \\
    \hline   
    \noalign{\smallskip}    
    SP1\_HGx7.25 &   1.53, 1.53, 1.52 &   17.6, 32.7, 95.4 &    1.7, 1.6, 1.6 &   1.1, 0.6, 0.8 \\
    SP1\_HGx4.95 &   1.05, 1.06, 1.04 &  16.2, 38.7, 101.0 &    2.0, 1.9, 1.9 &   0.3, 1.4, 0.9 \\
    SP2\_HGx7.25 &    1.71, 1.69,  * &     8.8, 18.9,  * &    1.7, 1.7,  * &   0.6, 0.2,  * \\
    SJC\_LGx0.66 & 0.019, 0.02, 0.019 & 31.6, 125.0, 126.3 & 63.2, 60.0, 63.2 & -0.7, 2.3, -0.3 \\
    SJC\_LGx1.85 &    0.054,  *,  * &     13.0,  *,  * &   29.6,  *,  * &  -0.2,  *,  * \\
    \hline  
    \end{tabular}
    \caption{Selected camera characterization results. For each property, we specify the median values computed across the detector area for working temperatures 5\degC, 15\degC, and 25\degC. We show only measurements done with four different conversion gain configurations, see the Measurement ID column; Dark curr. refers to dark current, ``R/O noise'' to readout noise and ``Non-lin.'' to non-linearity of the detector response. We use * to denote values that have not been measured. See the text for details.}
    \label{tab_cam_char}
\end{table}

\subsection{Flat-field and Spectrum-related Corrections}
\label{sec_spectroflat}

\paragraph*{Flat-field correction}
During the flight phase of \sunrisethree, a flat-field measurement matching the observation parameters (in particular the wavelength) is recorded before and after each science block \cite[for a description of science blocks see][]{Lagg25}. In flat-field mode the telescope pointing follows a Lissajous path, with a frequency ratio of 0.6 and a phase difference of 90 degrees between the azimuth and elevation axes, and covering an area with a radius of about 60\arcsec{} around solar disk center. During this movement, \ac{susi} continuously records frames, which are then later averaged to smear out solar structures. The flat-field measurement is used for flat-field, wavelength and spectral stray-light corrections, as well as for the verification of the spectral resolution, as described in further paragraphs of this section.  

Due to the required high frequency of flat-field measurements, the duration of each individual measurement is critical, given the limited mission time of \sunrisethree. An unnecessarily long flat-field measurement reduces the valuable time available for the science blocks. On the other hand, if the flat-field measurement is too short, the science data will be compromised by noise or residual solar structures (flatness error) in the flat-field image. Table~\ref{tab_flat} shows a reference case for \ac{susi} and the derived minimum flat-field integration times. In the case of \ac{susi}, which shows a rather low degree of instrumental polarization, the flat-field integration times are dictated by the photometric noise requirement in combination with solar flux and quiet-Sun intensity contrast. The integration time is wavelength dependent, because the solar flux as well as the intensity contrast vary with wavelength. In the \ac{susi} \ac{nuv} range, the intensity contrast is significantly higher than in the visible range \citep{Hirzberger10}, which indicates the need for a longer flat-field duration. Flat-field simulations, based on the Lissajous pattern traveled by the telescope, have shown that the intensity contrast of residual solar structures essentially scales with $\sigma / \sqrt{n}$, as expected for the ideal case of uncorrelated frames. Here, $\sigma$ is the \ac{rms} contrast of a single frame, and $n$ is the number of frames averaged.

\begin{table}
    \begin{tabular}{ll}
    \hline
    Maximum noise ratio$^1$ & $1.05$ \\
    Degree of instrumental polarization & $4 \cdot 10^{-2}$ \\
    Noise requirement for Stokes I & $5 \cdot 10^{-3}$ \\
    Noise requirement for Stokes Q/I, U/I, V/I & $1 \cdot 10^{-3}$ \\
    \hline
    \end{tabular}\\
    \smallskip    
    \begin{tabular}{llll}
    \hline
    & 309\,nm & 400\,nm \\
    \hline
    Quiet-Sun continuum intensity contrast & $2.3 \cdot 10^{-1}$ & $2.0 \cdot 10^{-1}$ \\
    Expected solar flux in e$^-$ per s and pixel & $4.3 \cdot 10^3$ & $2.8 \cdot 10^4$ \\
    Dominating effect & solar flux & contrast \\
    \hline
    \textbf{Min. flat-field integration time} & \textbf{92 s} & \textbf{34 s} \\
    \hline
    \end{tabular}
    \caption{Reference case for \ac{susi} flat-field measurements, showing the relevant parameters that determine the duration of flat-field measurements as well as the derived minimum flat-field integration times. The upper table shows the wavelength independent parameters. The lower table shows wavelength dependent parameters as well as the required minimum integration times for two example wavelengths within the \ac{susi} spectral range.}
       
    \label{tab_flat}
    \footnotesize $^1$Maximum noise ratio between the flat-field corrected and uncorrected images. 
\end{table}

\begin{figure}
\centerline{\includegraphics[width=\textwidth]{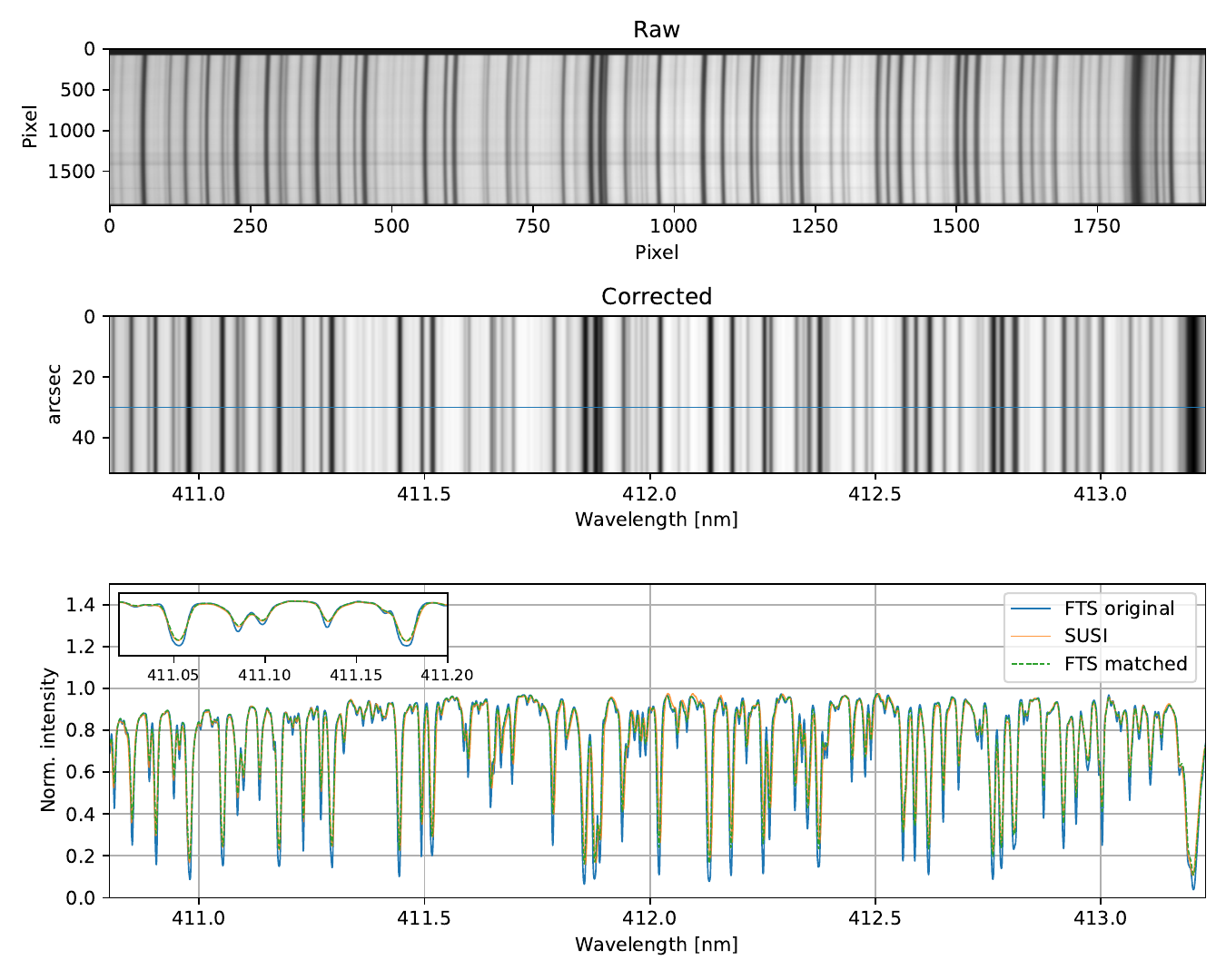}}
\small
\caption{Raw and corrected flat-field data for \ac{sp} camera no.~1. The flat-field measurement has been obtained during a pre-flight test with fully integrated telescope and gondola, and Sun pointing. \textit{Top panel}: raw image; \textit{Middle panel}: processed image, including dark, flat-field and spectral line curvature (smile) corrections, plus a wavelength calibration; the blue line denotes the cut shown in the bottom panel. \textit{Bottom panel}: Comparison of the measured \ac{susi} spectrum with the original (\ac{fts} original, see plot legend) and matched (\ac{fts} matched) Kitt Peak reference spectra. The latter has been degraded with the best matching \ac{susi} \ac{lsf} and stray-light parameters, see text for details. The inset shows a small magnified region of the spectra, for better comparison.}
\label{fig_spectrum_cam1}
\end{figure}

\begin{figure}
\centerline{\includegraphics[width=\textwidth]{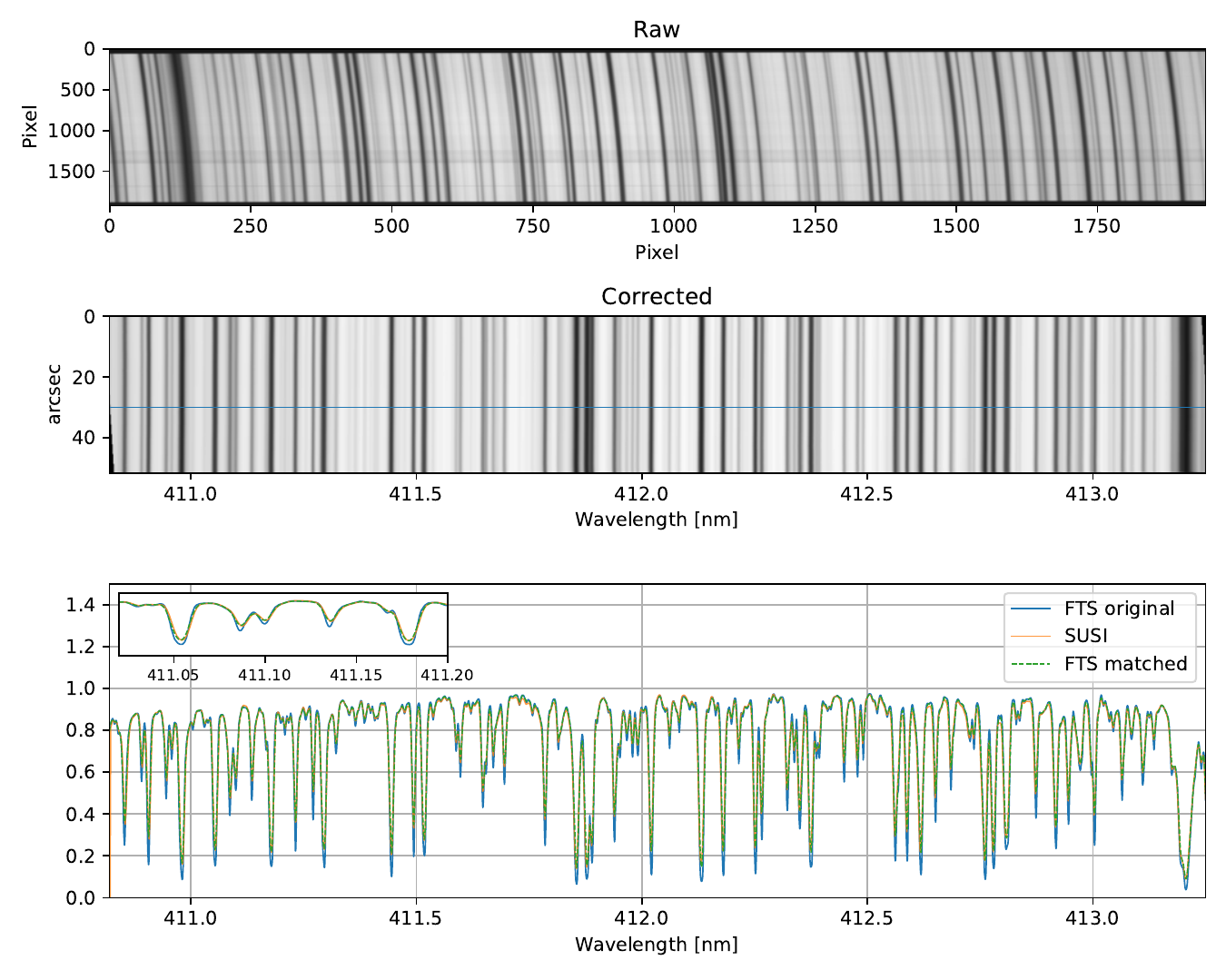}}
\small
\caption{Same as Figure~\ref{fig_spectrum_cam1} but for \ac{sp} camera no.~2. In addition to smile, the raw spectrum shows a rotation of about 0.5 degrees, which is corrected to match camera no.~1. Note also that the raw spectrum is mirrored compared to Figure~\ref{fig_spectrum_cam1}, because the \ac{pbs} introduces an additional reflection in the beam path.}
\label{fig_spectrum_cam2}
\end{figure}

\begin{figure}
\centerline{\includegraphics[width=1\textwidth]{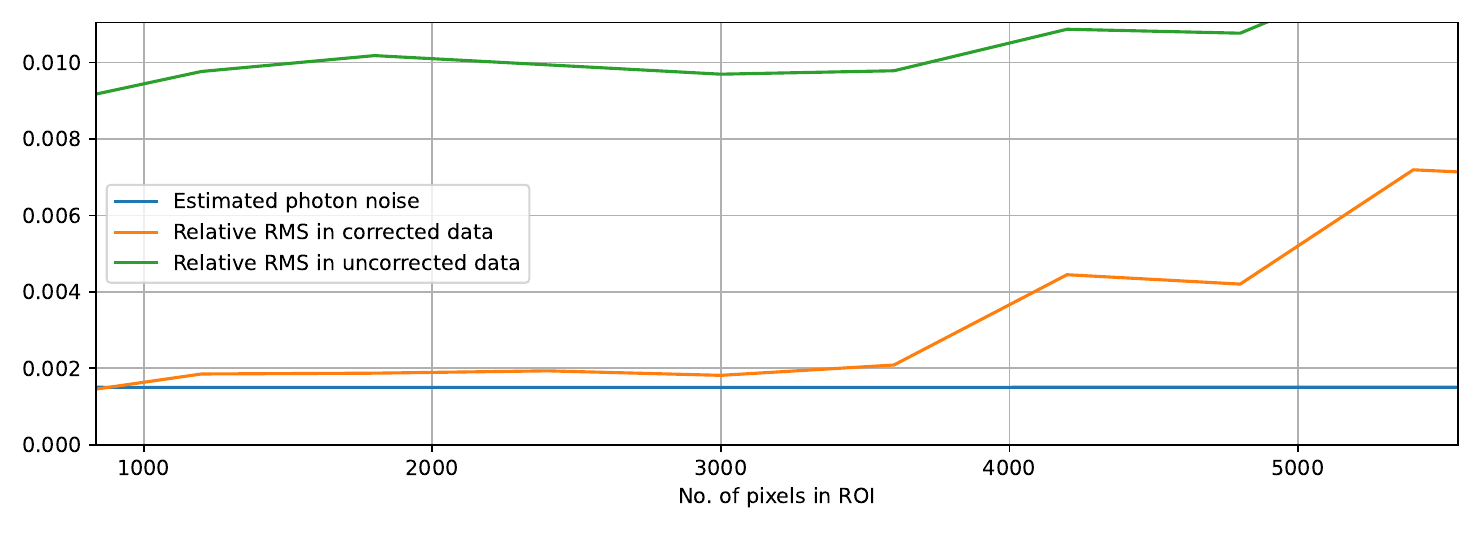}}
\small
\caption{Estimated photon noise vs. relative \ac{rms} for different \ac{roi} sizes of a \ac{susi} flat-field measurement recorded during a pre-flight test with Sun pointing in the 412\,nm region.}
\label{fig_photon_noise}
\end{figure}

To extract flat-field maps for sensor and slit features from the raw flat-field measurements we use the \texttt{spectroflat} code library as described in \cite{Hoelken23}. There we also show that the residual contrast can be suppressed down to the photon noise level. Figure~\ref{fig_photon_noise} shows estimated photon noise in comparison to the corrected and uncorrected relative \ac{rms} of a flat-field measurement obtained in the 412~nm region during a pre-flight test with Sun pointing. 

\paragraph*{Wavelength correction}
The wavelength calibration is carried out by comparing the spectrum of an averaged flat-field measurement with the Kitt Peak or Jungfraujoch solar atlas (cf. Section~\ref{sec_reqs}). For this step we use the \texttt{atlas-fit} library as described in \cite{Hoelken23}. The detected deviations from a linear relationship between pixel index and wavelength are removed from the \ac{susi} data by interpolating the wavelength corresponding to each pixel with sub-pixel precision. The calibrated wavelength range and the linear dispersion factor (in nm per pixel) are added to the metadata of the calibrated data frames.

\paragraph*{Spectral stray light}
We define spectral stray light as a gray (i.e. wavelength independent) or nearly gray intensity background in the observed spectra, which is caused by the instrument. By nearly gray we mean a slowly varying background level, such that the background within any given spectral window, covering a range up to 2.8~nm, can be considered gray. Correcting the data for spectral stray light is important, in particular for obtaining reliable inversion results, depending among others on true solar line depths.

Spectral stray light can have different origins. For example, light from the outside can leak into the instrument. Small ruling errors in the diffraction grating, or other optical surfaces with some residual roughness can lead to small amounts of diffusive reflection or transmission. For \ac{susi} the precise origins of spectral stray light have not been further assessed. The spectral stray-light contamination is at an acceptable level and can be corrected for in the data. We found that a basic local stray-light model can be used for both a stray-light estimate and for later correction, as described in the next paragraph about spectral resolution. The \ac{fts} spectra of the Kitt Peak atlas (cf. Section~\ref{sec_reqs}) are basically free of spectral stray light and are thus frequently used as a reference for stray-light characterization. 

\paragraph*{Spectral resolution}

The spectral resolution as well as the amount of stray light is estimated in a single-step procedure by fitting a matched Kitt Peak spectrum to a \ac{susi} flat-field spectrum. In the wavelength region covered by \ac{susi}, the Kitt Peak atlas has a comparatively high resolution such that the width of its own spectral \ac{lsf} can be neglected compared to the width of the \ac{susi} \ac{lsf}. Spatial averaging of solar structures in flat-field data leads to an overall spectral line broadening due to averaging over different velocities pointing towards and away from the observer, primarily due to granulation. The Kitt Peak atlas represents a large-scale spatial average as well (cf. Section~\ref{sec_reqs}), which allows for a fair comparison between \ac{susi} and Kitt Peak spectra. Besides being a good stray-light reference, the Kitt Peak atlas is thus a good reference for the characterization of the \ac{susi} spectral resolution as well.

The Kitt Peak spectrum is matched to the \ac{susi} flat-field spectrum using the following basic model:

\begin{eqnarray}
    I'_{\mathrm{FTS}}(\lambda) & = & I_{\mathrm{FTS}}(\lambda) * \mathrm{LSF}(\lambda) * G(\lambda, \sigma), \label{eq_FTS_res}\\
    I''_{\mathrm{FTS}}(\lambda) & = & (I'_{\mathrm{FTS}}(\lambda) + \rho \, \bar{I}'_{\mathrm{FTS}}) / (1 + \rho), \label{eq_FTS_stray}
\end{eqnarray}
with free parameters $\sigma$ and $\rho$ which denote the width of a Gaussian function $G$ and a stray-light factor, respectively.    

Equation~\ref{eq_FTS_res} describes the difference in resolution between the Kitt Peak and \ac{susi} spectra, which is modelled as the convolution of the original Kitt Peak spectrum $I_{\mathrm{FTS}}$ with the diffraction limited \ac{susi} \ac{lsf} and with an additional Gaussian function. The diffraction limited \ac{lsf} is modelled as the convolution of a Heaviside step function, representing the entrance slit, with the Airy function of an ideal circular aperture corresponding to the focal ratio on the \ac{sp} cameras (cf. Section~\ref{sec_optdesign}). The Gaussian function takes into account, in the most basic way, potential additional losses in resolution due to e.g. optical aberrations or instabilities (image jitter in the spectral dimension) during the measurement.    

Equation~\ref{eq_FTS_stray} describes the stray-light contamination of the \ac{susi} spectra. It is assumed that a fixed fraction of the average flux $\bar{I}'_{\mathrm{FTS}}$ within the camera spectral range contributes a flat (gray) background.

The matching is a least-squares optimization minimizing the straightforward error metric
\begin{equation}
L = \sum_i (I''_{\mathrm{FTS}}(\lambda_i) - I(\lambda_i))^2,
\end{equation}
where $I$ is the spatially averaged \ac{susi} flat-field spectrum. The index $i$ runs over all spectral sampling elements. A field dependent matching along the spatial direction has not been considered, as tests have shown that spatial variations in the resolution and stray-light parameters are negligible.

\begin{figure}
\centerline{\includegraphics[width=\textwidth]{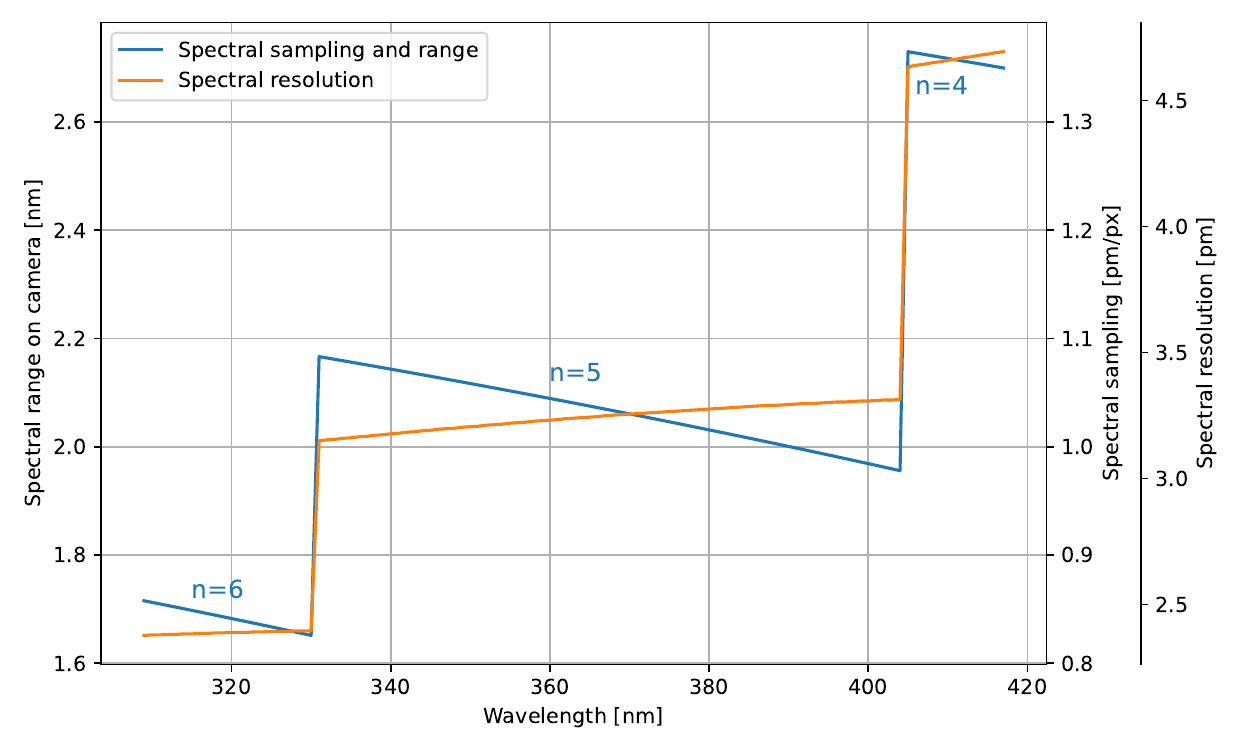}}
\small
\caption{Main \ac{susi} spectral characteristics versus wavelength: spectral sampling, wavelength range on the \ac{sp} cameras, and spectral resolution (\ac{lsf} \ac{fwhm}). The discontinuities mark the transitions between grating diffraction orders. The optimum orders to be used for each wavelength interval are labelled with $n=4 \ldots 6$ respectively.}
\label{fig_specres}
\end{figure}

Figures~\ref{fig_spectrum_cam1}~and~\ref{fig_spectrum_cam2} show an example flat-field spectrum recorded by \ac{sp} cameras no.~1 and 2 during a pre-flight test with Sun pointing in a spectral window around 412~nm. The matching yields $\sigma < 0.1$ pm 
for both \ac{sp} cameras, which shows that the \ac{susi} spectral resolution is practically diffraction limited. Figure~\ref{fig_specres} shows, among others, the diffraction limited \ac{fwhm} of the \ac{susi} \ac{lsf} as a function of wavelength. 

The fitted stray-light factors are 8.5\% and 5.5\% for \ac{sp} cameras SP1 and SP2 respectively. We haven't further assessed the sources of stray light, as the stray-light levels for both \ac{sp} cameras are within an acceptable range.         

\paragraph*{} Depending on the inversion code to be applied later, the stray-light factors and the \ac{susi} \ac{lsf} can either be provided as parameters to the inversion algorithm, or the measured spectra can be corrected prior to feeding them to the inversion. The first approach is typically beneficial in terms of noise characteristics, but it is not supported by all inversion codes.

\subsection{Polarimetric Calibration}
\label{sec_polcal}

\ac{susi} does not include a \ac{psg} onboard; therefore, its polarimetric demodulation matrix is estimated during pre-flight laboratory calibration measurements. The quality of this calibration is crucial to accurately demodulate the data after the flight and reach the instrument polarimetric accuracy goal. A detailed description of the polarimetric calibration is given in \cite{Iglesias25}, where we summarize the general procedure and main results.

The polarimetric calibration involves using \ac{led} light sources for different wavelength ranges, and a motorized \ac{psg} to generate 40 well-defined calibration Stokes parameters that are measured with \ac{susi}. The measurements are used to fit wavelength- and field-dependent \ac{susi} modulation matrices, along with additional unknowns: \ac{psg} waveplate retardance and position angle, and input intensity scaling coefficients. Given that the \ac{cmos} sensor used in the cameras has a rolling shutter, each sensor row samples a different \ac{pmu} rotation angle, and thus an independent fit is done for each pixel, or at least for each sensor row of the \ac{sp} cameras. We performed eight standalone calibration measurements at six different spectral ranges (324.00~nm - 326.00~nm; 326.81~nm - 328.80~nm; 348.73~nm -351.27~nm; 363.76~nm - 366.25~nm; 410.37~nm - 413.63~nm; 413.38~nm - 416.62~nm) before \ac{susi} integration into the \ac{pfi} unit, to asses the instrument functionality and its polarimetric response across the spectral working range. The standalone calibrations include the entire \ac{susi} \ac{sp} optical path, from the first scan mirror up to the \ac{sp} cameras. 

To ensure a target polarimetric accuracy of $1.0 \cdot 10^{-3}$, the absolute values of the elements of the polarimetric error matrix are required to stay below the following upper limits:
\begin{equation}
\label{eq:error_criterion}
   |\Delta X_{\mathrm{max}}|
    =
    \begin{bmatrix}
    - & 330 & 330 & 330 \\
    1.0 & 50 & 6.7 & 5.0 \\
    1.0 & 6.7 & 50 & 5.0 \\
    1.0 & 6.7 & 6.7 & 50 \\
    \end{bmatrix} \cdot 10^{-3}.   
\end{equation}
Following \cite{Ichimoto08}, the polarimetric error matrix is defined as
\begin{equation}
\Delta X = O_r^{-1} O - \mathbf{1},
\end{equation}
where $O_r$ is the modulation matrix used in the data reduction, $O$ is the true, but unknown, modulation matrix, and $\mathbf{1}$ is the identity matrix. $\Delta X_{\mathrm{max}}$ is based on conservative estimates for the maximum expected solar polarization signals (15\% in Stokes Q, U and 20\% in Stokes V) and on a relative scaling error of 5\% in all Stokes parameters.

The error matrix due to instabilities in the instrument and/or the calibration can be estimated by comparing the modulation matrices for a given wavelength $\lambda$ obtained at two different times (the latter denoted with indices 1 and 2): 
\begin{equation}
\Delta X_{\lambda}= {\hat{O}_{\lambda, 1}}^{-1} \, \hat{O}_{\lambda, 2} - \mathbf{1}.
\end{equation}
The modulation matrices $\hat{O}_{\lambda, 1}, \hat{O}_{\lambda, 2}$ are normalized to their first element. For measurements in the 365~nm range acquired five days apart, we get for each \ac{sp} camera:
\begin{equation}
   |\Delta X_{365}|_{\mathrm{SP1}}
    =
    \begin{bmatrix}
-&0.2\pm0.2&0.2\pm0.3&0.9\pm0.5\\
0.3\pm0.1&0.8\pm0.2&2.1\pm0.6&1.2\pm0.2\\
0.5\pm0.1&1.9\pm0.5&0.6\pm0.2&0.5\pm0.3\\
0.3\pm0.1&1.8\pm1.4&0.5\pm0.1&0.9\pm0.2&
    \end{bmatrix} \cdot 10^{-3},   
\end{equation}

\begin{equation}
   |\Delta X_{365}|_{\mathrm{SP2}}
    =
    \begin{bmatrix}
-&0.7\pm0.7&0.7\pm0.6&1.6\pm0.6\\
0.2\pm0.5&0.2\pm0.6&2.4\pm0.6&1.3\pm0.4\\
0.7\pm0.5&2.8\pm0.5&0.1\pm0.5&0.0\pm0.5\\
0.3\pm0.2&1.2\pm1.4&0.1\pm0.4&0.2\pm0.4
    \end{bmatrix} \cdot 10^{-3},   
\end{equation}

Then, for measurements in the 325~nm -- 327~nm range acquired approximately a month apart, we get:
\begin{equation}
   |\Delta X_{325}|_{\mathrm{SP1}}
    =
    \begin{bmatrix}
-&1.5\pm1.0&0.2\pm1.2&1.2\pm2.6\\
0.4\pm2.2&1.5\pm1.8&2.1\pm2.0&0.3\pm3.3\\
{0.9\pm2.1}&1.9\pm2.3&0.7\pm1.6&1.2\pm0.6\\
0.1\pm1.1&2.4\pm2.2&1.1\pm1.0&0.8\pm1.9
    \end{bmatrix} \cdot 10^{-3},   
\end{equation}

\begin{equation}
\label{eq:sta_325d_327_sp2}
   |\Delta X_{325}|_{\mathrm{SP2}}
    =
    \begin{bmatrix}
-&1.9\pm1.7&1.3\pm1.5&2.3\pm2.3\\
0.3\pm0.9&3.8\pm5.3&1.5\pm2.0&0.1\pm1.2\\
{0.6\pm0.6}&1.7\pm1.6&1.8\pm4.0&0.5\pm2.6\\
{0.6\pm1.0}&3.3\pm1.2&1.9\pm0.6&3.3\pm3.7&
    \end{bmatrix} \cdot 10^{-3},   
\end{equation}
where each element indicates the mean $\pm$ the \ac{rms} computed over the sensor area. Note that the  mean values of all elements in both channels are smaller than $|\Delta X_{\mathrm{max}}|$.

Some $I\to Q,U,V$ elements exceeding the error requirements\footnote{Such excessive errors are in particular due to instrument jitter and residual camera non-linearity, in combination with strong spectral or spatial intensity gradients. The polarimetric calibration measurements carried out with artificial illumination are not sensitive enough to these types of errors.} can be calibrated post-facto by using magnetically insensitive spectral lines and/or spectral continuum data which are known to present no solar Zeeman polarization signatures. 

\begin{figure}
\centering
\includegraphics[scale=0.31]{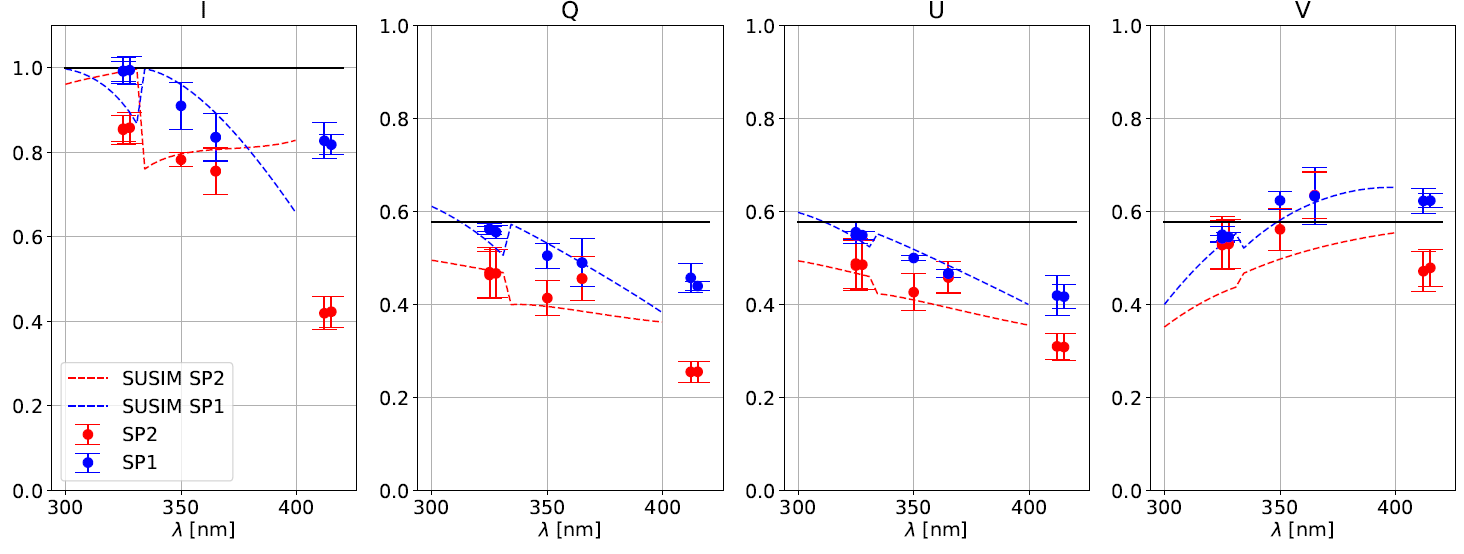}
\caption{{ Measured (\textit{dots}), modelled (\textit{dashed lines}) and ideally balanced (\textit{black lines}) polarimetric efficiencies vs. wavelength for each camera, see the legend. SUSIM in the legend refers to the \ac{susi} instrument model, a numerical model including among others the polarimetric properties of the instrument. For the measurements we show the mean values (\textit{dots}) and the 3 $\cdot$ \ac{rms} values (\textit{error bars}) computed over the sensor area. See the text for additional details.}}
\label{fig_pol_eff}
\end{figure}

Finally, the measured polarimetric efficiencies are within $\sim10\%$ of the design values, with the SP2 camera channel presenting up to $\sim20\%$ lower efficiencies in Q, U and V (see Figure~\ref{fig_pol_eff}).

To retrieve accurate solar atmospheric quantities, the polarimetric response of the full \sunrisethree{} beam path, including the telescope and \ac{islid} optics, must be calibrated. We have performed additional polarimetric calibrations at the F1 and F2 focal positions, which are currently under analysis. Their results are to be combined with the standalone \ac{susi} calibration, to derive and validate the end-to-end calibration methodology to be applied to \ac{susi} scientific data. Such an analysis will be reported in a future paper, including a validation using solar observations acquired during pre-flight ground tests with Sun pointing and during the flight.

\subsection{Photon Budget}
\label{sec_throu_char}

Given the low solar flux in the \ac{nuv} in combination with high resolution observations, a careful assessment of the photon budget is crucial. The photon flux predictions have limited accuracy. Near-UV observations with SuFI during previous \sunrise{} flights have been a valuable source of information for constraining the residual atmospheric extinction at flight altitude, but some substantial uncertainty remains. Lab verification measurements with calibrated light sources at different wavelengths are consistent with the predicted instrument throughput to within some $\pm$ 30\%. The same is true for solar flux measurements at wavelengths around 400\,nm, acquired in the \ac{sp} channel during pre-flight tests including the full telescope beam path, and despite substantial weather-dependent uncertainties in atmospheric extinction at ground level. The measurement results, supporting our predictions at selected wavelengths, give us sufficient confidence in the expected photon budget for the entire \ac{susi} spectral working range. The level of confidence is acceptable for realistic observations planning, in particular in terms of defining \ac{sp} scan speeds and integration times. 

In the case of the \ac{sj} channel, a certain confidence in the signal level is crucial to avoid unrecoverable detector saturation\footnote{Detector saturation is unrecoverable if the amount of photo-charges, collected during the shortest possible exposure time, exceeds the pixel full-well.}. Again, the level of confidence is acceptable here, guaranteeing sufficient margin in terms of camera exposure times.

\begin{figure}
\centerline{\includegraphics[width=\textwidth]{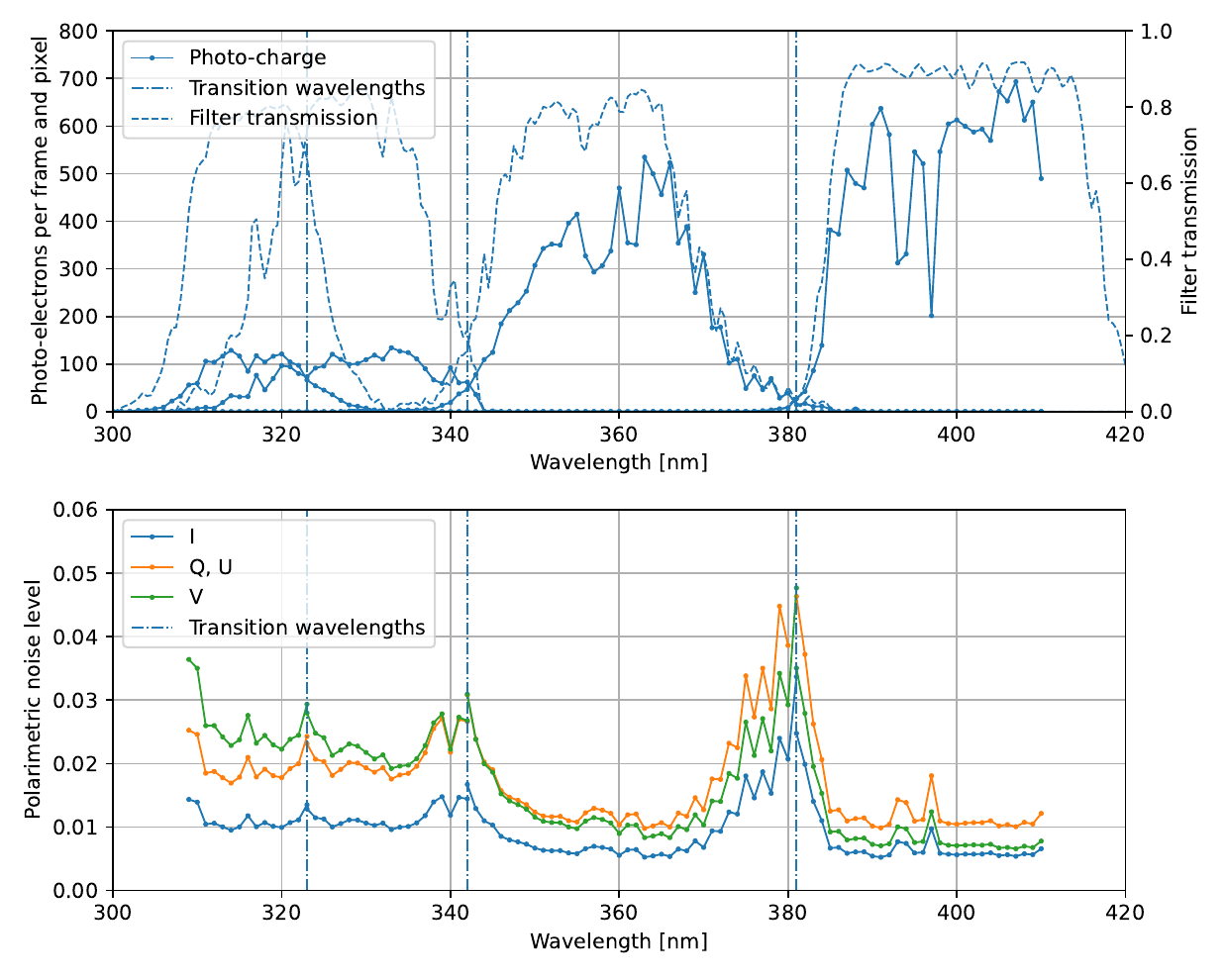}}
\small
\caption{\textit{Upper panel:} expected amount of photo-electrons collected per pixel and frame in a single \ac{sp} camera, running at nominal frame rate (\textit{solid curves}). For reference, the transmission curves of the four \acp{osf} are overplotted (\textit{dashed curves}). The \textit{vertical dash-dotted lines} mark the optimum transition wavelengths between \acp{osf}. \textit{Lower panel:} expected polarimetric noise levels for the different Stokes parameters, after 1s integration, and at the intrinsic \ac{sp} spectral and spatial sampling (no data binning).}
\label{fig_phot_budget}
\end{figure}

\begin{table}
    \begin{center}
    \begin{tabular}{llllll}    
    & \textbf{Symbol} & \multicolumn{2}{c}{\textbf{Value}} & \textbf{Unit} & \textbf{Comments} \\
    & & \textbf{SP} & \textbf{\ac{sj}} &  &  \\
    \hline   
    \noalign{\smallskip}    
     Wavelength & $\lambda$ & 360 & 325 & nm & \\
     Solar spectral irrad. & $I_0$ & 1.16 & 0.794 & W m$^{-2}$ nm$^{-1}$ & 1 \\
     Limb darkening & $L$ & 0.708 & 0.691 &  & 2 \\
     Telescope area & $A$ & \multicolumn{2}{c}{0.753} & m$^2$ & \\
     Spatial sampling & $\Delta x_0$ & 0.034 & 0.031 & \arcsec & \\
      & $\Delta y_0$ & 0.035 & 0.035 & \arcsec & \\
     Slit width & $s$ & \multicolumn{2}{c}{0.07} & \arcsec & \\
     Spectral sampling & $\Delta \lambda_0$ & 1.18 & 910 & pm & 3 \\
     Atmosph. extinction & $E$ & 0.89 & 0.82 & & 4 \\
     Grey throughput & $T$ & 0.13 & 0.18 & & 5 \\
     Grating apert. fact. & $G$ & 0.76 & n/a & & 6\\
     Grating reflectivity & $T_g$ & 0.7 & n/a & & \\ 
     OSF transm. & $T_o$ & 0.79 & n/a & \\
     PBC transm. & $T_p$ & 0.47 & n/a & \\
     Detector QE & $Q$ & 0.62 & 0.46 & \\
     Exp. time & $\Delta t_0$ & 21.3 & 1 & ms & \\
    \hline 
    Coll. photo-charge & $C$ & $6.0 \cdot 10^2$ & $3.7 \cdot 10^4$ & e- & 7 \\
    \hline
    Nominal gain & $g$ & 1.5 & 0.054 & DN/e- & \\
    Detector response & $S$ & $9.0 \cdot 10^2$ & $2.0 \cdot 10^3$ & DN & 8 \\
    \end{tabular}
    \end{center}
    \underline{Comments}\\
    \vspace{-\baselineskip}
    \begin{enumerate}
    \renewcommand{\labelenumi}{\arabic{enumi}.}
    \item Based on the ATLAS~3 spectrum \citep[see e.g.][]{Thuillier04}
    \item Ratio between disk integrated and disk center intensity \cite[][Table~14.17]{Cox99}
    \item For the \ac{sj} channel this corresponds to the equivalent width of the bandpass filter profile
    \item For a median Sun elevation of 22.5$^{\circ}$; estimate based on solar flux measured with \ac{sufi} during previous \sunrise{} flights. 
    \item Includes all optical components with a low wavelength dependence of their reflectivity or transmission.
    \item Fraction of the effective pupil area covered by the grating, taking into account diffraction at the \ac{sp} entrance slit
    \item Per pixel and frame
    \item The maximum value is 4095 DN (12 bit ADC). The practical limit, defined by linearity considerations (cf. Section~\ref{sec_camera_characterization}), is lower. The values stated here are in a safe range, with sufficient margin. 
    \end{enumerate}
    \caption{Detailed photon budget for the \ac{sj} channel and for one example wavelength of the \ac{sp} channel.}
    \label{tab_phot_budget}
\end{table}

Figure~\ref{fig_phot_budget} shows the expected amount of photo-electrons collected per pixel and frame in each \ac{sp} camera, at the intrinsic spatial and spectral resolution, and with the camera running at its nominal frame rate. Table~\ref{tab_phot_budget} shows the detailed photon budget for one example \ac{sp} wavelength. The table also shows the expected amount of photo-charges collected in the \ac{sj} camera, which is operated with a fixed bandpass filter. The photo-charge spectrum is based on ATLAS~3 solar spectral irradiance data \cite[cf.][]{Thuillier04} that have been interpolated to a consistent sampling of 1\,nm. The flux in the cores of narrow spectral lines is not properly represented at this sampling, and more detailed assessments are necessary for the planning of science observations which crucially depend on line core signal-to-noise ratios. The photo-charge spectrum shows three areas of increased sensitivity which are essentially delimited by the \ac{osf} transmission spectra (and to a lesser extent by the grating diffraction orders). The first area, centered around 325\,nm, corresponds to the combined coverage of \acp{osf} no.~1 and 2. The other two areas, centered around 360\,nm and 400\,nm, are covered by \acp{osf} no.~3 and 4 respectively.

At the lower end of the spectral working range, and for spectral lines with a depth of order 10\% of the neighbouring continuum intensity, the photo-charge collected in the line core approaches the noise-equivalent charge of the camera readout noise. In this case, which can be seen as a limiting case, the number of accumulated frames has to be increased by a factor of two in order to reach the same noise level as in the ideal photon limited case. The integration overhead, which we define as the ratio between the frame number $n$ required to reach a given noise level in the presence of camera noise, and the frame number $n_p$ in the photon-limited case, is given by $n/n_p = 1 + \sigma^2/C$, where $C$ represents the amount of photo-charges and $\sigma$ the readout noise.

Figure~\ref{fig_phot_budget} also shows the expected polarimetric noise levels after 1s integration, and at the intrinsic \ac{sp} resolution. The values are based on the \ac{susi} polarimetric efficiencies (cf. Section~\ref{sec_polcal}) and it is assumed that the images of both \ac{sp} cameras are combined (dual-beam polarimetry). 

To reach the required polarimetric sensitivity, the data have to be binned in one or several dimensions (temporal, spatial, or spectral). Many-line inversions (cf. Section~\ref{sec_reqs}) can be considered as a special case of spectral binning in this context. We note here that \ac{susi} records raw data at the intrinsic resolution of the instrument. Binning will be performed during post-flight data reduction, and can be handled in a flexible way, depending on the needs of a specific science case (cf. Section~\ref{sec_data_reduction}). The only relevant parameter to be set during observation is the scan speed. 

In general, the polarimetric noise spectrum shown in Figure~\ref{fig_phot_budget} can be re-scaled to different resolutions and solar flux values\footnote{E.g. taking into account line core intensities from a solar spectrum representative of the \ac{susi} spectral resolution. Further, in the context of this scaling rule, we assume constant photo-charge distribution across the resolution elements. In practice this assumption is approximately valid, if the resolution elements are small compared to spatial or spectral structures on the Sun with significant contrast.} as:
\begin{equation}
\sigma = \sigma_0 \cdot \left( \frac{I_0}{I} \cdot \frac{\Delta \lambda_0}{\Delta \lambda} \cdot \frac{\Delta x_0}{\Delta x} \cdot \frac{\Delta y_0}{\Delta y} \cdot \frac{1}{\Delta t} \right)^{1/2},
\end{equation}
The original and target polarimetric noise values are denoted by $\sigma_0$ and $\sigma$ respectively. The integration time $\Delta t$ is in units of seconds. The other quantities and their units are as defined in Table~\ref{tab_phot_budget}. The scan speed $v$ to be set for a given observing program is $v = \Delta x / \Delta t$. For some (if not most) science cases it is appropriate to take into account the coupling between spatial and temporal resolution due to solar evolution. The scan speed $v$ is then also to be interpreted as the signal speed in the solar atmosphere which is most characteristic for the considered science case. Example values for $v$ are the sound speed in the photosphere or in chromospheric layers. Assuming $\Delta x = \Delta y = v \, \Delta t$, the noise then scales as 
\begin{equation}
\sigma = \sigma_0 \cdot \left( \frac{I_0}{I} \cdot \frac{\Delta \lambda_0}{\Delta \lambda} \cdot \frac{\Delta x_0 \Delta y_0}{v^2} \right)^{1/2} \cdot \Delta t^{-3/2}.
\label{eq_noise_solar_evolution}
\end{equation}
For a given target noise level and expected signal speed on the Sun, we can thus define an appropriate and unique integration time $\Delta t$, and derive therefrom the later spatial binning to be performed in the data reduction. We also note that for a given target noise level, the appropriate spatial sampling scales with $v^{1/3}$, as can be seen from Equation~\ref{eq_noise_solar_evolution}. This means for example that the spatial resolution that can be achieved in observations of the chromosphere is lower than for photospheric observations, not only due to the typically lower flux in chromospheric lines, but also due to faster solar evolution. Table~\ref{tab_noise_cases} shows two example observations and the resolution that can be realistically achieved, based on the above considerations.      

\begin{table}
    \begin{center}
    \begin{tabular}{lp{30ex}p{30ex}l}    
    & \textbf{Case 1} & \textbf{Case 2} &  \\
    \hline   
    Description & Zeeman diagnostics in the wings of most lines within a spectral window centered at 328\,nm. & Observations of scattering polarization in the CaII K line core. & \\
    \hline
    & \multicolumn{2}{c}{Values} & Unit \\
    \hline
    $\sigma$ & $2 \cdot 10^{-3}$ & $1 \cdot 10^{-3}$ & \\
    $\lambda$ & 328.0 & 393.3 & nm \\
    $I_0 / I$ & 1 & 2 & \\
    $v$ & 0.01, 0.07 & 0.01, 0.07 & \arcsec~s$^{-1}$\\
    \hline 
    $\Delta t$ & 11.6, 3.2 & 12.4, 3.4 & s \\
    $\Delta x \, (=\Delta y$) & 0.12, 0.22 & 0.12, 0.24 & \arcsec \\
    \hline  
    \end{tabular}
    \end{center}
  
    \caption{Example observations and derived temporal and spatial sampling to reach a given target polarimetric noise level, taking into account solar evolution. The sampling is scaled with respect to the largest noise value of Stokes Q, U and V (cf. Figure~\ref{fig_phot_budget}). Further, the sampling is computed for two different solar signal speeds (values separated by commas), representative of typical photospheric and chromospheric conditions. The spectral sampling is kept at its intrinsic value.}
    \label{tab_noise_cases}
\end{table}

\section{Instrument Control and Operations}
\label{sec_operation}

\subsection{Instrument Control}
\label{sec_control}

The main \ac{susi} instrument control tasks are: (1) To control the four mechanisms of the \ac{osf}, grating, scanner, and \ac{pmu} units. (2) To handle the data streams from the three \ac{susi} cameras in real time. (3) To compress the camera data and to transfer the compressed images to the \acf{ics}, which stores the data on its onboard \ac{dss} \citep{Lagg25}. Since this task does not differ considerably from the one of the \ac{sufi} instrument flown during the first two \sunrise{} science flights \citep{Barthol11} and because the development of instrument software requires a sizeable amount of manpower, the aim was to provide a system design that allows for a re-use or an easy refactoring of large parts of the \ac{sufi} software. A major challenge is the fact that the raw data rate of \ac{susi} is 140 times higher than the one of \ac{sufi}.

Although the technological advances of recent years led to significantly more powerful embedded PCs with much higher transfer rates, the limitations of a balloon mission with respect to mass, electrical power consumption, and available financial budget did not change considerably compared to the first two \sunrise{} flights. These boundary conditions finally led to the decision not to run the \ac{susi} software on a separate dedicated instrument computer (as it was the case for \ac{sufi} in \sunriseonetwo~and as it is still the case for \ac{scip} and \ac{tumag} in \sunrisethree), but to run the  system control software of the observatory and the \ac{susi} software in parallel on the \acf{icu}, an embedded PC of type Supermicro X10SDV-7TP8F. The PC is equipped with an Intel processor Xeon D-1587 with 16 cores \citep{Lagg25}. The mainboard possesses sufficient computational power and memory to compress the \ac{susi} data in real time, to accomplish all other tasks of the system control and \ac{susi} software, and to still have enough margin to compensate for fluctuations in the needed computational power. A framegrabber of type KY-FGK-CLHS developed by the company Kaya Instruments in the form of a PCI express plug-in card is used for connecting the three \ac{susi} cameras via fiber optic cables.

Because the used hardware is based on commercial off-the-shelf components, the thermal requirements were most easily fulfilled by housing the hardware inside a pressurized vessel. Hence all interfaces to the outside world require vacuum feedthroughs. A 10-Gbps Ethernet interface between \ac{susi} and the \ac{icu} and the corresponding feedthroughs are not needed because both software packages run on the same hardware. This change in the hardware design is transparent to the software because the data transfer is programmed using the socket interface \citep{Stevens97}.

An additional novelty in the design is the fact that the \ac{susi} mechanisms are not connected directly to the instrument computer, but an additional intermediate hardware layer in the form of the \ac{scu} was introduced. The \ac{scu} is based on a small Raspberry Pi single-board computer and receives commands for controlling the mechanisms (grating, scanner, filter wheel) from the \ac{susi} software via a 100-Mbps Ethernet interface and it sends housekeeping data of the mechanisms to the \ac{susi} software. The integrated intelligence of the \ac{scu} translates the high-level commands received from the \ac{susi} software into low-level commands that are needed to control the actual mechanism hardware.

\subsection{Software Design}
\label{sec_software}

The major tasks of the \ac{susi} software comprise the following items:
\begin{itemize}
\item Initialize and configure the three \ac{susi} cameras
\item Initialize and configure the four mechanisms for \acp{osf}, grating, scanner and \ac{pmu}
\item Receive and compress the camera data in real time
\item Acquire and collect all housekeeping data
\item Create thumbnails (extracts of images having a reduced amount of data) on demand
\item Transfer the compressed camera data, housekeeping data, and thumbnails to the \ac{icu}
\item Receive and execute commands sent by the \ac{icu}
\end{itemize}

For a nearly autonomous operation of an observatory over six days, robustness and reliability of the software are absolutely mandatory. We therefore use the \ac{ace} library whose maturity has been proven for many applications on multiple platforms over more than two decades \citep{Schmidt01, Schmidt02}. Core of the \ac{ace} library
is a Thread Pool Reactor \citep{Schmidt00} responsible for the data transfer between \ac{susi} and \ac{icu} via the TCP/IP protocol. Science data, housekeeping data, and thumbnails are transferred from the \ac{susi} software to the system control software (local host communication), where housekeeping and thumbnail data are downlinked to the ground station if telemetry bandwidth is available. Science images and housekeeping data are stored onboard on the \ac{dss}. In the opposite direction, commands from the system control software are transferred to the \ac{susi} software. There are two possible command sources, either from a pre-defined observing program stored on the \ac{icu} and processed in the timeline framework or from the ground operator's manual input transferred
over the telemetry channels. Commands can have one of two priority levels, i.e., there are high and low priority commands. In case the execution of a low priority command needs much time and hence blocks the command channel, the concept allows for bypassing the blocking via the high priority command channel. Eight threads are available for the Thread Pool Reactor in order to process multiple transfer channels in parallel and keeping the \ac{susi} software responsive at any time.

Almost all the computational power required by the \ac{susi} software is needed for the lossless compression of the camera data. A higher efficiency has been reached by replacing the zlib compression algorithm used in \ac{sufi} (average compression factor of about 1.6) by the Anacrunch algorithm developed by R.~A.~Shine \citep[][average compression factor of about 2]{Shine92}. The compression part of the software is explained in more details below.

\begin{figure}
\centering
\includegraphics[width=\textwidth]{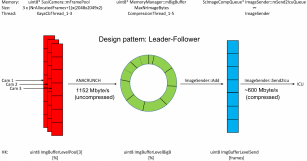}
\caption{Schematic illustration of the Leader-Follower design pattern for compressing the \ac{susi} data and transferring the compressed data to the \ac{icu}. See main text for details.}
\label{fig_leader_follower}
\end{figure}

During the initialization each of the three \ac{susi} cameras is assigned memory for 145 images (red blocks in Figure~\ref{fig_leader_follower}). The framegrabber driver creates a new thread for each of the three cameras that stores the image in the next free block. A pool of threads organized according to the design pattern Leader-Follower \citep{Schmidt00} provides sufficient resources for compressing the images in real time. The next free worker thread of the pool compresses the next image via the Anacrunch algorithm and stores the compressed data in a big ring buffer (green blocks in Figure~\ref{fig_leader_follower}). Additionally, the worker thread adds a pointer to the compressed data to the queue coloured in blue in Figure~\ref{fig_leader_follower}. A further thread named ImageSender transfers the data to the \ac{icu} via the Thread Pool Reactor. Every three seconds the filling level of the buffers and queue is sent to the ground station and displayed to the operator. Fluctuations of the data rate are compensated by the green ring buffer, that is designed big enough to store the entire data stream for 17 seconds.

Since the development of a concurrent and distributed application is error-prone, we developed a tool that detects memory allocation issues already in the pre-flight phase.
A second tool named Callstack-Logger can identify deadlocks and race conditions. The performance of the software has been optimized by a Profiler by measuring the execution time
of critical software parts. The development of the Callstack-Logger and Profiler is based on the technique of method call interception \citep{Sayfan05}.

\subsection{Co-Alignment between Science Instruments}
\sunrisethree{} can simultaneously observe the same solar features with its three instruments. Before launch, the co-alignment between the two slit-based instruments, \ac{susi} and \ac{scip}, is determined. An offset between the center positions of their slits is corrected by commanding an offset of opposite sign to either scan unit. The \ac{scip} scan unit can be offset in both the slit and scan directions. The \ac{susi} scan unit allows a correction along the scan direction only. The angle between the slits is minimized once during the integration of the instruments in the \ac{pfi}, and cannot be actively corrected during operation. The scan speeds of both instruments are equated by commanding a speed correction factor to the \ac{susi} scan unit. 

In the commissioning phase after launch, co-alignment and scan speed are checked again, and deviations from pre-flight values are corrected by updating the commanded offset and speed parameters. The in-flight co-alignment procedure uses grid target images recorded simultaneously by both instruments. The \ac{susi} images are re-binned to match the larger plate scale of \ac{scip}. The images are then cross-correlated, and alignment is verified by blinking the two images and taking their difference. The offset and the angle between the slits are computed after image alignment. Several images, recorded at defined times during a grid target scan, are used to determine the scan speeds. This procedure results in an offset accuracy of about 0.1\arcsec{} at all times during \ac{susi} and \ac{scip} co-scanning. The accuracy is within the 0.3\arcsec{} requirement. The residual angle error between the slits is 0.55°, which keeps the maximum offset at both slit ends below 0.3\arcsec. 

\subsection{Data Reduction and Data Products}
\label{sec_data_reduction}

\begin{figure}
\centerline{\includegraphics[width=\textwidth]{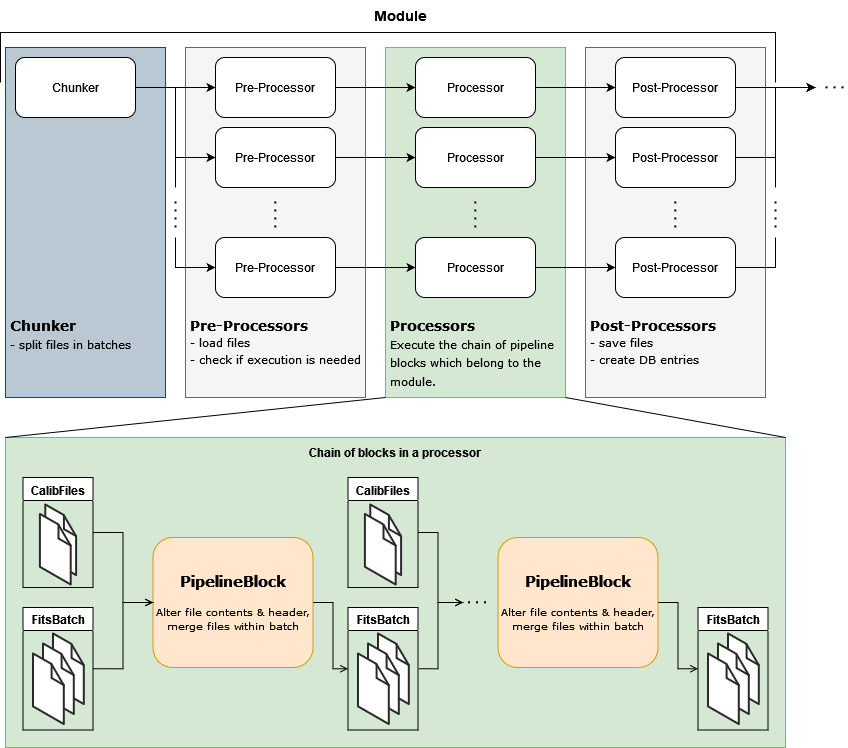}}
\small
\caption{Structural diagram of the \ac{susi} data reduction pipeline. For each module the \acs{fits} files to be processed are split into batches. Batch sizes are typically tailored to the capabilities of the computing infrastructure and processing requirements. For each batch a parallel processing chain is started where the blocks of the module are executed sequentially. Each block takes the batch of files and (optional) auxiliary files as input and returns a batch of \acs{fits} files.}
\label{fig_pipeline}
\end{figure}

The \ac{susi} data reduction pipeline is implemented as a \textsf{Python} based processing chain, which is also designed for efficient processing of large amounts of data\footnote{Large in relation to the typical computing environment available to the authors for data reduction tasks, at the time of writing.}. We note that a typical science block with e.g. a duration of 60 minutes, results in about 4 TB of uncompressed data to be fed into the pipeline. 

The processing chain consists of different modules. Each module can run in parallel on multiple batches of \ac{susi} data frames. Modules consist of an arbitrary number of processing blocks that will be applied in sequence within one batch. 
All pipeline blocks read a batch of data frames, in the form of files in the widely used \ac{fits} format \citep{Wells81}, and optional auxiliary input data such as a gain table, a pixel offset map for distortion corrections, or a demodulation matrix. 
The output is again a batch of data frames (\ac{fits} files) with altered data and metadata. The processing history is tracked in the metadata, typically in form of \ac{fits} headers, or in a separate database. The sizes of the input and output data may differ. For example, upon demodulation or temporal averaging, the number of files in the batch is reduced. Figure~\ref{fig_pipeline} shows a structural diagram of the pipeline. Some example blocks are described in Table~\ref{tab_pipeline_blocks}.

\begin{table}
\caption{
Some generic example blocks from the \ac{susi} pipeline. For \ac{sj} and \ac{sp} data, additional specialized blocks may be applied, depending on the requirements of a given science case.
}
\label{tab_pipeline_blocks}
\centering 
\begin{tabular}{p{.75in} | c | p{2.15in} | p{1in} } 
Name & ID & Description  & Auxiliary Files \\
\hline
Binning & \textbf{B} & Performs data binning in the temporal, spatial or spectral dimension, with the main goal to reduce noise  & - \\
\hline
Camera \mbox{Calibration} & \textbf{C} & 
Includes basic calibration steps, i.e. dark field, response non-linearity. Also includes cropping of unwanted areas in the data frames, e.g. shielded camera pixels &
DARK FIELD \\
\hline
Flat \mbox{Fielding} & \textbf{F} & Applies a flat-field correction & GAIN TABLE \\
\hline
Demodulation & \textbf{D} & Polarimetric demodulation of the data to yield Stokes images & DEMODULATION MATRIX
\end{tabular}
\end{table}

After each module, the resulting files are saved as a specific data product level. The metadata of the given product level are checked before execution of the next module and only those files are processed which are not available in the next product level yet. This flexible implementation allows for the definition of multiple pipelines, with partially overlapping modules, or for recurrent data processing with partially modified pipelines while keeping the usage of computing resources at a minimum. Limiting the data (re-)processing to altered parts of a pipeline is particularly important in the early phase of working with \ac{susi} data, when algorithm and pipeline adjustments are expected to be frequent. 

A given pipeline is represented in configuration files by a sequence of characters, and vertical dashes, e.g. ``\texttt{C}\(|\)\texttt{FBD}''. Each character denotes a block. The dashes act as separators between modules and thus also define the different reduction levels where cached data are written out. Future improvements can be conveniently added as new blocks to the pipeline. A versioning scheme of pipelines and data products allows for transparent change tracking, and is also reflected in the metadata of a given data product to enable a clear traceability of the data reduction process. The versioning convention reflects breaking and non-breaking (i.e. backwards compatible) changes, in accordance with common practices in software engineering. All pipelines share the level 0 data product, which basically represents raw data after applying camera calibration (block~\texttt{C}). The other data product levels are pipeline specific, depending on the block(s) applied in each module. 

An overview of the planned pipelines and resulting available data products can be found in Table~\ref{tab_pipeline_definitions}. Some of those pipelines are already implemented and tested, particularly those needed for pre-flight test-data reduction. The remaining pipelines will be further developed after a successful science flight. A full description of available blocks, modules, pipelines and intermediate internal levels can be found in the software documentation. The \ac{susi}-specific data reduction software is currently accessible upon request only, a public release is planned at a later stage.  

\begin{table}
\caption{Planned \ac{susi} standard pipelines and resulting data products}
\label{tab_pipeline_definitions}
\centering 
\begin{tabular}{p{.83in} | c | p{3.3in}} 
Name & ID & Description\\
\hline
Quick-Look \mbox{Polarimeter} & QP &
Produce highly binned, partially calibrated and unrestored single-beam Stokes images, keeping computing resources low. The main purpose of this data product is to allow for a quick assessment of general data quality or suitability, before applying more expensive pipelines yielding data products for actual science analysis. \\
\hline 
Quick-Look \mbox{Slit-Jaw} & QS &
Produce highly binned, partially calibrated and unrestored SJ context images. Otherwise same purpose as for QP. \\
\hline 
Low-Resolution Polarimeter & LP &
Produce fully calibrated but unrestored dual-beam Stokes images \\
\hline 
Low-Resolution Slit-Jaw & LS &
Produce fully calibrated unrestored SJ context images \\
\hline 
High-Resolution Polarimeter & HP &
Produce fully calibrated and restored  dual-beam Stokes images with highest spatial resolution \\
\hline 
High-Resolution Slit-Jaw & HS &
Produce fully calibrated and restored slit-jaw context images with highest spatial resolution 
\end{tabular}
\end{table}

For flat fielding and for removing the spectrographic curvature effect, we rely on the \textsf{spectroflat} library \citep[see][]{Hoelken23}. For the image restoration of the \ac{sj} context data and of the SP spectral scans we plan to use the \ac{momfbd} and \textsf{specrestore} codes \citep[see][respectively]{vanNoort05, vanNoort17}. A \ac{susi} specific implementation based on those code packages is currently in progress and will be described in a later publication. 

Besides the \ac{susi} data reduction pipeline, we have created an online observation database where for each calibration measurement and each observation an entry will be made. Each entry contains the most relevant information such as start and end time, scan mode, scan range and speed, wavelength, NOAA region number (if any) etc. Further any event can be logged and will be linked to the observation(s) carried out during the event. It is planned to make this database public along with the data release to provide transparent access to any observational information related to the data. The database also allows for fast browsing and searching for interesting data sets. 
The data reduction pipeline can be connected to the observation log database via a Representational State Transfer Application Programming Interface (REST API). This allows to retrieve relevant information in an automated way, reducing human errors.

\section{Conclusions}\label{sec_conclusions}

\ac{susi} is designed to significantly extend the diagnostic potential of existing solar instrumentation. A unique feature of \ac{susi} is the combination of high spatial and spectral resolution, and polarization information, in the \ac{nuv} region which is densely populated with spectral lines covering a large atmospheric height range. 
This enables \ac{susi} to study the complex interplay of magnetic fields, plasma motions and heat transport mechanisms in the solar atmosphere in an underexplored spectral region, bearing large potential for new discoveries. 
The combined and complementary information of simultaneous and co-spatial observations with \ac{scip} and \ac{tumag} will be another key advantage. 

The extensive experience with ground-based solar spectropolarimeters and the valuable lessons learned from the previous \sunriseonetwo{} missions have definitely benefited the development of \ac{susi}. The instrument design, integration and testing, as well as the in-flight operation has been carried out at the \ac{mps}. A large number of subsystems have been developed at the \ac{mps} as well, as mentioned in the respective sections of this work. Numerous verification measurements have confirmed that \ac{susi} meets our requirements in terms of optical, spectral and polarimetric performance. Successful environmental tests at subsystem and system levels give us great confidence that the instrument will perform reliably under the challenging conditions of a stratospheric flight. Autonomous and coordinated observing by means of scripted timelines and observing programs as well as general in-flight operation has been extensively tested as well, in close co-operation with the teams of the other science instruments, and including gondola/pointing and image stabilization \citep{Lagg25}.

Although the first flight of \sunrisethree{} did not reach its science phase due to a technical failure in the gondola elevation axis mechanism \citep{Lagg25}, the functional tests performed during the ascent phase and during the early commissioning phase at final float altitude have further increased our confidence in the reliability of the instrument. The second flight attempt in July 2024 was a great success, and at the time of writing the science data reduction process has just begun.

\begin{fundinginformation}
 The German contribution to \sunrisethree{} and \ac{susi} is funded  by the Strategic Innovations Fund of the President of the Max Planck Society (MPG), and by the Max-Planck-Förderstiftung with private donations by supporting members of the MPG. \ac{susi} has also received funding from  the Bundesministerium für Wirtschaft und Klimaschutz through Deutsches Zentrum für Luft- und Raumfahrt e.V. (DLR), grant. no.~50 OO 1608, and from the European Research Council (ERC) under the European Union's Horizon 2020 research and innovation programme (grant agreement no.~101097844, project WINSUN).
 The contributions of J.~H\"olken, D.~Vukadinovi\'{c} and K.~Sant are supported by the International Max Planck Research School (IMPRS) for solar system science.
 F.~Iglesias is a member of the ``Carrera del Investigador Cient\'ifico" of CONICET and supported by the MPG through the Max Planck Partner Group between MPS and the University of Mendoza, Argentina. The Japanese contribution has been funded by the ISAS/JAXA Small Mission-of-Opportunity program and JSPS KAKENHI JP18H05234/JP23H01220. The Spanish contribution has been funded by the Spanish MCIN/AEI under projects RTI2018-096886-B-C5, and PID2021-125325OB-C5; and from the ``Center of Excellence Severo Ochoa'' awards to IAA-CSIC (SEV-2017-0709, CEX2021-001131-S), all co-funded by European REDEF funds, ``A way of making Europe''.
 \end{fundinginformation}

\bibliographystyle{spr-mp-sola}
\bibliography{main}  

\begin{thebibliography}{41}
% BibTex style file: spr-mp-sola.bst, v2.04, 2021-11-17
\ifx\bisbn     \undefined \def\bisbn  #1{ISBN #1}\fi
\ifx\binits    \undefined \def\binits#1{#1}\fi
\ifx\bauthor   \undefined \def\bauthor#1{#1}\fi
\ifx\batitle   \undefined \def\batitle#1{#1}\fi
\ifx\bjtitle   \undefined \def\bjtitle#1{\textit{#1}}\fi
\ifx\bvolume   \undefined \def\bvolume#1{\textbf{#1}}\fi
\ifx\byear     \undefined \def\byear#1{#1}\fi
\ifx\bissue    \undefined \def\bissue#1{#1}\fi
\ifx\bfpage    \undefined \def\bfpage#1{#1}\fi
\ifx\blpage    \undefined \def\blpage #1{#1}\fi
\ifx\burl      \undefined \def\burl#1{#1}\fi
\ifx\href      \undefined \def\href#1#2{#2}\fi
\ifx\betal     \undefined \def\betal{et al.}\fi
\ifx\bctitle   \undefined \def\bctitle#1{#1}\fi
\ifx\beditor   \undefined \def\beditor#1{#1}\fi
\ifx\bbtitle   \undefined \def\bbtitle#1{\textit{#1}}\fi
\ifx\bedition  \undefined \def\bedition#1{#1}\fi
\ifx\bseriesno \undefined \def\bseriesno#1{\textbf{#1}}\fi
\ifx\blocation \undefined \def\blocation#1{#1}\fi
\ifx\bsertitle \undefined \def\bsertitle#1{\textit{#1}}\fi
\ifx\bsnm      \undefined \def\bsnm#1{#1}\fi
\ifx\bsuffix   \undefined \def\bsuffix#1{#1}\fi
\ifx\bparticle \undefined \def\bparticle#1{#1}\fi
\ifx\barticle  \undefined \def\barticle#1{}\fi
\ifx\binstitute  \undefined \def\binstitute#1{#1}\fi
\ifx\bpublisher  \undefined \def\bpublisher#1{#1}\fi
\ifx\doiurl    \undefined \def\doiurl#1{\href{#1}{DOI}}\fi
\makeatletter
\def\safeHref#1#2#3{\in@{http}{#2}\ifin@\href{#2}{#3}\else\href{#1#2}{#3}\fi}
\makeatother
\ifx\adsurl    \undefined
  \def\adsurl#1{\safeHref{https://ui.adsabs.harvard.edu/abs/}{#1}{ADS}}\fi
\ifx\arxivurl  \undefined
  \def\arxivurl#1{\safeHref{http://arxiv.org/abs/}{#1}{arXiv}}\fi
\ifx\botherref \undefined \def\botherref#1{}\fi
\ifx\url       \undefined \def\url#1{#1}\fi
\ifx\bchapter  \undefined \def\bchapter#1{}\fi
\ifx\bbook     \undefined \def\bbook#1{}\fi
\ifx\bcomment  \undefined \def\bcomment#1{#1}\fi
\ifx\oauthor   \undefined \def\oauthor#1{#1}\fi
\ifx\citeauthoryear \undefined\def \citeauthoryear#1{#1}\fi
\def\endbibitem {}
\ifx\bconflocation  \undefined \def\bconflocation#1{#1} \fi

\bibitem[\protect\citeauthoryear{{Bail{\'e}n} et~al.}{2024}]{Bailen24}
\begin{barticle}
\bauthor{\bsnm{{Bail{\'e}n}}, \binits{F.J.}}, \betal:
\byear{2024},
\batitle{{Determination of the SO/PHI-HRT wavefront degradation using multiple
  defocused images}}.
\bjtitle{\aap}
\bvolume{681},
\bfpage{A58}.
\doiurl{https://doi.org/10.1051/0004-6361/202346019}.
\adsurl{2024A&A...681A..58B}.
\end{barticle}
\endbibitem

\bibitem[\protect\citeauthoryear{{Barthol} et~al.}{2011}]{Barthol11}
\begin{barticle}
\bauthor{\bsnm{{Barthol}}, \binits{P.}}, \betal:
\byear{2011},
\batitle{{The \sunrise{} Mission}}.
\bjtitle{\solphys}
\bvolume{268},
\bfpage{1}.
\doiurl{https://doi.org/10.1007/s11207-010-9662-9}.
\adsurl{2011SoPh..268....1B}.
\end{barticle}
\endbibitem

\bibitem[\protect\citeauthoryear{{Berkefeld} et~al.}{2025}]{Berkefeld25}
\begin{botherref}
\oauthor{\bsnm{{Berkefeld}}, \binits{T.}}, et al.:
2025,
\sunrisethree: The Wavefront Correction System.
\textit{\solphys}
\textbf{300}.
Part of the Topical Collection "The Sunrise III Solar Observatory".
\end{botherref}
\endbibitem

\bibitem[\protect\citeauthoryear{{Bernasconi} et~al.}{2025}]{Bernasconi25}
\begin{botherref}
\oauthor{\bsnm{{Bernasconi}}, \binits{P.}}, et al.:
2025,
The Gondola for the \sunrisethree{} Balloon-Borne Solar Observatory.
\textit{\solphys}
\textbf{300}.
Part of the Topical Collection "The Sunrise III Solar Observatory".
\end{botherref}
\endbibitem

\bibitem[\protect\citeauthoryear{{Bruns} et~al.}{2021}]{Bruns21}
\begin{bchapter}
\bauthor{\bsnm{{Bruns}}, \binits{S.}}, \betal:
\byear{2021},
\bctitle{{UV bandpass filters based on Ta$_2$O$_5$ and ZrO$_2$ for solar
  observation}}.
In: \beditor{\bsnm{Cugny}, \binits{B.}},
\beditor{\bsnm{Sodnik}, \binits{Z.}},
\beditor{\bsnm{Karafolas}, \binits{N.}} (eds.)
\bbtitle{International Conference on Space Optics — ICSO 2020}
\bseriesno{11852},
\bpublisher{SPIE},
\bfpage{118521P}.
\doiurl{https://doi.org/10.1117/12.2599291}.
\end{bchapter}
\endbibitem

\bibitem[\protect\citeauthoryear{{Cox}}{1999}]{Cox99}
\begin{bbook}
\bauthor{\bsnm{{Cox}}, \binits{A.N.}}:
\byear{1999},
\bbtitle{{Allen's astrophysical quantities}},
\bedition{4}th edn.
\bpublisher{{Springer}},
\blocation{{New York}}.
\end{bbook}
\endbibitem

\bibitem[\protect\citeauthoryear{{de Wijn}}{2018}]{Dewijn2019}
\begin{barticle}
\bauthor{\bsnm{{de Wijn}}, \binits{A.G.}}:
\byear{2018},
\batitle{Characterization of Cameras for the COSMO K-coronagraph}.
\bjtitle{\aj}
\bvolume{157},
\bfpage{8}.
\doiurl{https://doi.org/10.3847/1538-3881/aaedc0}.
\end{barticle}
\endbibitem

\bibitem[\protect\citeauthoryear{{del Toro Iniesta}
  et~al.}{2025}]{delToroIniesta25}
\begin{botherref}
\oauthor{\bsnm{{del Toro Iniesta}}, \binits{J.C.}}, et al.:
2025,
TuMag: the tunable magnetograph for the \sunrisethree{} mission.
\textit{\solphys}
\textbf{300}.
Part of the Topical Collection "The Sunrise III Solar Observatory".
\end{botherref}
\endbibitem

\bibitem[\protect\citeauthoryear{{Delbouille} and
  {Roland}}{1995}]{Delbouille95}
\begin{bchapter}
\bauthor{\bsnm{{Delbouille}}, \binits{L.}},
\bauthor{\bsnm{{Roland}}, \binits{C.}}:
\byear{1995},
\bctitle{{Jungfraujoch Solar Atlases}}.
In: \beditor{\bsnm{{Sauval}}, \binits{A.J.}},
\beditor{\bsnm{{Blomme}}, \binits{R.}},
\beditor{\bsnm{{Grevesse}}, \binits{N.}} (eds.)
\bbtitle{Laboratory and Astronomical High Resolution Spectra},
\bsertitle{ASP Conference Series}
\bseriesno{81},
\bfpage{32}.
\adsurl{1995ASPC...81...32D}.
\end{bchapter}
\endbibitem

\bibitem[\protect\citeauthoryear{{Delbouille}, {Roland}, and
  {Neven}}{1973}]{Delbouille73}
\begin{bbook}
\bauthor{\bsnm{{Delbouille}}, \binits{L.}},
\bauthor{\bsnm{{Roland}}, \binits{G.}},
\bauthor{\bsnm{{Neven}}, \binits{L.}}:
\byear{1973},
\bbtitle{Atlas photometrique du spectre solaire de $\lambda=3000$~\AA{} \`a
  $\lambda=10000$~\AA},
\bpublisher{Institut d'Astrophysique de l'Universit{\'e} de Li{\`e}ge},
\blocation{Li{\`e}ge, Belgium}.
\adsurl{1973apds.book.....D}.
\end{bbook}
\endbibitem

\bibitem[\protect\citeauthoryear{{Fernández-Soler}
  et~al.}{2020}]{FernandezSoler20}
\begin{bchapter}
\bauthor{\bsnm{{Fernández-Soler}}, \binits{A.}}, \betal:
\byear{2020},
\bctitle{{Thermal Analysis of SUSI-O on \sunrisethree}}.
In: \bbtitle{Proceedings for the 2020 International Conference on Environmental
  Systems}.
\end{bchapter}
\endbibitem

\bibitem[\protect\citeauthoryear{{Gandorfer}}{2005}]{Gandorfer05}
\begin{bbook}
\bauthor{\bsnm{{Gandorfer}}, \binits{A.}}:
\byear{2005},
\bbtitle{{The Second Solar Spectrum: A high spectral resolution polarimetric
  survey of scattering polarization at the solar limb in graphical
  representation. Volume III: 3160 {\r{A}} to 3915 {\r{A}}}},
\bpublisher{vdf Hochschulverlag},
\blocation{Zurich, Switzerland}.
\adsurl{2005sss..book.....G}.
\end{bbook}
\endbibitem

\bibitem[\protect\citeauthoryear{{Gandorfer} et~al.}{2004}]{Gandorfer04}
\begin{barticle}
\bauthor{\bsnm{{Gandorfer}}, \binits{A.}}, \betal:
\byear{2004},
\batitle{{Solar polarimetry in the near UV with the Zurich Imaging Polarimeter
  ZIMPOL II}}.
\bjtitle{\aap}
\bvolume{422},
\bfpage{703}.
\doiurl{https://doi.org/10.1051/0004-6361:20040254}.
\adsurl{2004A&A...422..703G}.
\end{barticle}
\endbibitem

\bibitem[\protect\citeauthoryear{{Gandorfer} et~al.}{2011}]{Gandorfer11}
\begin{barticle}
\bauthor{\bsnm{{Gandorfer}}, \binits{A.}}, \betal:
\byear{2011},
\batitle{{The Filter Imager SuFI and the Image Stabilization and Light
  Distribution System ISLiD of the \sunrise{} Balloon-Borne Observatory:
  Instrument Description}}.
\bjtitle{\solphys}
\bvolume{268},
\bfpage{35}.
\doiurl{https://doi.org/10.1007/s11207-010-9636-y}.
\adsurl{2011SoPh..268...35G}.
\end{barticle}
\endbibitem

\bibitem[\protect\citeauthoryear{{Hirzberger} et~al.}{2010}]{Hirzberger10}
\begin{barticle}
\bauthor{\bsnm{{Hirzberger}}, \binits{J.}}, \betal:
\byear{2010},
\batitle{{Quiet-sun Intensity Contrasts in the Near-ultraviolet as Measured
  from \sunrise}}.
\bjtitle{\apjl}
\bvolume{723},
\bfpage{L154}.
\doiurl{https://doi.org/10.1088/2041-8205/723/2/L154}.
\adsurl{2010ApJ...723L.154H}.
\end{barticle}
\endbibitem

\bibitem[\protect\citeauthoryear{{Hölken} et~al.}{2024}]{Hoelken23}
\begin{barticle}
\bauthor{\bsnm{{Hölken}}, \binits{J.}}, \betal:
\byear{2024},
\batitle{Spectroflat: A generic spectrum and flat-field calibration library for
  spectro-polarimetric data}.
\bjtitle{\aap}
\bvolume{687},
\bfpage{A22}.
\doiurl{https://doi.org/10.1051/0004-6361/202348877}.
\end{barticle}
\endbibitem

\bibitem[\protect\citeauthoryear{{Ichimoto} et~al.}{2008}]{Ichimoto08}
\begin{barticle}
\bauthor{\bsnm{{Ichimoto}}, \binits{K.}},
\bauthor{\bsnm{{Lites}}, \binits{B.}},
\bauthor{\bsnm{{Elmore}}, \binits{D.}},
\bauthor{\bsnm{{Suematsu}}, \binits{Y.}},
\bauthor{\bsnm{{Tsuneta}}, \binits{S.}},
\bauthor{\bsnm{{Katsukawa}}, \binits{Y.}},
\bauthor{\bsnm{{Shimizu}}, \binits{T.}},
\bauthor{\bsnm{{Shine}}, \binits{R.}},
\bauthor{\bsnm{{Tarbell}}, \binits{T.}},
\bauthor{\bsnm{{Title}}, \binits{A.}},
\bauthor{\bsnm{{Kiyohara}}, \binits{J.}},
\bauthor{\bsnm{{Shinoda}}, \binits{K.}},
\bauthor{\bsnm{{Card}}, \binits{G.}},
\bauthor{\bsnm{{Lecinski}}, \binits{A.}},
\bauthor{\bsnm{{Streander}}, \binits{K.}},
\bauthor{\bsnm{{Nakagiri}}, \binits{M.}},
\bauthor{\bsnm{{Miyashita}}, \binits{M.}},
\bauthor{\bsnm{{Noguchi}}, \binits{M.}},
\bauthor{\bsnm{{Hoffmann}}, \binits{C.}},
\bauthor{\bsnm{{Cruz}}, \binits{T.}}:
\byear{2008},
\batitle{{Polarization Calibration of the Solar Optical Telescope onboard
  Hinode}}.
\bjtitle{\solphys}
\bvolume{249},
\bfpage{233}.
\doiurl{https://doi.org/10.1007/s11207-008-9169-9}.
\adsurl{2008SoPh..249..233I}.
\end{barticle}
\endbibitem

\bibitem[\protect\citeauthoryear{{Iglesias} et~al.}{2025}]{Iglesias25}
\begin{botherref}
\oauthor{\bsnm{{Iglesias}}, \binits{F.A.}}, et al.:
2025,
The \sunrise{} Ultraviolet Spectropolarimeter and Imager: standalone
  polarimetric calibration.
\textit{\solphys}
\textbf{300}.
Part of the Topical Collection "The Sunrise III Solar Observatory".
\end{botherref}
\endbibitem

\bibitem[\protect\citeauthoryear{{Iglesias} and {Feller}}{2019}]{Iglesias19}
\begin{barticle}
\bauthor{\bsnm{{Iglesias}}, \binits{F.A.}},
\bauthor{\bsnm{{Feller}}, \binits{A.}}:
\byear{2019},
\batitle{{Instrumentation for solar spectropolarimetry: state of the art and
  prospects}}.
\bjtitle{\opteng}
\bvolume{58},
\bfpage{082417}.
\doiurl{https://doi.org/10.1117/1.OE.58.8.082417}.
\adsurl{2019OptEn..58h2417I}.
\end{barticle}
\endbibitem

\bibitem[\protect\citeauthoryear{{Ishikawa} et~al.}{2015}]{Ishikawa15}
\begin{barticle}
\bauthor{\bsnm{{Ishikawa}}, \binits{S.}}, \betal:
\byear{2015},
\batitle{{Development of a Precise Polarization Modulator for UV
  Spectropolarimetry}}.
\bjtitle{\solphys}
\bvolume{290},
\bfpage{3081}.
\doiurl{https://doi.org/10.1007/s11207-015-0774-0}.
\adsurl{2015SoPh..290.3081I}.
\end{barticle}
\endbibitem

\bibitem[\protect\citeauthoryear{{Katsukawa} et~al.}{2025}]{Katsukawa25}
\begin{botherref}
\oauthor{\bsnm{{Katsukawa}}, \binits{Y.}}, et al.:
2025,
The Sunrise Chromospheric Infrared Spectro-Polarimeter SCIP: an Instrument for
  \sunrisethree.
\textit{\solphys}
\textbf{300}.
Part of the Topical Collection "The Sunrise III Solar Observatory".
\end{botherref}
\endbibitem

\bibitem[\protect\citeauthoryear{{Keller}}{1996}]{Keller96}
\begin{barticle}
\bauthor{\bsnm{{Keller}}, \binits{C.U.}}:
\byear{1996},
\batitle{{Recent Progress in Imaging Polarimetry}}.
\bjtitle{\solphys}
\bvolume{164},
\bfpage{243}.
\doiurl{https://doi.org/10.1007/BF00146637}.
\adsurl{1996SoPh..164..243K}.
\end{barticle}
\endbibitem

\bibitem[\protect\citeauthoryear{{Korpi-Lagg} et~al.}{2025}]{Lagg25}
\begin{botherref}
\oauthor{\bsnm{{Korpi-Lagg}}, \binits{A.}}, et al.:
2025,
\sunrisethree: Overview of Observatory and Instruments.
\textit{\solphys}
\textbf{300}.
Part of the Topical Collection "The Sunrise III Solar Observatory".
\end{botherref}
\endbibitem

\bibitem[\protect\citeauthoryear{{Kurucz} et~al.}{1984}]{kurucz84}
\begin{bbook}
\bauthor{\bsnm{{Kurucz}}, \binits{R.L.}}, \betal:
\byear{1984},
\bbtitle{{Solar flux atlas from 296 to 1300 nm}}.
\adsurl{1984sfat.book.....K}.
\end{bbook}
\endbibitem

\bibitem[\protect\citeauthoryear{{L{\"o}fdahl} and
  {Scharmer}}{1994}]{loefdahl94}
\begin{barticle}
\bauthor{\bsnm{{L{\"o}fdahl}}, \binits{M.G.}},
\bauthor{\bsnm{{Scharmer}}, \binits{G.B.}}:
\byear{1994},
\batitle{{Wavefront sensing and image restoration from focused and defocused
  solar images.}}
\bjtitle{\aaps}
\bvolume{107},
\bfpage{243}.
\adsurl{1994A&AS..107..243L}.
\end{barticle}
\endbibitem

\bibitem[\protect\citeauthoryear{{Neckel}}{1999}]{Neckel99}
\begin{barticle}
\bauthor{\bsnm{{Neckel}}, \binits{H.}}:
\byear{1999},
\batitle{Announcement: Spectral Atlas of Solar Absolute Disk-averaged and
  Disk-center Intensity From 3290 to 12510~\AA{} (Brault and Neckel, 1987) Now
  Available From Hamburg Observatory Anonymous FTP Site}.
\bjtitle{\solphys}
\bvolume{184},
\bfpage{421}.
\doiurl{https://doi.org/10.1023/A:1017165208013}.
\adsurl{1999SoPh..184..421N}.
\end{barticle}
\endbibitem

\bibitem[\protect\citeauthoryear{{Riethm{\"u}ller} and
  {Solanki}}{2019}]{Riethmueller19}
\begin{barticle}
\bauthor{\bsnm{{Riethm{\"u}ller}}, \binits{T.L.}},
\bauthor{\bsnm{{Solanki}}, \binits{S.K.}}:
\byear{2019},
\batitle{{The potential of many-line inversions of photospheric
  spectropolarimetric data in the visible and near UV}}.
\bjtitle{\aap}
\bvolume{622},
\bfpage{A36}.
\doiurl{https://doi.org/10.1051/0004-6361/201833379}.
\adsurl{2019A&A...622A..36R}.
\end{barticle}
\endbibitem

\bibitem[\protect\citeauthoryear{{Sant}}{2022}]{Sant22}
\begin{botherref}
\oauthor{\bsnm{{Sant}}, \binits{K.}}:
2022,
{CMOS Sensors and Polarimetric Studies on the Sun}.
PhD thesis,
Technische Universit{\"a}t Braunschweig.
\end{botherref}
\endbibitem

\bibitem[\protect\citeauthoryear{{Sayfan}}{2005}]{Sayfan05}
\begin{barticle}
\bauthor{\bsnm{{Sayfan}}, \binits{G.}}:
\byear{2005},
\batitle{{Method Call Interception. Entering and leaving method calls during
  development}}.
\bjtitle{C/C++ Users Journal}
\bvolume{4},
\bfpage{32}.
\end{barticle}
\endbibitem

\bibitem[\protect\citeauthoryear{{Schmidt} and {Huston}}{2001}]{Schmidt01}
\begin{bbook}
\bauthor{\bsnm{{Schmidt}}, \binits{D.C.}},
\bauthor{\bsnm{{Huston}}, \binits{S.D.}}:
\byear{2001},
\bbtitle{{C++ Network Programming, Volume 1: Mastering Complexity with ACE and
  Patterns}},
\bpublisher{Addison-Wesley},
\blocation{Amsterdam}.
\end{bbook}
\endbibitem

\bibitem[\protect\citeauthoryear{{Schmidt} and {Huston}}{2002}]{Schmidt02}
\begin{bbook}
\bauthor{\bsnm{{Schmidt}}, \binits{D.C.}},
\bauthor{\bsnm{{Huston}}, \binits{S.D.}}:
\byear{2002},
\bbtitle{{C++ Network Programming, Volume 2: Systematic Reuse with ACE and
  Frameworks}},
\bpublisher{Addison-Wesley},
\blocation{Amsterdam}.
\end{bbook}
\endbibitem

\bibitem[\protect\citeauthoryear{{Schmidt} et~al.}{2000}]{Schmidt00}
\begin{bbook}
\bauthor{\bsnm{{Schmidt}}, \binits{D.C.}},
\bauthor{\bsnm{{Stal}}, \binits{M.}},
\bauthor{\bsnm{{Rohnert}}, \binits{H.}},
\bauthor{\bsnm{{Buschmann}}, \binits{F.}}:
\byear{2000},
\bbtitle{{Pattern-Oriented Software Architecture: Volume 2: Patterns for
  Concurrent and Distributed Objects}},
\bpublisher{John Wiley \& Sons},
\blocation{West-Sussex}.
\end{bbook}
\endbibitem

\bibitem[\protect\citeauthoryear{{Shine} and {Majani}}{1992}]{Shine92}
\begin{barticle}
\bauthor{\bsnm{{Shine}}, \binits{R.A.}},
\bauthor{\bsnm{{Majani}}, \binits{E.E.}}:
\byear{1992},
\batitle{{Data Compression Experiments with High Resolution Solar Images}}.
\bjtitle{Am. Astron. Soc. Meeting Abstracts}
\bvolume{181},
\bfpage{81.13}.
\adsurl{1992AAS...181.8113S}.
\end{barticle}
\endbibitem

\bibitem[\protect\citeauthoryear{{Solanki} et~al.}{2010}]{Solanki10}
\begin{barticle}
\bauthor{\bsnm{{Solanki}}, \binits{S.K.}}, \betal:
\byear{2010},
\batitle{{\sunrise: Instrument, Mission, Data, and First Results}}.
\bjtitle{\apjl}
\bvolume{723},
\bfpage{L127}.
\doiurl{https://doi.org/10.1088/2041-8205/723/2/L127}.
\adsurl{2010ApJ...723L.127S}.
\end{barticle}
\endbibitem

\bibitem[\protect\citeauthoryear{{Solanki} et~al.}{2017}]{Solanki17}
\begin{barticle}
\bauthor{\bsnm{{Solanki}}, \binits{S.K.}}, \betal:
\byear{2017},
\batitle{{The Second Flight of the \sunrise{} Balloon-borne Solar Observatory:
  Overview of Instrument Updates, the Flight, the Data, and First Results}}.
\bjtitle{\apjs}
\bvolume{229},
\bfpage{2}.
\doiurl{https://doi.org/10.3847/1538-4365/229/1/2}.
\adsurl{2017ApJS..229....2S}.
\end{barticle}
\endbibitem

\bibitem[\protect\citeauthoryear{{Stevens}}{1997}]{Stevens97}
\begin{bbook}
\bauthor{\bsnm{{Stevens}}, \binits{W.R.}}:
\byear{1997},
\bbtitle{{UNIX Network Programming: Volume 1: Network APIs: Sockets and XTI --
  2nd ed.}},
\bpublisher{{Prentice-Hall}},
\blocation{{London}}.
\end{bbook}
\endbibitem

\bibitem[\protect\citeauthoryear{{Thuillier} et~al.}{2004}]{Thuillier04}
\begin{bchapter}
\bauthor{\bsnm{{Thuillier}}, \binits{G.}}, \betal:
\byear{2004},
\bctitle{{Solar Irradiance Reference Spectra}}.
In: \beditor{\bsnm{{Pap}}, \binits{J.M.}},
\beditor{\bsnm{{Fox}}, \binits{P.}},
\beditor{\bsnm{{Frohlich}}, \binits{C.}},
\beditor{\bsnm{{Hudson}}, \binits{H.S.}},
\beditor{\bsnm{{Kuhn}}, \binits{J.}},
\beditor{\bsnm{{McCormack}}, \binits{J.}},
\beditor{\bsnm{{North}}, \binits{G.}},
\beditor{\bsnm{{Sprigg}}, \binits{W.}},
\beditor{\bsnm{{Wu}}, \binits{S.T.}} (eds.)
\bbtitle{Solar Variability and its Effects on Climate},
\bsertitle{Geophysical Monograph Series}
\bseriesno{141},
\bfpage{171}.
\doiurl{https://doi.org/10.1029/141GM13}.
\adsurl{2004GMS...141..171T}.
\end{bchapter}
\endbibitem

\bibitem[\protect\citeauthoryear{{van Noort}}{2017}]{vanNoort17}
\begin{barticle}
\bauthor{\bsnm{{van Noort}}, \binits{M.}}:
\byear{2017},
\batitle{{Image restoration of solar spectra}}.
\bjtitle{\aap}
\bvolume{608},
\bfpage{A76}.
\doiurl{https://doi.org/10.1051/0004-6361/201731339}.
\adsurl{2017A&A...608A..76V}.
\end{barticle}
\endbibitem

\bibitem[\protect\citeauthoryear{{Van Noort}, {Rouppe Van Der Voort}, and
  {L{\"o}fdahl}}{2005}]{vanNoort05}
\begin{barticle}
\bauthor{\bsnm{{Van Noort}}, \binits{M.}},
\bauthor{\bsnm{{Rouppe Van Der Voort}}, \binits{L.}},
\bauthor{\bsnm{{L{\"o}fdahl}}, \binits{M.G.}}:
\byear{2005},
\batitle{{Solar Image Restoration By Use Of Multi-frame Blind De-convolution
  With Multiple Objects And Phase Diversity}}.
\bjtitle{\solphys}
\bvolume{228},
\bfpage{191}.
\doiurl{https://doi.org/10.1007/s11207-005-5782-z}.
\adsurl{2005SoPh..228..191V}.
\end{barticle}
\endbibitem

\bibitem[\protect\citeauthoryear{{Vukadinovi\'{c}}
  et~al.}{2024}]{vukadinovic24}
\begin{botherref}
\oauthor{\bsnm{{Vukadinovi\'{c}}}, \binits{D.}}, et al.:
2024,
globin: A spectropolarimetric inversion code for the coupled inference of
  atomic line parameters.
\textit{\aap}.
\doiurl{https://doi.org/10.1051/0004-6361/202347752}.
\end{botherref}
\endbibitem

\bibitem[\protect\citeauthoryear{{Wells}, {Greisen}, and
  {Harten}}{1981}]{Wells81}
\begin{barticle}
\bauthor{\bsnm{{Wells}}, \binits{D.C.}},
\bauthor{\bsnm{{Greisen}}, \binits{E.W.}},
\bauthor{\bsnm{{Harten}}, \binits{R.H.}}:
\byear{1981},
\batitle{{FITS - a Flexible Image Transport System}}.
\bjtitle{\aaps}
\bvolume{44},
\bfpage{363}.
\adsurl{1981A&AS...44..363W}.
\end{barticle}
\endbibitem

\end{thebibliography}

\end{article} 

\newpage

\appendix
\section{Acronyms}
%!TEX root = main.tex
\begin{acronym}
\setlength{\itemsep}{-4ex}

%instruments
\acro{susi}[\textsc{SUSI}]{\sunrise{} UV Spectropolarimeter and Imager}
\acro{imax}[\textsc{IMaX}]{Imaging Magnetograph eXperiment}
\acro{tumag}[\textsc{TuMag}]{Tunable Magnetograph}
\acro{scip}[\textsc{SCIP}]{\sunrise{} Chromospheric Infrared spectro-Polarimeter}
\acro{sufi}[\textsc{SuFI}]{\sunrise{} Filtergraph and Imager}
\acro{pfi}[PFI]{Post Focus Instrumentation platform}
\acro{islid}[\textsc{ISLiD}]{Image Stabilization and Light Distribution unit}
\acro{cws}[CWS]{Correlating Wavefront Sensor}
\acro{icu}[ICU]{Instrument Control Unit}
\acro{ics}[ICS]{Instrument Control System}
\acro{dss}[DSS]{Data Storage System}
\acro{sgt}[SGT]{Sun Guider Telescope}
\acro{imu}[IMU]{Inertial Measurement Unit}
\acro{ps}[PS]{Pointing System}
\acro{elink}[E-Link]{Esrange Airborne Data Link}
\acro{evtm}[EVTM]{Ethernet Via Telemetry}
\acro{tdrss}[TDRSS]{Tracking \& Data Relay Satellite System}
\acro{starlink}[Starlink]{Starlink (a division of SpaceX)}
\acro{iridium}[Iridium Pilot]{Iridium Pilot\textsuperscript{\textregistered}}
\acro{gusto}[GUSTO]{Galactic/Extragalactic ULDB Spectroscopic Terahertz Observatory}
\acro{pi}[PI]{Principal Investigator}
\acro{coi}[CoI]{Co-Investigator}
\acro{goc}[GOC]{Göttingen Operations Center}
\acro{eoc}[EOC]{Esrange Operations Center}
\acro{occ}[OCC]{Operations Control Center}
\acro{vpn}[VPN]{virtual private network}
\acro{mhd}[MHD]{magneto-hydrodynamic}
\acro{MHD}[MHD]{Magneto-Hydrodynamic}
\acro{muram}[MURaM]{MPS/University of Chicago Radiative MHD}
\acro{muramce}[MURaM-CE]{MURaM Chromospheric Extension}
\acro{dkist}[DKIST]{Daniel K. Inouye Solar Telescope}
\acro{sdo}[SDO]{Solar Dynamics Observatory}
\acro{sswg}[SSWG]{\sunrise{} Science Working Group}
\acro{nvst}[NVST]{New Vacuum Solar Telescope}
\acro{gregor}[GREGOR]{GREGOR}
\acro{gst}[GST]{Goode Solar Telescope}
\acro{bbso}[BBSO]{Big Bear Solar Observatory} 
\acro{vtt}[VTT]{German Vacuum Tower Telescope} 
\acro{smart}[SMART]{Solar Magnetic Activity Research Telescope}
\acro{dst}[DST]{Dunn Solar Telescope}
\acro{sst}[SST]{Swedish Solar Telescope}
\acro{alma}[ALMA]{Atacama Large Millimeter/sub-millimeter Array}
\acro{mast}[MAST]{Multi-Application Solar Telescope}
\acro{imtek}[IMTEK]{Department of Microsystems Engineering}
\acro{iris}[IRIS]{Interface Region Imaging Spectrograph}
\acro{hmi}[HMI]{Helioseismic Magnetic Imager}
\acro{phi}[PHI]{Polarimetric and Helioseismic Imager}
\acro{sophi}[SO/PHI]{Solar Orbiter Polarimetric and Helioseismic Imager}
\acro{solo}[SolO]{Solar Orbiter}
\acro{sot}[SOT]{Solar Optical Telescope}
\acro{sotsp}[SP]{Spectro-Polarimeter}
\acro{sotnfi}[NFI]{Narrowband Filter Imager}
\acro{ramon}[RAMON]{RAdiation MONitor}
\acro{fits}[FITS]{Flexible Image Transport System}
\acro{lcvr}[LCVR]{liquid crystal variable retarder}
\acro{psf}[PSF]{point spread function}
\acro{wfs}[WFS]{wavefront sensor}
\acro{ct}[CT]{correlation tracker}
\acro{smm}[SMM]{scan mirror mechanism}
\acro{olr}[OLR]{outgoing longwave radiation}
\acro{led}[LED]{light-emitting diode}
\acro{lldpe}[LLDPE]{linear low-density polyethylene}
\acro{hrw}[HRW]{Heat Rejection Wedge}
\acro{erack}[E-rack]{electronic rack}
\acro{sip}[SIP]{Support Instrumentation Package}
\acro{sipm}[SiPM]{Silicon PhotoMultiplier}
\acro{gps}[GPS]{Global Positioning System}
\acro{ssm}[SSM]{second surface mirror}
\acro{tmtc}[TM/TC]{Telemetry and Telecommand}
\acro{tm}[TM]{Telemetry}
\acro{tc}[TC]{Telecommand}
\acro{mli}[MIL]{multilayer insulation}
\acro{vda}[VDA]{vapor deposited aluminum}
\acro{hoc}[HOC]{hot operational case}
\acro{coc}[COC]{cold operational case}
\acro{cfrp}[CFRP]{carbon fiber-reinforced polymer}
\acro{cmos}[CMOS]{complementary metal oxide semiconductor}
\acro{hnoc}[HNOC]{hot non-operational case}
\acro{cnoc}[CNOC]{cold non-operational case}
\acro{fps}[fps]{frames per second}
\acro{rtc}[RTC]{real-time clock}
\acro{pcb}[PCB]{printed circuit board}
\acro{fsm}[FSM]{finite state machine}
\acro{ao}[AO]{adaptive optics}
\acro{ssd}[SSD]{solid state disk}
\acro{ldb}[LDB]{long-distance balloon}
\acro{pbs}[PBS]{polarising beam-splitter}
\acro{pmu}[PMU]{Polarization Modulation Unit}
\acro{pmu-rot}[PMU-ROT]{PMU rotating waveplate}
\acro{scu}[SCU]{\textsc{SUSI} Controller Unit}
\acro{egse}[EGSE]{electrical ground support equipment}
\acro{hvps}[HVPS]{high voltage power supply}
\acro{ace}[ACE]{Adaptive Communication Environment}
\acro{ld}[LD]{launch date}
\acro{pcu}[PCU]{Power Converter Unit}
\acro{pdu}[PDU]{Power Distribution Unit}
\acro{poe}[PoE]{Power over Ethernet}
\acro{ppd}[PPD]{PFI Power Distribution Unit}
\acro{mtc}[MTC]{Main Telescope Controller}
\acro{ss}[SS]{Science Stack}
\acro{ssh}[SSH]{Secure Shell}
\acro{nic}[NIC]{network interface controller}
\acro{sp}[SP]{Spectrograph}
\acro{osf}[OSF]{Order-sorting Filter}
\acro{sj}[SJ]{Slit-Jaw}
\acro{pid}[PID]{Phase-Diversity Image Doubler}
\acro{o-unit}[O-unit]{optics unit}
\acro{atlas}[ATLAS]{Atmospheric Laboratory for Applications and Science}
\acro{fpga}[FPGA]{field programmable gate array}
\acro{lvds}[LVDS]{low voltage differential signaling}
\acro{dn}[DN]{digital numbers}
\acro{adc}[ADC]{analog to digital converter}
\acro{mtf}[MTF]{modulation transfer function}
\acro{fts}[FTS]{Fourier-transform spectrograph}
\acro{lsf}[LSF]{line spread function}
\acro{roi}[ROI]{region of interest}
\acro{psg}[PSG]{polarization state generator}

%institutes
\acro{mpae}[MPAe]{Max-Planck-Institut für Aeronomie}
\acro{mps}[MPS]{Max Planck Institute for Solar System Research}
\acro{mpg}[MPG]{Max-Planck-Gesellschaft}
\acro{apl}[JHUAPL]{Johns Hopkins University Applied Physics Laboratory}
\acro{nasa}[NASA]{National Aeronautics and Space Administration}
\acro{kis}[KIS]{Institut für Sonnenphysik}
\acro{iaa}[IAA]{Instituto de Astrofísica de Andalucía}
\acro{iac}[IAC]{Instituto de Astrofísica de Canarias}
\acro{inta}[INTA]{Instituto Nacional de Técnica Aeroespacial}
\acro{kbsi}[KBSI]{Korea Basic Science Institute}
\acro{unival}[UV]{Universitat de València}
\acro{upm}[UPM]{Universidad Politécnica de Madrid}
\acro{s3pc}[S$^3$PC]{Spanish Space Solar Physics Consortium}
\acro{jaxa}[JAXA]{Japan Aerospace Exploration Agency}
\acro{naoj}[NAOJ]{National Astronomical Observatory of Japan}
\acro{ssc}[SSC]{Swedish Space Corporation}
\acro{mit}[MIT]{Massachusetts Institute of Technology}
\acro{esrange}[Esrange]{Esrange Space Center}
\acro{csbf}[CSBF]{Columbia Scientific Ballooning Facility}
\acro{bpo}[BPO]{Balloon Program Office}

%other
\acro{2d}[2D]{two-dimensional}
\acro{usaf}[USAF]{U.S. Air Force}
\acro{mrr}[MRR]{Mission Readiness Review}
\acro{frr}[FRR]{Flight Readiness Review}
\acro{ssct}[SSCT]{Science/Support Compatibility Test}    
\acro{uv}[UV]{ultraviolet}
\acro{nuv}[NUV]{near-ultraviolet}
\acro{ir}[IR]{infrared}
\acro{nir}[near-IR]{near-infrared}
\acro{los}[LoS]{line-of-sight}
\acro{fov}[FoV]{field-of-view}
\acro{rms}[rms]{root-mean-square}
\acro{s2n}[S/N]{signal-to-noise}
\acro{lte}[LTE]{local thermodynamic equilibrium}
\acro{nlte}[non-LTE]{non-local thermodynamic equilibrium}
\acro{tv}[TV]{thermal-vacuum}
\acro{fram}[FRAM]{ferroelectric random access memory}
\acro{fwhm}[FWHM]{full width at half maximum}
\acro{cwl}[CWL]{central wavelength}
\acro{od}[OD]{optical density}
\acro{momfbd}[MOMFBD]{Multi-Object Multi-Frame Blind Deconvolution}

\end{acronym}

\end{document}